\documentclass[aps,prl,twocolumn,superscriptaddress,nofootinbib]{revtex4-1}

\usepackage{empheq}
\usepackage{gensymb}
\usepackage{graphicx}
\usepackage{dcolumn}
\usepackage{bm}
\usepackage{adjustbox}
\usepackage{amssymb}   
\usepackage{hyperref}
\usepackage[mathlines]{lineno}
\usepackage[caption=false]{subfig}
\usepackage{xcolor}
\usepackage{subfig}
\usepackage{amsmath}
\usepackage{subfiles}
\usepackage{placeins}
\usepackage{array}
\usepackage{titlesec}
\usepackage[normalem]{ulem}

\hypersetup{
    pdfnewwindow=true,      
    colorlinks=true,       
    linkcolor=blue,          
    citecolor=blue,        
    filecolor=blue,      
    urlcolor=blue           
}

\begin{document}

\title{Neutrino-induced coherent $\pi^{+}$ production in C, CH, Fe and Pb at $\langle E_{\nu}\rangle \sim 6$ GeV}

\newcommand{\Rutgers}{Rutgers, The State University of New Jersey, Piscataway, New Jersey 08854, USA}
\newcommand{\Hampton}{Hampton University, Dept. of Physics, Hampton, VA 23668, USA}
\newcommand{\Dortmund}{Institute of Physics, Dortmund University, 44221, Germany }
\newcommand{\Otterbein}{Department of Physics, Otterbein University, 1 South Grove Street, Westerville, OH, 43081 USA}
\newcommand{\JMU}{James Madison University, Harrisonburg, Virginia 22807, USA}
\newcommand{\Florida}{University of Florida, Department of Physics, Gainesville, FL 32611}
\newcommand{\UCIrvine}{Department of Physics and Astronomy, University of California, Irvine, Irvine, California 92697-4575, USA}
\newcommand{\CBPF}{Centro Brasileiro de Pesquisas F\'{i}sicas, Rua Dr. Xavier Sigaud 150, Urca, Rio de Janeiro, Rio de Janeiro, 22290-180, Brazil}
\newcommand{\PUCP}{Secci\'{o}n F\'{i}sica, Departamento de Ciencias, Pontificia Universidad Cat\'{o}lica del Per\'{u}, Apartado 1761, Lima, Per\'{u}}
\newcommand{\INRM}{Institute for Nuclear Research of the Russian Academy of Sciences, 117312 Moscow, Russia}
\newcommand{\Jlab}{Jefferson Lab, 12000 Jefferson Avenue, Newport News, VA 23606, USA}
\newcommand{\Pittsburgh}{Department of Physics and Astronomy, University of Pittsburgh, Pittsburgh, Pennsylvania 15260, USA}
\newcommand{\Guanajuato}{Campus Le\'{o}n y Campus Guanajuato, Universidad de Guanajuato, Lascurain de Retana No. 5, Colonia Centro, Guanajuato 36000, Guanajuato M\'{e}xico.}
\newcommand{\Athens}{Department of Physics, University of Athens, GR-15771 Athens, Greece}
\newcommand{\Tufts}{Physics Department, Tufts University, Medford, Massachusetts 02155, USA}
\newcommand{\WM}{Department of Physics, William \& Mary, Williamsburg, Virginia 23187, USA}
\newcommand{\FNAL}{Fermi National Accelerator Laboratory, Batavia, Illinois 60510, USA}
\newcommand{\Purdue}{Department of Chemistry and Physics, Purdue University Calumet, Hammond, Indiana 46323, USA}
\newcommand{\MCLA}{Massachusetts College of Liberal Arts, 375 Church Street, North Adams, MA 01247}
\newcommand{\UMD}{Department of Physics, University of Minnesota -- Duluth, Duluth, Minnesota 55812, USA}
\newcommand{\Northwestern}{Northwestern University, Evanston, Illinois 60208}
\newcommand{\UNI}{Facultad de Ciencias, Universidad Nacional de Ingenier\'{i}a, Apartado 31139, Lima, Per\'{u}}
\newcommand{\Rochester}{Department of Physics and Astronomy, University of Rochester, Rochester, New York 14627 USA}
\newcommand{\Austin}{Department of Physics, University of Texas, 1 University Station, Austin, Texas 78712, USA}
\newcommand{\USM}{Departamento de F\'{i}sica, Universidad T\'{e}cnica Federico Santa Mar\'{i}a, Avenida Espa\~{n}a 1680 Casilla 110-V, Valpara\'{i}so, Chile}
\newcommand{\Geneva}{University of Geneva, 1211 Geneva 4, Switzerland}
\newcommand{\Chicago}{Enrico Fermi Institute, University of Chicago, Chicago, IL 60637 USA}
\newcommand{\hired}{}
\newcommand{\OregonState}{Department of Physics, Oregon State University, Corvallis, Oregon 97331, USA}
\newcommand{\oxford}{Oxford University, Department of Physics, Oxford, OX1 3PJ United Kingdom}
\newcommand{\umiss}{University of Mississippi, Oxford, Mississippi 38677, USA}
\newcommand{\upenn}{Department of Physics and Astronomy, University of Pennsylvania, Philadelphia, PA 19104}
\newcommand{\AMU}{Department of Physics, Aligarh Muslim University, Aligarh, Uttar Pradesh 202002, India}
\newcommand{\wroclaw}{University of Wroclaw, plac Uniwersytecki 1, 50-137 Wroa\l{}aw, Poland}
\newcommand{\Mohali}{Department of Physical Sciences, IISER Mohali, Knowledge City, SAS Nagar, Mohali - 140306, Punjab, India}
\newcommand{\CINVESTAV}{Departamento de Fisica Col. San Pedro Zacatenco, 07360 Mexico, DF, Av. Instituto PolitÃ©cnico Nacional, Mexico}
\newcommand{\york}{York University, Department of Physics and Astronomy, Toronto, Ontario, M3J 1P3 Canada}
\newcommand{\ND}{Department of Physics, University of Notre Dame, Notre Dame, Indiana 46556, USA}
\newcommand{\ICL}{The Blackett Laboratory,  Imperial College London,  London SW7 2BW, United Kingdom}
\newcommand{\warwick}{Department of Physics, University of Warwick, Coventry, CV4 7AL, UK}

\newcommand{\mascencioThanks}{Now at Iowa State University, Ames, IA 50011, USA}
\newcommand{\amitbashyalThanks}{Now at  High Energy Physics/Center for Computational Excellence Department, Argonne National Lab, 9700 S Cass Ave, Lemont, IL 60439}
\newcommand{\ricfregianThanks}{now at Department of Physics and Astronomy, University of California at Davis, Davis, CA 95616, USA}
\newcommand{\finerThanks}{Now at Los Alamos National Laboratory, Los Alamos, New Mexico 87545, USA}
\newcommand{\kleykampThanks}{now at Department of Physics and Astronomy, University of Mississippi, Oxford, MS 38677}
\newcommand{\bamThanks}{Now at University of Minnesota, Minneapolis, Minnesota 55455, USA}
\newcommand{\byaeggyThanks}{Now at Department of Physics, University of Cincinnati,  Cincinnati, Ohio 45221, USA}

\author{M.A.~Ram\'{i}rez}                 \affiliation{\upenn}  \affiliation{\Guanajuato}
\author{S.~Akhter}                        \affiliation{\AMU}
\author{Z.~~Ahmad~Dar}                    \affiliation{\WM}  \affiliation{\AMU}
\author{F.~Akbar}                         \affiliation{\AMU}
\author{V.~Ansari}                        \affiliation{\AMU}
\author{M.~V.~Ascencio}\thanks{\mascencioThanks}  \affiliation{\PUCP}
\author{M.~Sajjad~Athar}                  \affiliation{\AMU}
\author{A.~Bashyal}\thanks{\amitbashyalThanks}  \affiliation{\OregonState}
\author{L.~Bellantoni}                    \affiliation{\FNAL}
\author{A.~Bercellie}                     \affiliation{\Rochester}
\author{M.~Betancourt}                    \affiliation{\FNAL}
\author{A.~Bodek}                         \affiliation{\Rochester}
\author{J.~L.~Bonilla}                    \affiliation{\Guanajuato}
\author{A.~Bravar}                        \affiliation{\Geneva}
\author{H.~Budd}                          \affiliation{\Rochester}
\author{G.~Caceres}\thanks{\ricfregianThanks}  \affiliation{\CBPF}
\author{T.~Cai}                           \affiliation{\Rochester}
\author{G.A.~D\'{i}az~}                   \affiliation{\Rochester}
\author{H.~da~Motta}                      \affiliation{\CBPF}
\author{S.A.~Dytman}                      \affiliation{\Pittsburgh}
\author{J.~Felix}                         \affiliation{\Guanajuato}
\author{L.~Fields}                        \affiliation{\ND}
\author{A.~Filkins}                       \affiliation{\WM}
\author{R.~Fine}\thanks{\finerThanks}     \affiliation{\Rochester}
\author{H.~Gallagher}                     \affiliation{\Tufts}
\author{A.~Ghosh}                         \affiliation{\USM}  \affiliation{\CBPF}
\author{S.M.~Gilligan}                    \affiliation{\OregonState}
\author{R.~Gran}                          \affiliation{\UMD}
\author{E.Granados}                       \affiliation{\Guanajuato}
\author{D.A.~Harris}                      \affiliation{\york}  \affiliation{\FNAL}
\author{S.~Henry}                         \affiliation{\Rochester}
\author{D.~Jena}                          \affiliation{\FNAL}
\author{S.~Jena}                          \affiliation{\Mohali}
\author{J.~Kleykamp}\thanks{\kleykampThanks}  \affiliation{\Rochester}
\author{A.~Klustov\'{a}}                  \affiliation{\ICL}
\author{M.~Kordosky}                      \affiliation{\WM}
\author{D.~Last}                          \affiliation{\upenn}
\author{A.~Lozano}                        \affiliation{\CBPF}
\author{X.-G.~Lu}                         \affiliation{\warwick}  \affiliation{\oxford}
\author{E.~Maher}                         \affiliation{\MCLA}
\author{S.~Manly}                         \affiliation{\Rochester}
\author{W.A.~Mann}                        \affiliation{\Tufts}
\author{C.~Mauger}                        \affiliation{\upenn}
\author{K.S.~McFarland}                   \affiliation{\Rochester}
\author{B.~Messerly}\thanks{\bamThanks}   \affiliation{\Pittsburgh}
\author{J.~Miller}                        \affiliation{\USM}
\author{O.~Moreno}                        \affiliation{\WM}  \affiliation{\Guanajuato}
\author{J.G.~Morf\'{i}n}                  \affiliation{\FNAL}
\author{D.~Naples}                        \affiliation{\Pittsburgh}
\author{J.K.~Nelson}                      \affiliation{\WM}
\author{C.~Nguyen}                        \affiliation{\Florida}
\author{A.~Olivier}                       \affiliation{\Rochester}
\author{V.~Paolone}                       \affiliation{\Pittsburgh}
\author{G.N.~Perdue}                      \affiliation{\FNAL}  \affiliation{\Rochester}
\author{K.-J.~Plows}                      \affiliation{\oxford}
\author{R.D.~Ransome}                     \affiliation{\Rutgers}
\author{D.~Ruterbories}                   \affiliation{\Rochester}
\author{H.~Schellman}                     \affiliation{\OregonState}
\author{H.~Su}                            \affiliation{\Pittsburgh}
\author{M.~Sultana}                       \affiliation{\Rochester}
\author{V.S.~Syrotenko}                   \affiliation{\Tufts}
\author{E.~Valencia}                      \affiliation{\WM}  \affiliation{\Guanajuato}
\author{N.H.~Vaughan}                     \affiliation{\OregonState}
\author{A.V.~Waldron}                     \affiliation{\ICL}
\author{B.~Yaeggy}\thanks{\byaeggyThanks}  \affiliation{\USM}
\author{L.~Zazueta}                       \affiliation{\WM}

\collaboration{The MINER$\nu$A Collaboration}
\noaffiliation
\date{\today}

\begin{abstract}
MINERvA has measured the $\nu_{\mu}$-induced coherent $\pi^{+}$ cross section simultaneously in
hydrocarbon (CH), graphite (C), iron (Fe) and lead (Pb) targets using neutrinos from 2 to 20 GeV.
The measurements exceed the predictions of the Rein-Sehgal and Berger-Sehgal PCAC based models at
multi-GeV $\nu_{\mu}$ energies and at produced $\pi^{+}$ energies and angles, $E_{\pi}>1$ GeV and 
$\theta_{\pi}<10^{\circ}$. Measurements of the cross-section ratios of Fe and Pb relative to CH 
reveal the effective $A$-scaling to increase from an approximate $A^{1/3}$ scaling at few GeV to 
an $A^{2/3}$ scaling for $E_{\nu}>10$ GeV.
\end{abstract}

\maketitle

In neutrino-induced coherent pion production the nucleons in the nucleus recoil in phase under the
impact of an incident neutrino. The nucleus remains in its initial quantum state and recoils with 
an energy below the detection threshold of most neutrino detectors. A $\pi$ meson and a lepton are
created, both with relatively small angles with respect to the incoming neutrino. Both charged (CC)
and neutral current (NC) interactions can occur, induced by a neutrino or anti-neutrino of any flavor,
according to $\nu_{l}+A\rightarrow l+\pi+A$, where $\nu_{l}$ is a neutrino of flavor $l$, $A$ is the 
nucleus, and $l$ and $\pi$, a lepton and a pion of the proper charge, respectively. The four-momentum
transfer to the nucleus,

\begin{align}
        \lvert t\rvert=\left\vert\left(p_{\nu}{-}p_{l}{-}p_{\pi}\right)^{2}\right\vert\approx\left(\sum_{i=l,\pi}\textbf{p}_{i}^{T}\right)^{2}\!{+}\!\left(\sum_{i=l,\pi}E_{i}{-}p_{i}^{L}\right)^{2},
        \label{eq:1}
\end{align}

\noindent must be between $\lvert t_{min}\rvert\simeq\left[\left(Q^{2}+m^{2}_{\pi}\right)/2E_{\pi}\right]^{2}$
\cite{Paschos:2005km}, and $\lvert t_{max}\rvert=1/R_{N}^{2}$ \cite{CHARM-II:1993xmz} for the interaction
to happen, where $p_{\nu}$, $p_{l}$ and $p_{\pi}$ are the neutrino, lepton and pion four-momenta, respectively;
$\textbf{p}^{T}$ and $p^{L}$ are the lepton's or pion's transverse and longitudinal momenta, respectively; $E$
is the lepton's or pion's total energy, $Q^{2}$ is the square of the four-momentum transferred by the neutrino,
$m_{\pi}$ is the pion mass, and $R_{N}$ is the nuclear radius.

\par Historically, most experiments \cite{CHARM-II:1993xmz, Faissner:1983ng, Isiksal:1984vh, WA59:1984wvj, CHARM:1985bva, SKAT:1985uch, Baltay:1986cv, BEBCWA59:1986jkg, E632:1988dqx, BEBCWA59:1988sqc, E632:1992ydt, K2K:2005uiu, MiniBooNE:2008mmr, SciBooNE:2008bzb, NOMAD:2009idt, SciBooNE:2010lca} used the Rein-Sehgal model (R-S) \cite{Rein:1982pf} to simulate coherent 
$\pi$ production. It is based on Adler's Partially Conserved Axial Current (PCAC) theorem \cite{Adler:1964yx},
which relates the neutrino-nucleus inelastic cross section to the pion-nucleus elastic cross section, assuming
the incoming neutrino and the outgoing lepton are parallel (when $Q^{2}=0$), and neglecting the lepton mass.
The CC channel differential cross section is

\begin{eqnarray}
        \left.\frac{d^{3}\sigma^{CC}_{coh}}{dQ^{2}dy d\lvert t\rvert}\right\vert_{Q^{2}=0}=
        \frac{G^{2}_{F}f^{2}_{\pi}}{2\pi^{2}}\frac{1-y}{y}\frac{d\sigma^{\pi^{\pm}A}}
        {d\lvert t\rvert},
        \label{eq:3}
\end{eqnarray}

\noindent where $y=\nu/E_{\nu}=\left(E_{\nu}-E_{\mu}\right)/E_{\nu}\approx E_{\pi}/E_{\nu}$, $f^{2}_{\pi}$ is the pion decay constant, 
$E_{\nu}$ is the neutrino energy, and $d\sigma^{\pi^{\pm}A}/d\lvert t\rvert$ is the pion-nucleus 
elastic cross section. The model extrapolates Eq. (\ref{eq:3}) to $Q^{2}\!>\!0$ with a form factor $\left[m^{2}_{A}/\left(m^{2}_{A}{+}Q^{2}\right)\right]^{2}$, where $m_{A}\approx1$ GeV is the axial-vector mass. 

\par CC coherent pion production is an important background for CC quasi-elastic interactions in
$\nu_{\mu}$-disappearance measurements \cite{MINOS:2006foh}, when the $\pi^{+}$ is mis-reconstructed as a
proton. It is also a significant fraction of MINERvA's own CC1$\pi^{+}$ sample \cite{MINERvA:2022djk}. Both
are very valuable for upcoming neutrino oscillation analyses in the few-GeV region \cite{DUNE:2015lol,
Hyper-KamiokandeProto-:2015xww}.

\par By using $\lvert t\rvert$ to isolate signal-like events, MINERvA was the first experiment to 
observe the CC coherent $\pi^{\pm}$ in that energy region, using $\nu_{\mu}$ and $\overline{\nu}_{\mu}$
beams on a hydrocarbon (CH) target \cite{MINERvA:2014ani, MINERvA:2017ipy}. These and two later 
publications \cite{ArgoNeuT:2014uwh, T2K:2016soz} used an improved version of the R-S model that
includes the lepton mass \cite{Adler:2005ada, Rein:2006di}.

\par Prior to this work, all published results on coherent pion production used a single target with
mass number $A\leq40$ ($A\leq80$ for NC) \cite{CHARM-II:1993xmz, Faissner:1983ng, Isiksal:1984vh, WA59:1984wvj, CHARM:1985bva, SKAT:1985uch, Baltay:1986cv, BEBCWA59:1986jkg, E632:1988dqx, BEBCWA59:1988sqc, E632:1992ydt, K2K:2005uiu, MiniBooNE:2008mmr, SciBooNE:2008bzb, NOMAD:2009idt, SciBooNE:2010lca, MINERvA:2014ani, MINERvA:2017ipy, ArgoNeuT:2014uwh, T2K:2016soz}.
Compared to the previous MINERvA measurement, the present work uses data from a more energetic and
more intense beam \cite{Ainsworth:2020mqq}, and from a longer exposure, representing an increase of
the protons on target (POT), from ${\sim}3{\times}10^{20}$ to ${\sim}10.5{\times}10^{20}$. This paper
presents measurements carried out simultaneously on four different samples: hydrocarbon (CH), graphite
(C), steel (Fe) and lead (Pb). Absolute cross sections and ratios to scintillator (CH) are reported
for nuclei with a wide range of $A$ values: $12$, $56$ and $207$.

\par These measurements are obtained using the NuMI beam line at the Fermi National Accelerator Laboratory 
\cite{Adamson:2015dkw} where 120-GeV protons colliding on a graphite target, create hadrons which are 
focused using a pair of magnetic horns, and sent to a decay pipe where they create a beam of muon-neutrinos,
with $\langle E_{\nu}\rangle{\sim}6.0$ GeV \cite{Ainsworth:2020mqq}, made of ${\sim}95\%\;\nu_{\mu}$, and
${\sim}5\%$ of $\overline{\nu}_{\mu}$, $\nu_{e}$ and $\overline{\nu}_{e}$ \cite{MINERvA:2017dzh}. The neutrino
beam is simulated with a {\small Geant4} model \cite{GEANT4:2002zbu, MINERvA:2016iqn}.

\par The MINERvA detector consists of an inner detector made of an upstream ``nuclear target'' and a 
downstream ``tracker'' region, and an outer detector composed of electromagnetic (ECAL) and hadronic
(HCAL) calorimeters \cite{MINERvA:2013zvz}. The nuclear target region is ${\sim}1.4$ m long with five
different passive materials: solid C, Fe, and Pb; and liquid He and H$_{2}$O, all installed in seven
targets. Following the beam direction, solid targets are labeled from 1 to 5. Targets 1, 2 and 5 had
segments of Fe and Pb, and thickness of ${\sim}2.6$ cm in targets 1 and 2, and ${\sim}1.3$ cm in target
5. Target 3 had C, Fe, and Pb segments, with thickness of ${\sim}7.6$ cm, ${\sim}2.9$ cm and ${\sim}2.6$
cm, respectively. Target 4 was made of Pb with a thickness of ${\sim}0.8$ cm. Eight planes of tracking 
plastic scintillator (CH) were placed between the targets (only four between targets 4 and 5). Different
target positions and thicknesses tried to equalize mass, acceptance and particle containment; maximize
event rates, vertex and track resolution; and minimize the energy threshold of particles exiting the
passive materials. The tracker region is ${\sim}2.7$ m long and made of 120 scintillator planes. Planes
consist of 127 triangular prism scintillator strips with 33-mm base, $17$-mm height, and varying length
to form an hexagonal plane. Planes are rotated by $60^{\circ}$ with respect to adjacent ones, enabling
three-dimensional reconstruction. The detector's single hit position resolution is ${\sim}3$ mm and the
time resolution is 3 ns \cite{MINERvA:2013zvz}. The ECAL surrounds the inner detector, and the HCAL
surrounds the ECAL. The former (latter) consists of planes of lead (iron) and scintillator to contain
and track electromagnetically (strongly) interacting particles. Located 2 m downstream of MINERvA, the
MINOS near detector \cite{MINOS:1998kez, MINOS:2008hdf} served as a magnetized spectrometer to determine
muon charge and momentum.

\par The signal process is CC coherent interactions on C, Fe and Pb, induced by a $\nu_{\mu}$ from 2 to
20 GeV. Events with pion angle larger than 70 degrees cannot be tracked and have zero efficiency. Events
in the nuclear target (tracker) region with muon angle larger than 13 degrees have an efficiency of 
$\sim$1\% ($\sim$4\%) due to MINOS acceptance. The percentage of simulated signal events in these 
categories is $\sim$5\% for CH and C, and $\sim$2\% for Fe and Pb. 

\par
\begin{figure}[h]
\centering
\includegraphics[width=1.0\linewidth]{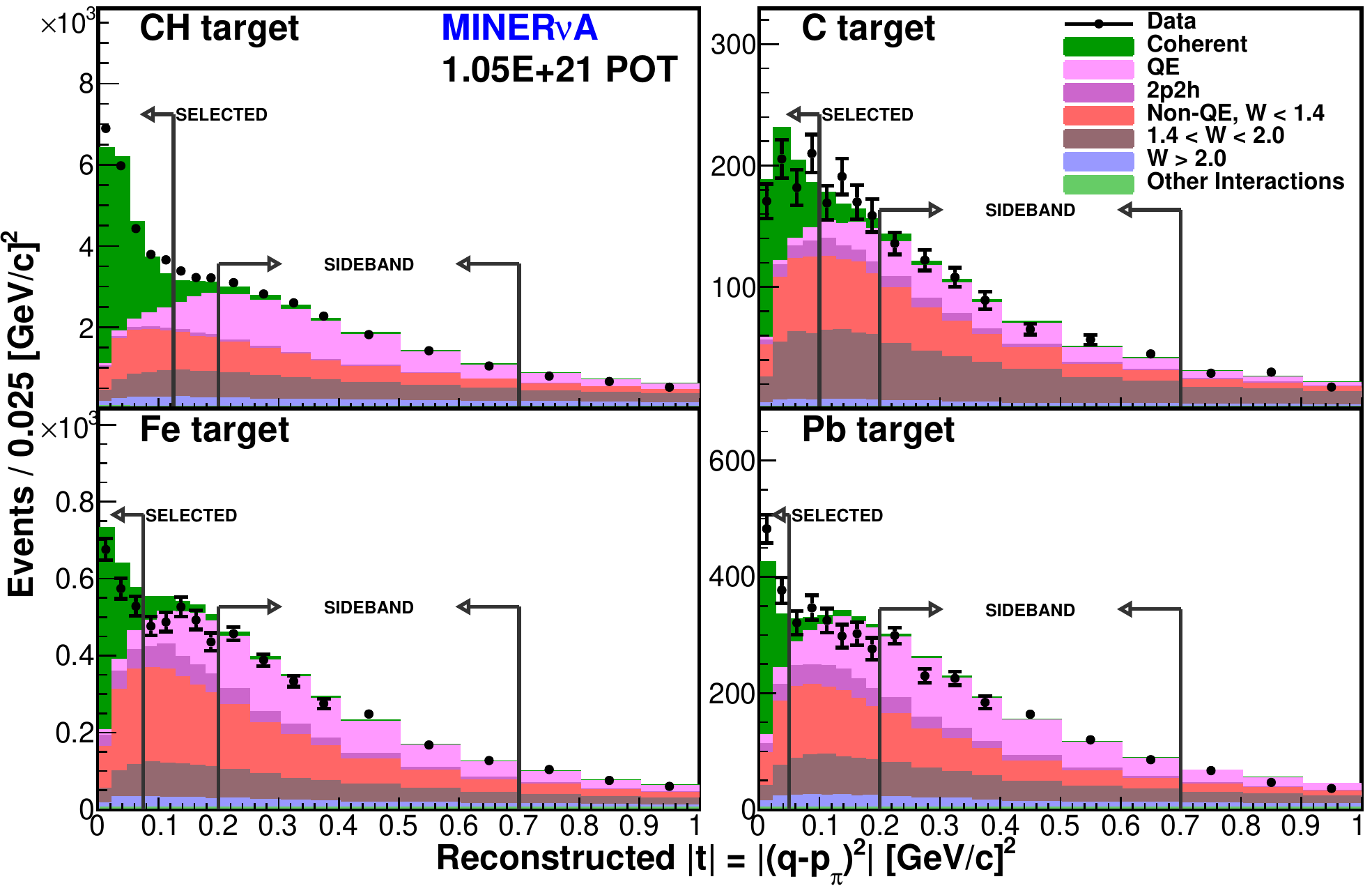}
\caption{Reconstructed $\lvert t\rvert$ distributions after $E_{vtx}$ cut and background tuning: CH, C, Fe 
and Pb, in reading order. Regions in between arrows are the sidebands for background tuning. Events to the 
left of the lower-$\lvert t\rvert$ arrow, are selected.}
\label{fig:tcut}
\end{figure}

 Neutrino interactions are simulated using a modified version of the {\small GENIE} event generator
v2.12.6 \cite{Andreopoulos:2009rq, Andreopoulos:2015wxa}. The signal's cross section is given by the R-S
model with the lepton mass correction. Background processes (Fig. \ref{fig:tcut}) in increasing hadronic
invariant mass $W$, are: CC quasielastic (QE), correlated pairs of nucleons (2p2h), resonant $\pi^{+}$ 
production (Non-QE, $W\!<\!1.4$ or RES), inelastic scattering ($1.4\!<\!W\!<\!2.0$ or INE) and deep 
inelastic scattering ($W\!>\!2.0$ or DIS). Quasielastic scattering is simulated using the Llewellyn-Smith
model \cite{LlewellynSmith:1971uhs} with an axial-vector form factor from a z-expansion fit to deuterium
data \cite{Meyer:2016oeg} and a correction from the Valencia Random Phase Approximation (RPA) 
\cite{Nieves:2004wx}. The 2p2h process is simulated with the Valencia model \cite{Nieves:2011pp, Gran:2013kda, Schwehr:2016pvn}
and modified according to a ``low recoil'' fit by MINERvA \cite{MINERvA:2015ydy}. Resonant pion production
uses the Rein-Sehgal model \cite{Rein:1980wg} with its normalization increased $15\%$ based on fits from
a deuterium data analysis \cite{Rodrigues:2016xjj}, plus an additional {\it ad hoc} suppression for 
$Q^{2}\!<\!0.7$ [GeV/c]$^{2}$ due to collective nuclear effects \cite{MINERvA:2019kfr}. Inelastic 
interactions use a tuned model of discrete baryon resonances \cite{Rein:1980wg}, and the Bodek-Yang model
for the transition region to DIS, as well as non-resonant pion production across the full $W$ range 
\cite{Bodek:2004pc}, that was reduced by $43\%$ based on a tune to the same deuterium data \cite{Rodrigues:2016xjj}. 
These tunes to {\small GENIE} are labeled as the MINERvA tune v4.4.1 \cite{MINERvA:2022mnw}.

\par Final state particles coming from the {\small GENIE} simulation are propagated through the detector
using a {\small Geant4} simulation of the detector's geometry and material composition, light yield and
energy deposition of the particles in the scintillator, and their hadronic and electromagnetic interactions
\cite{Kaidalov:1982xg, Guthrie:1968ue}. The detector's energy scale was established by making sure that 
simulated through-going muons agreed with data in both light yield and reconstructed energy deposition. 
The detector's simulated response to different particles is validated in a test beam measurement 
\cite{MINERvA:2015yej}, and the effects of accidental activity, electronics charge and time resolution
were also included \cite{MINERvA:2013zvz}. 

\par Scintillator strips with deposited energy greater than 1 MeV are grouped per plane according to their
position and time, into ``clusters''. These are grouped with clusters in adjacent planes to form tracks. 
Backwards-projected tracks find interaction vertices. Angles are measured between the simulated beam direction
and the direction of the track in its first planes downstream of the vertex.

\par This analysis isolates events with two tracks from a common vertex. The reconstructed momentum of the 
muon candidate is the addition of the momentum determined by range inside MINERvA plus its momentum determined
by range or curvature inside MINOS. The pion candidate has to be fully contained inside MINERvA, so 
$\lvert t\rvert$ can be measured. The pion total energy is reconstructed calorimetrically from all the energy not
associated with the muon, given the assumption $E_{\nu}\approx E_{\pi}+E_{\mu}$ from Eq. (\ref{eq:1}), where
$E_{\mu}$ is the muon's total energy. 

\par The reconstructed interaction vertex is defined as the upstream end of the muon track, and it is required
to be inside the fiducial volume under study. The CH fiducial volume is $108$ planes long (${\sim}2.4$ m)
centered in the tracker region with the area of a $0.85$-m apothem hexagon \cite{footnote1}. The fiducial volume
in the passive targets, is the area times the thickness of the segment of interest. For the passive materials, 
the vertex is projected into the $\it z$ center of the target, where the $\left(\it x, y\right)$ coordinate 
determines the segment (material). Events from different targets but same material, are combined into a single
sample.

\par The reconstructed neutrino energy must be between 2 and 20 GeV to remove events with mis-reconstructed 
muon energy \cite{MINERvA:2016iqn}. To reject protons from quasielastic and resonance production backgrounds, 
$dE/dx$-based $\chi^{2}$ compared to pion and proton hypotheses of the pion candidate track are built. A log 
likelihood ratio \cite{Neyman:1933wgr} between the hypotheses removes (keeps) ${\sim}70\%$ (${\sim}87\%$) of 
protons (pions) according to the simulation.

The energy of the vertex region ($E_{vtx}$), defined as a $200$-mm radius, $7-$plane height cylinder centered
at the interaction vertex, must be consistent with the energy deposited by one minimum-ionizing charged pion
and one muon. The $E_{vtx}$ distribution of simulated signal events is fit to a Gaussian function, and events
within $\pm1\sigma$ of the mean are selected. Due to different target thickness, $\langle E_{vtx}\rangle$ is 
target-dependent, varying from ${\sim}60$ to ${\sim}95$ MeV. This cut removes (keeps)  
${\sim}86\%$ (${\sim}60\%$) of the background (signal).

\par Due to their proximity to tracking scintillator planes, the C, Fe and Pb samples have contamination from
events occurring in scintillator upstream and downstream of the passive material. These events are considered
background, and are tuned using the plastic regions between passive targets as sidebands. There is an 
``upstream'' and a ``downstream'' plastic sideband for each passive material. The tuned plastic backgrounds
represent ${\sim}13\%$, ${\sim}14\%$ and ${\sim}21\%$ of the C, Fe and Pb selected samples, respectively.
 
 \par After removing events with high $E_{vtx}$, and subtracting the plastic background, all samples in Fig. 
 \ref{fig:tcut} show a signal dominance at low $\lvert t\rvert$. For heavier nuclei, the signal shrinks to a 
 lower $\lvert t\rvert$ region as $R_{N}$ increases. The C distribution has a significant excess of RES and 
 INE events from ${\sim}0.025$ to ${\sim}0.5$ [GeV/c]$^{2}$ compared to CH, despite both being interactions 
 on Carbon. This is due to the ${\sim}7.6$-cm thickness of the C segment, where one or more pions from those
 backgrounds are absorbed inside the passive material, which allows the event to pass the $E_{vtx}$ cut.

\par A high $\lvert t\rvert$ sideband ($0.2\!<\!\lvert t\rvert\!<\!0.7$ [GeV/c]$^2$) is used to tune the QE, 
RES (Non-QE, $W\!<\!1.4$), INE ($1.4\!<\!W\!<\!2.0$) and DIS ($W>2.0$) backgrounds. Due to their small content,
``Coherent'' and ``Other Interactions'' (NC-, $\overline{\nu}_{\mu}$- or $\overset{\fontsize{2pt}{2pt}\selectfont(-)}{\nu_{e}}$-induced) are not tuned, and 2p2h is considered QE during the tuning. Because the C target has limited statistics, two 
modifications were made to the fit for that target only: RES and INE were combined, and the QE and DIS scale 
factors were replaced by their CH counterparts. The scale factors for each of the backgrounds are in the 
supplemental material. 

\par Events with $\lvert t\rvert\!<$ (0.1, 0.125, 0.075, and 0.05) [GeV/c]$^{2}$ were selected for C, CH, Fe 
and Pb, respectively. More than 99\% of {\small GENIE} true signal events are below those cuts. After the 
$\lvert t\rvert$ cut and background subtraction, there are 14855$\pm$433 CH, 303$\pm$41 C, 726$\pm$89 Fe and 
492$\pm$41 Pb candidate events.

\par An iterative unfolding approach \cite{DAgostini:1994fjx} was used to correct the background-subtracted
distributions for resolution effects. The unfolded distributions were then efficiency-corrected. The cross
sections were extracted according to the expression $\sigma = N^{DATA\;eff}/\left(\Phi T\right)$, where
$N^{DATA\;eff}$ is the background-subtracted, unfolded and efficiency-corrected data, $\Phi$ is the incident
neutrino flux, and $T$ the number of C, Fe or Pb nuclei. The largest sources of inefficiency come
predominantly from well-understood random processes, which supports the assumption that the non-detected
events have the same relative background composition.

\par The extracted cross sections are compared to the R-S model ({\small GENIE} v2.12.6) and to the Berger-Sehgal
(B-S) model ({\small GENIE} v3.0.6) \cite{GENIE:2021zuu, Stowell_2017, Berger:2008xs}. The latter is also 
PCAC-based, and also includes the muon mass correction, but uses pion-carbon data \cite{Paschos:2005km} to 
model the elastic pion-nucleus cross section, instead of pion-deuterium data as the R-S model. 

\par
\begin{figure}[h!]
\includegraphics[width=1.\linewidth]{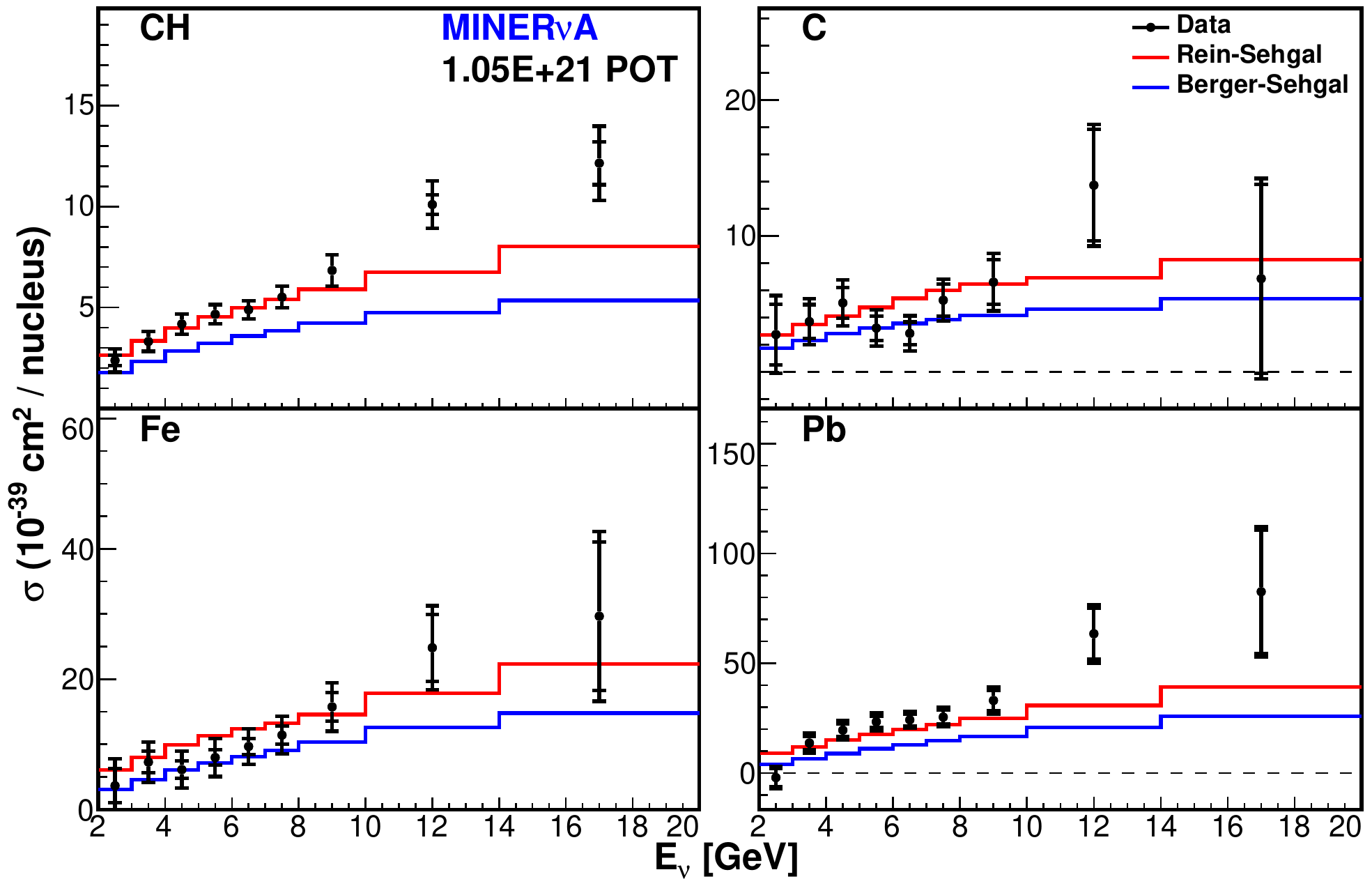}
\caption{Total cross sections as function of $E_{\nu}$: CH, C, Fe, and Pb, in reading order. Data is 
compared to the R-S (red) and B-S (blue) models.}
\label{fig:xsecenu}
\end{figure}

\begin{figure}[h!]
\includegraphics[width=1.\linewidth]{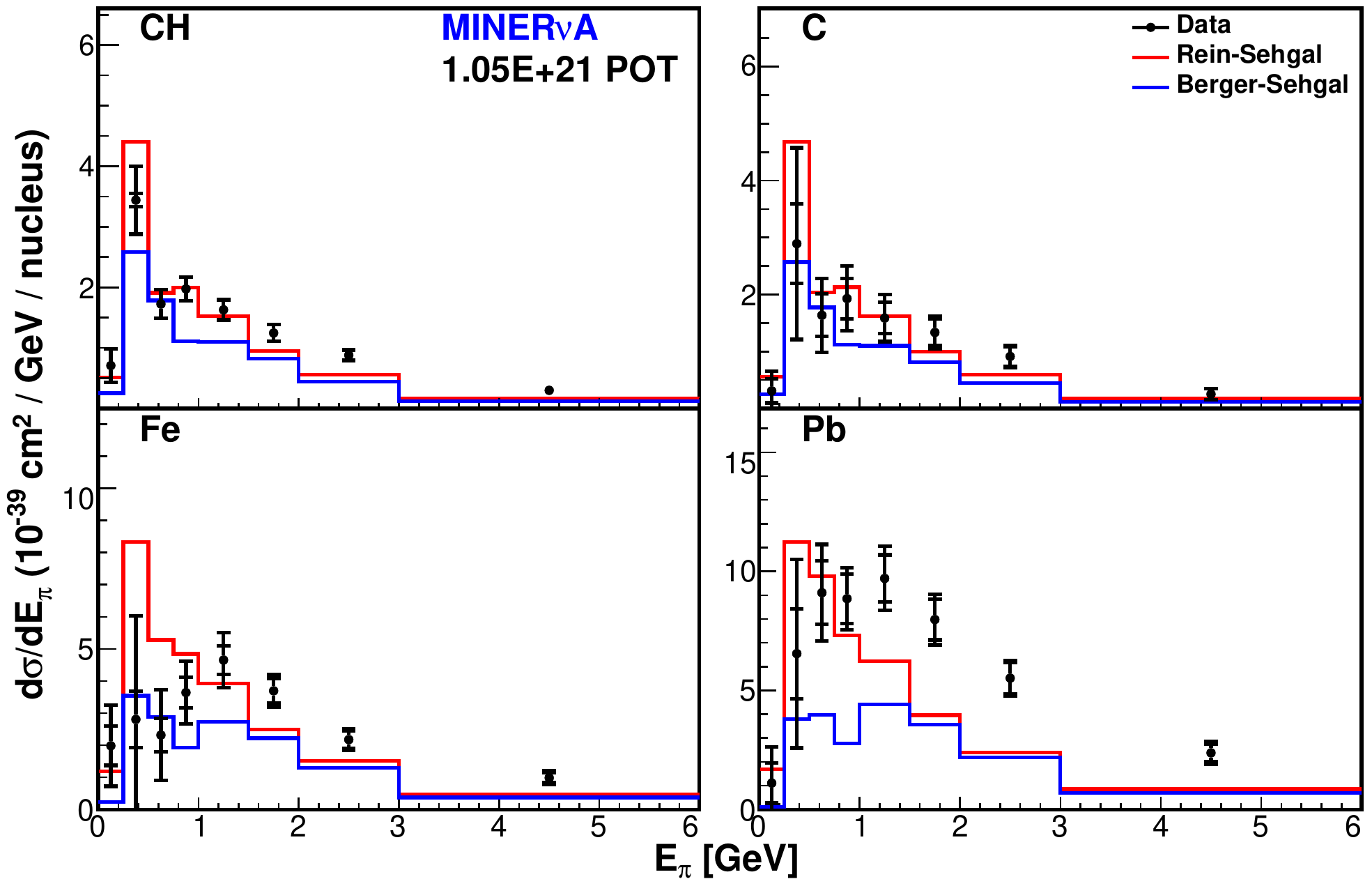}
\caption{Differential cross sections as function of $E_{\pi}$: CH, C, Fe, and Pb, in reading order. Data 
is compared to the R-S (red) and B-S (blue) models.}
\label{fig:xsecepi}
\end{figure}

\begin{figure}[h!]
\includegraphics[width=1.\linewidth]{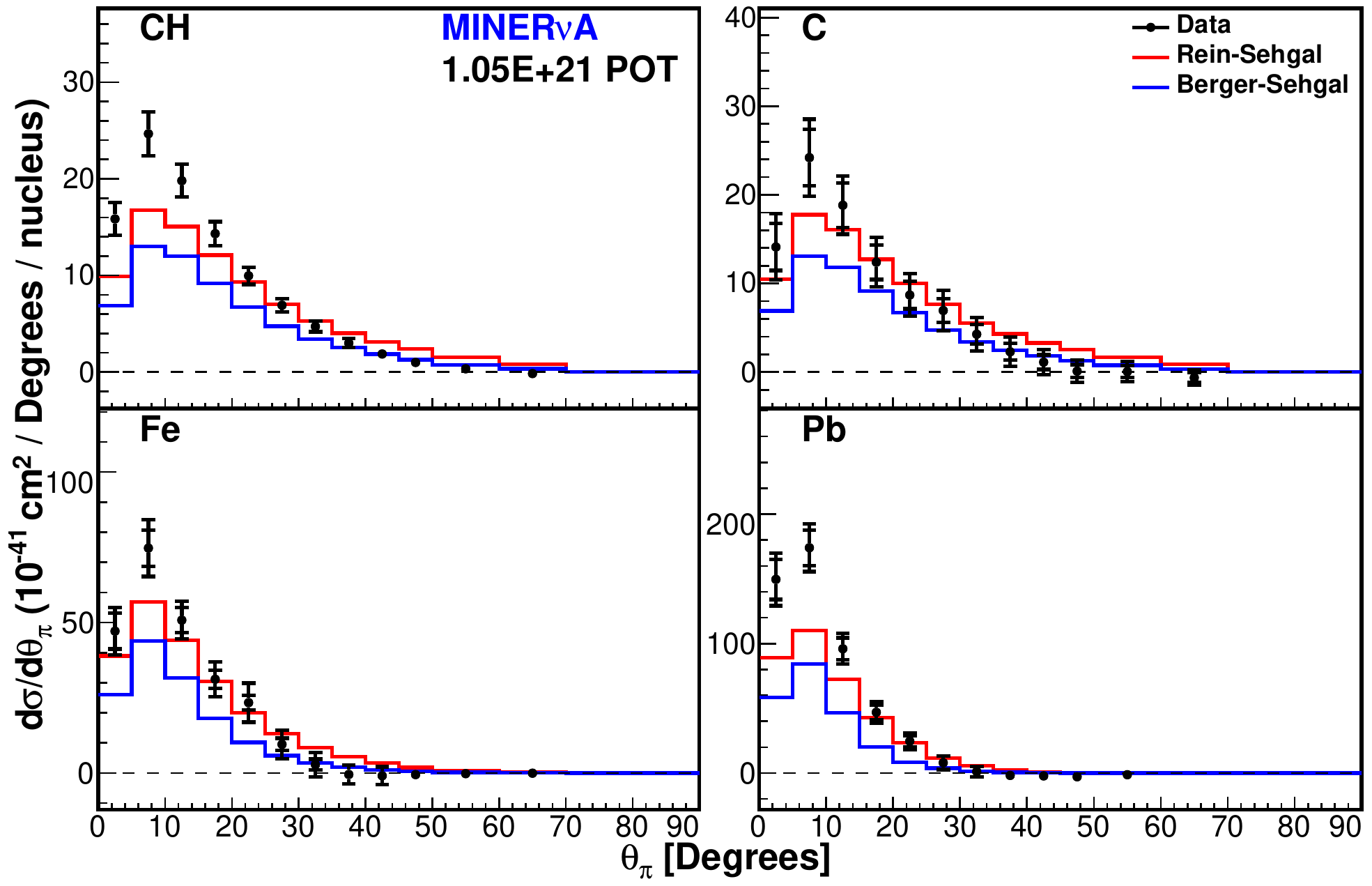}
\caption{Differential cross sections as function of $\theta_{\pi}$: CH, C, Fe, and Pb, in reading order. 
Data is compared to the R-S (red) and B-S (blue) models.}
\label{fig:xsecthetapi}
\end{figure}

 Figure \ref{fig:xsecenu} shows the total cross section as a function of $E_{\nu}$, with the flux
integrated per bin, where both models under-predict the reaction rate at high neutrino energies in the four
materials. Inner (outer) error bars are the statistical (statistical+systematic) uncertainties. The
differential cross sections with respect to $E_{\pi}$ and $\theta_{\pi}$, are flux-averaged from
$2<E_{\nu}<20$ GeV. In $d\sigma/dE_{\pi}$ (Fig. \ref{fig:xsecepi}) there is a clear disagreement between the
models and the data of the two heavier nuclei, for low (high) $E_{\pi}$ in iron (lead). Figure
\ref{fig:xsecthetapi} shows that the models also under-predict the $d\sigma/d\theta_{\pi}$ cross section at
very forward angles in all materials. Notably, forward pion production in the heavier nuclei is enhanced
relative to scattering on carbon, where for lead, the cross section becomes negligible for
$\theta_{\pi}>30^{\circ}$.

\par
\begin{figure}[h!]
\includegraphics[width=1.0\linewidth]{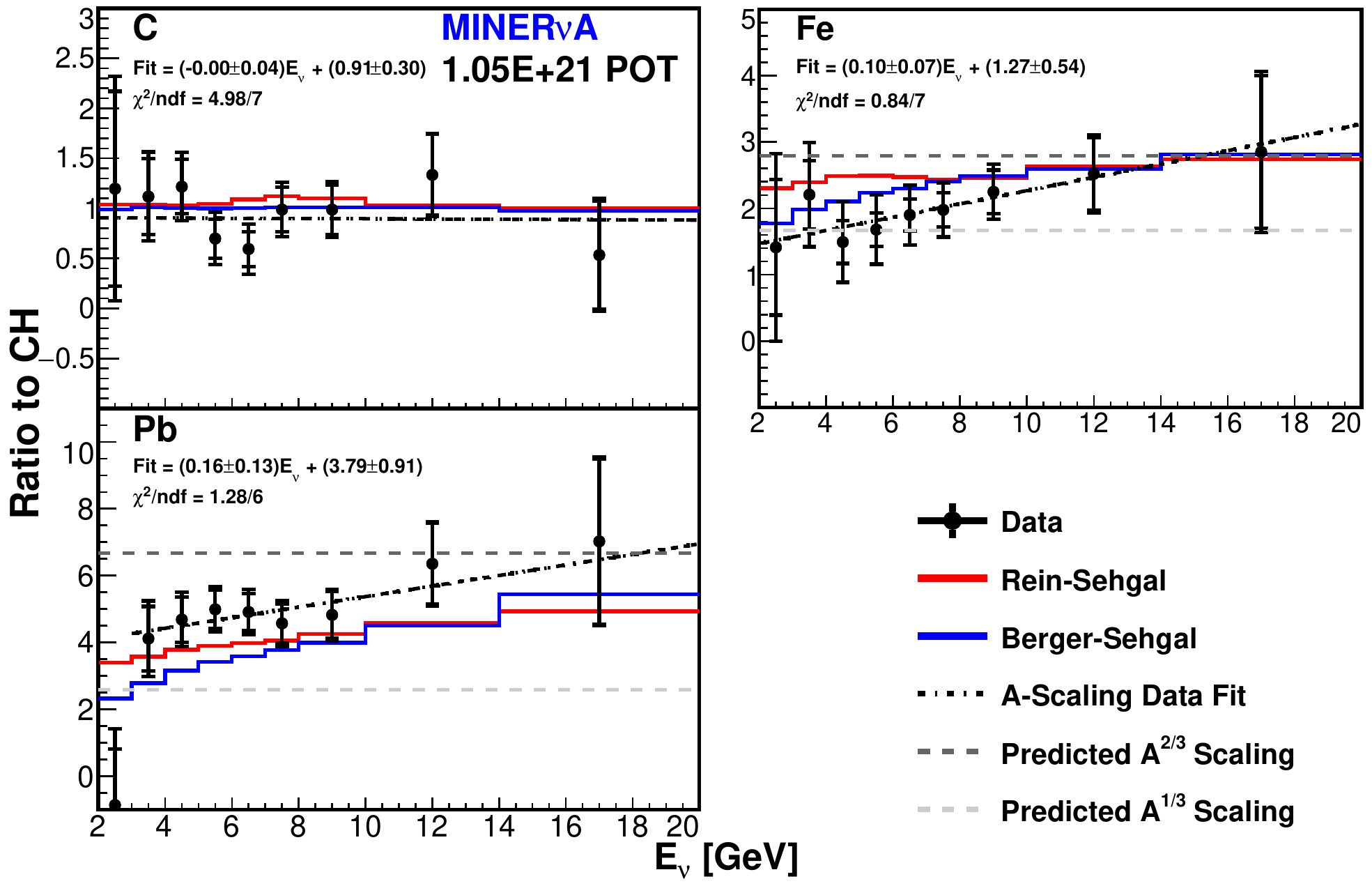}
\caption{Cross section ratios as function of $E_{\nu}$: $\sigma_{C}/\sigma_{CH}$, $\sigma_{Fe}/\sigma_{CH}$, 
and $\sigma_{Pb}/\sigma_{CH}$, in reading order. The upper (lower) dashed line is the ratio predicted by an 
$A^{2/3}$ ($A^{1/3}$) scaling. The slope is the best $A$-scaling fit. The 2-3 GeV bin is not included in the 
$\sigma_{Pb}/\sigma_{CH}$ fit due to the null cross section in lead in that bin (Fig. \ref{fig:xsecenu}.)}
\label{fig:enuratios}
\end{figure}

 The simultaneous neutrino exposure of the various targets enables precise measurement of cross section 
ratios thanks to the same beam configuration in all targets at any given time. Figure \ref{fig:enuratios} 
shows the cross section ratios as a function of $E_{\nu}$: $\sigma_{C}/\sigma_{CH}$, $\sigma_{Fe}/\sigma_{CH}$ 
and $\sigma_{Pb}/\sigma_{CH}$. As expected, the former is consistent with unity \cite{footnote2}. 
The CH cross sections used to calculate the ratios, were reweighted to use a flux that matched the flux used 
to calculate the C, Fe or Pb cross sections \cite{MINERvA:2022djk}.

The R-S and B-S models predict a scaling of the cross-section with respect to the mass number $A$ and
practically energy independent (horizontal dashed lines in Fig. \ref{fig:enuratios}), $\sim A^{1/3}$
\cite{Rein:1982pf, Lackner:1979ax} and $\sim A^{2/3}$ \cite{Faissner:1983ng, Hufner:1975ys}, respectively. The
PCAC-based Belkov-Kopeliovich (B-K) model predicts a scaling close to $A^{1/3}$ at low pion energy but close
to $A^{2/3}$ at high pion energy \cite{Kopeliovich:2011xw, Kopeliovich:2012tu}. In terms of neutrino energy,
the B-K model predicts a scaling of ${\sim}A^{1/3}$ (${\sim}A^{2/3}$) at neutrino energies below (above)
$\sim$10 GeV \cite{Belkov:1986hn}.

\par The measured $\sigma_{Fe}/\sigma_{CH}$ resembles the trend predicted by B-K, where below $\sim$8 GeV 
there is a clear agreement with the $A^{1/3}$ scaling, and a better agreement with the $A^{2/3}$ scaling 
above $\sim$10 GeV, with a constant increase in between. A similar trend occurs for the measured 
$\sigma_{Pb}/\sigma_{CH}$ but with an $A$-scaling larger than predicted below 10 GeV.

\par
\begin{figure}[h!]
\includegraphics[width=1.0\linewidth]{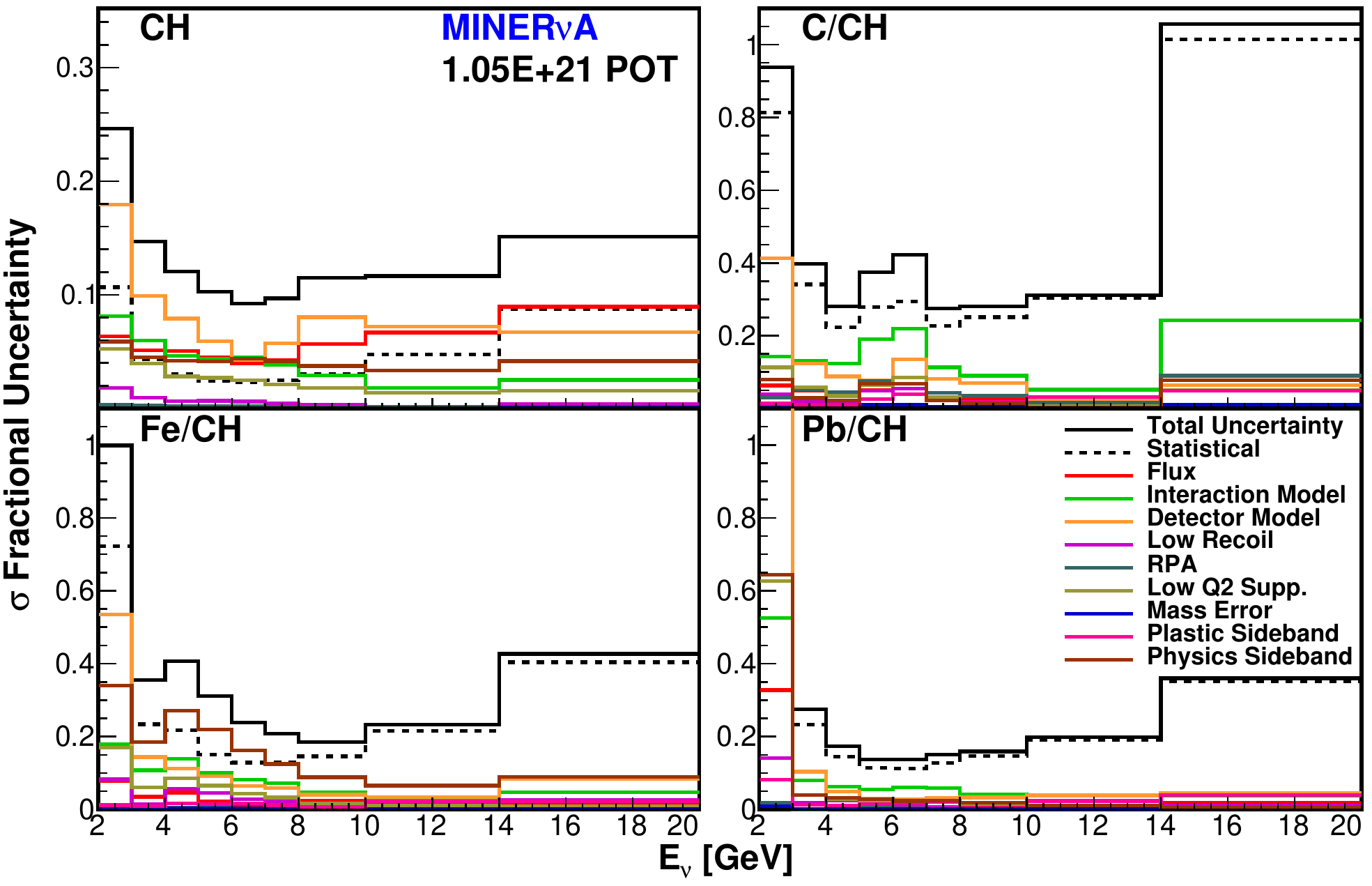}
\caption{Uncertainties in the total cross section as a function of $E_{\nu}$: CH, C/CH, Fe/CH, and Pb/CH, in 
reading order. The systematic uncertainties are described in the text.}
\label{fig:enuunc}
\end{figure}

 The statistical uncertainty of the total cross section dominates in the three passive materials 
(Fig. \ref{fig:enuunc}). The largest systematic uncertainties are related to the detector's geometry and 
particles interacting in it (Detector Model), like the muon energy deposition in MINERvA and MINOS 
\cite{MINERvA:2019gsf}. Uncertainties associated with the ``Interaction Model'', come from {\small GENIE} 
and the uncertainties from the MINERvA tune v.4.4.1. The ``Physics Sideband'' is the uncertainty on the 
backgrounds scale factors, plus a ``per-bin'' uncertainty covering for the remaining disagreement between 
data and the simulation in the high $\lvert t\rvert$ sideband.

\par The ``Flux'' uncertainty comes from the uncertainty on the beam line parameters, and hadron interactions 
\cite{MINERvA:2016iqn}. It was further constrained from $7.6\%$ to $3.9\%$ using a neutrino-electron scattering 
measurement \cite{MINERvA:2019hhc}. \par Other sources of uncertainty are the discrepancy in the detector mass; 
modifications to the QE-like background (Low Recoil and RPA); low $Q^{2}$ suppression of resonant pion production; 
and the uncertainty on the plastic background scale factors (Plastic Sideband). They contribute less than 
${\sim}5\%$ (${\sim}15\%$) to the total cross section uncertainty in CH (C, Fe and Pb). The CH sample provides 
the most precise measurement of the interaction so far, reducing the total uncertainty from ${\sim}25\%$ to 
${\sim}15\%$ compared to the previous MINERvA measurement \cite{MINERvA:2017ipy}. Cross section ratios have a 
further reduction of some systematic uncertainties, in particular the flux, reduced by ${\sim}75\%$ of itself 
(Fig. \ref{fig:enuunc}).

\par The measurements in this letter represent the first simultaneous measurement of the interaction in multiple 
materials and the first measurement in nuclei with $A>40$ ($^{56}$Fe and $^{207}$Pb), from which cross section 
ratios with respect to CH are measured. The data indicates that the R-S and B-S PCAC models do not accurately  
describe the angular dependence on $\theta_{\pi}$, the energy-dependence on $E_{\pi}$, or the $A$-dependence.
While the $\sigma_{Fe}/\sigma_{CH}$ qualitatively agrees with the B-K model's energy-dependent $A$-scaling, 
$\sigma_{Pb}/\sigma_{CH}$ does not, at least at low $E_{\nu}$. 
\par The estimate of the cross sections A-scaling provided in this letter could be used to extrapolate to
materials where measurements do not exist or the statistics are limited, like H$_{2}$O for Hyper-K or Ar for
DUNE. For the latter, pion production will make up around three quarters of the detected neutrino-induced
events.

This document was prepared by members of the MINERvA Collaboration using the resources of the Fermi National 
Accelerator Laboratory (Fermilab), a U.S. Department of Energy, Office of Science, HEP User Facility. Fermilab 
is managed by Fermi Research Alliance, LLC (FRA), acting under Contract No. DE-AC02-07CH11359. These resources 
included support for the MINERvA construction project, and support for construction also was granted by the 
United States National Science Foundation under Award No. PHY-0619727 and by the University of Rochester. 
Support for participating scientists was provided by NSF and DOE (USA); by CAPES and CNPq (Brazil); by CoNaCyT 
(M\'{e}xico); by Proyecto Basal FB 0821, CONICYT PIA ACT1413, and Fondecyt 3170845 and 11130133 (Chile); by 
CONCYTEC (Consejo Nacional de Ciencia, Tecnolog\'{i}a e Innovaci\'{o}n Tecnol\'{o}gica), DGI-PUCP (Direcci\'{o}n 
de Gesti\'{o}n de la Investigaci\'{o}n - Pontificia Universidad Cat\'{o}lica del Per\'{u}), and VRI-UNI 
(Vice-Rectorate for Research of National University of Engineering) (Per\'{u}); NCN Opus Grant No. 2016/21/B/ST2/01092
(Poland); by Science and Technology Facilities Council (UK); by EU Horizon 2020 Marie Sk lodowska-Curie Action; 
by a Cottrell Postdoctoral Fellowship from the Research Corporation for Scientific Advancement; by an Imperial 
College London President’s PhD Scholarship. We thank the MINOS Collaboration for use of its near detector data. 
Finally, we thank the staff of Fermilab for support of the beam line, the detector, and computing infrastructure. 
M.A. Ram\'{i}rez specially acknowledges support from a Postdoctoral Fellowship from the University of Pennsylvania.

\onecolumngrid
\pagebreak
\appendix{Supplemental Material}

\section{Background scale factors} \label{sec1:bkgds}

\begin{table}[h!]
\setlength{\tabcolsep}{6pt} 
\renewcommand{\arraystretch}{1.3} 
\begin{tabular}{c||c|c|c|c|c||c|c|c|c}
Material & \multicolumn{5}{c||}{Incoherent backgrounds}                                     & \multicolumn{4}{c}{Plastic scintillator backgrounds}              \\
\hline\hline
         & $\alpha$ QE   & $\alpha$ RES  & $\alpha$ INE  & $\alpha$ DIS  & $\chi^{2}$/ndf & $\alpha$ US    & $\chi^{2}$/ndf & $\alpha$ DS    & $\chi^{2}$/ndf \\
CH       & 1.22$\pm$0.02 & 1.29$\pm$0.04 & 0.60$\pm$0.02 & 0.65$\pm$0.03 & 281/30         & ---            & ---            & ---            & ---            \\
C        & 1.22$\pm$0.02 & 1.17$\pm$0.05 & 1.17$\pm$0.05 & 0.65$\pm$0.03 & 105/30         & 1.14$\pm$0.01  & 10/4           & 1.10$\pm$0.008 & 57/5           \\
Fe       & 1.40$\pm$0.07 & 1.15$\pm$0.09 & 0.58$\pm$0.06 & 1.08$\pm$0.14 & 89/30          & 1.16$\pm$0.008 & 36/8           & 1.18$\pm$0.004 & 153/19         \\
Pb       & 1.09$\pm$0.05 & 0.66$\pm$0.07 & 0.50$\pm$0.05 & 0.98$\pm$0.14 & 119/30         & 1.16$\pm$0.005 & 90/13          & 1.16$\pm$0.006 & 75/19         
\end{tabular}
\caption{Background scale factors: Quasi-Elastic ($\alpha_{QE}$), Resonance ($\alpha_{RES}$), Inelastic ($\alpha_{INE}$), DIS ($\alpha_{DIS}$), Upstream ($\alpha_{US}$) and Downstream ($\alpha_{DS}$) plastic.}
\label{tab:scalefactors}
\end{table}

\section{$d\sigma/dQ^{2}$, $d\sigma/d\theta_{\mu}$ and $d\sigma/dE_{\mu}$ cross sections in CH, C, Fe and Pb}
\label{section:Q2EmuThetamuXSec}

\begin{figure}[h!]
\includegraphics[width=0.65\linewidth]{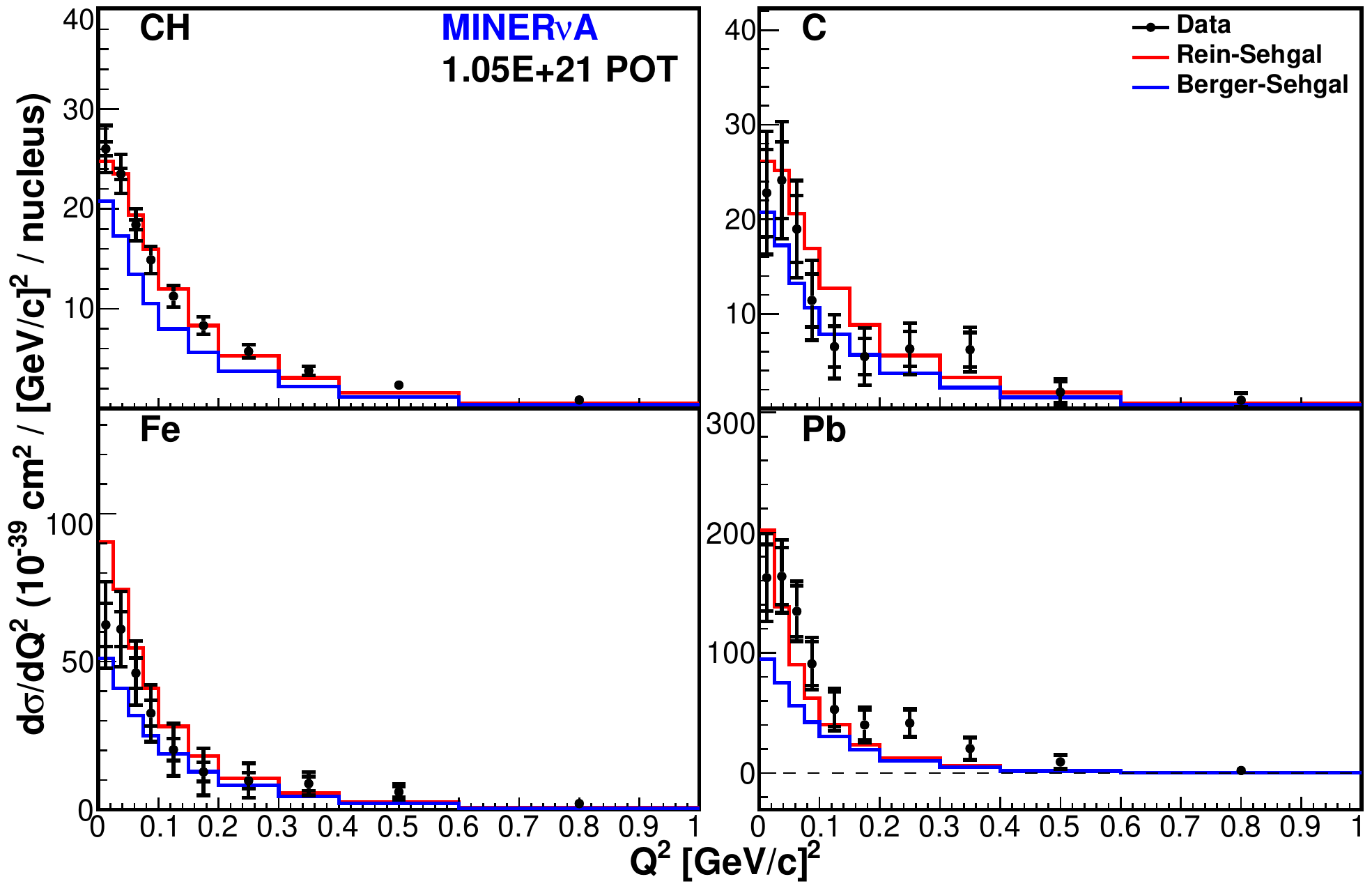}
\caption{Differential cross sections as function of $Q^{2}$: CH, C, Fe, and Pb, in reading order. Data is compared to the R-S (red) and B-S (blue) models.}
\label{fig:xsecq2}
\end{figure}

\begin{figure}[h!]
\includegraphics[width=0.65\linewidth]{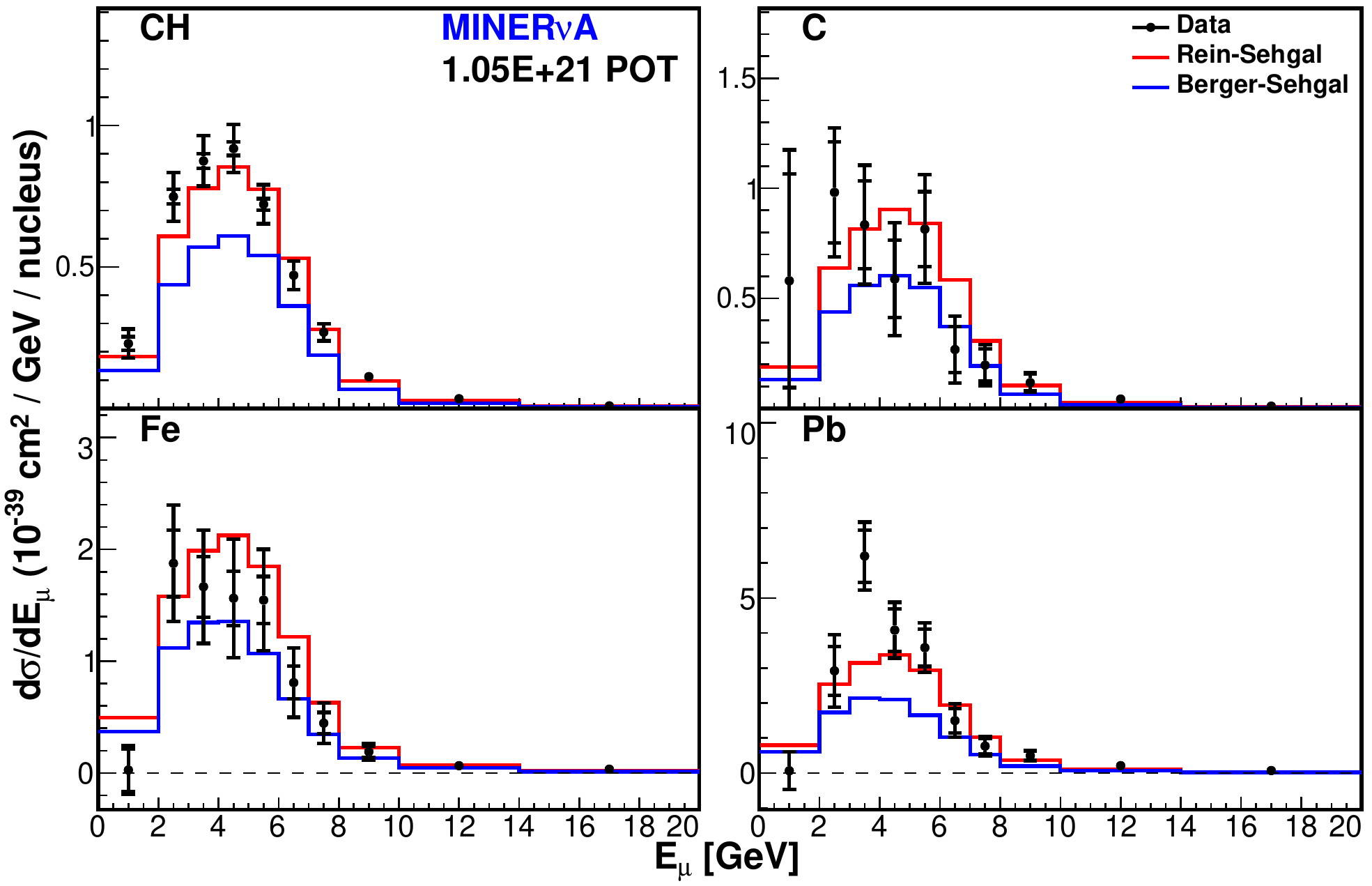}
\caption{Differential cross sections as function of $E_{\mu}$: CH, C, Fe, and Pb, in reading order. Data is compared to the R-S (red) and B-S (blue) models.}
\label{fig:xsecemu}
\end{figure}

\begin{figure}[h!]
\includegraphics[width=0.65\linewidth]{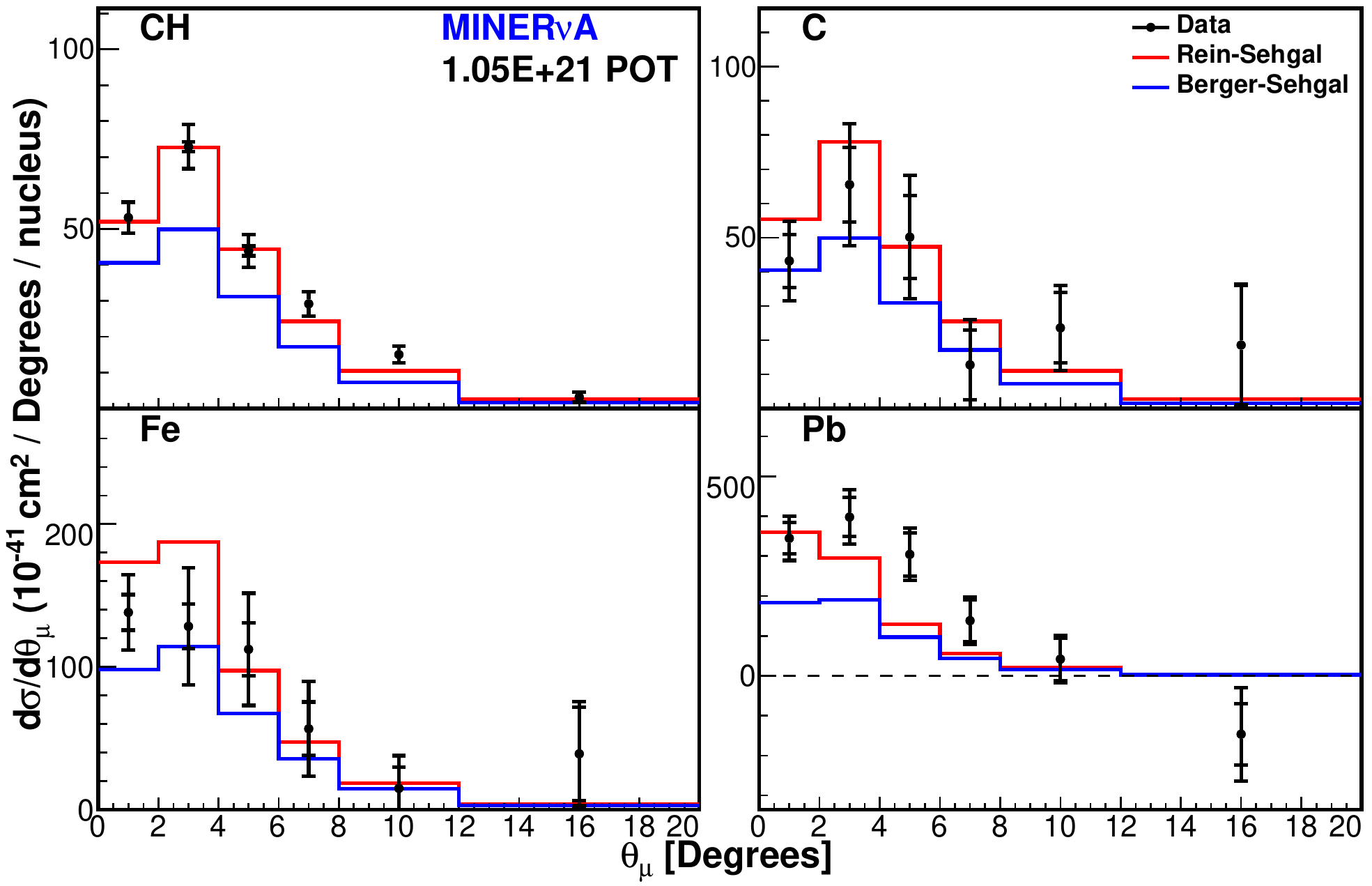}
\caption{Differential cross sections as function of $\theta_{\mu}$: CH, C, Fe, and Pb, in reading order. Data is compared to the R-S (red) and B-S (blue) models.}
\label{fig:xsecthetamu}
\end{figure}
\clearpage
\pagebreak[4]

\section{Cross section ratios of C, Fe and Pb relative to CH, as a function of $E_{\pi}$, $\theta_{\pi}$, $Q^{2}$, $E_{\mu}$, and $\theta_{\mu}$}
\label{section:extraXSecRatios}

\begin{figure}[h!]
        \includegraphics[width=0.65\linewidth]{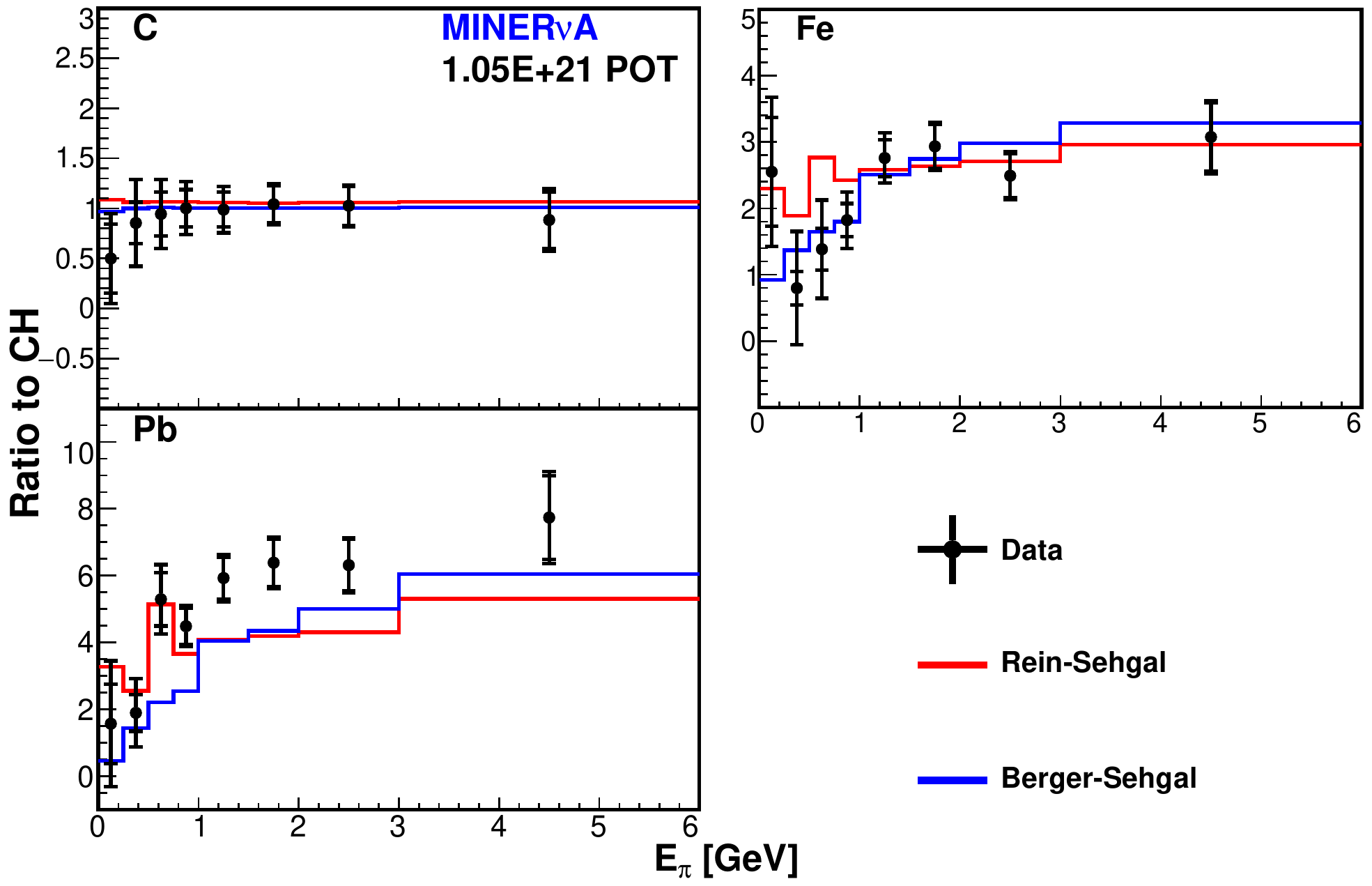}
        \caption{Ratios of the differential cross section as a function of $E_{\pi}$. C, Fe and Pb with respect to CH, in reading order. Data is compared to the Rein-Sehgal (red) and Berger-Sehgal (blue) models.}
\label{fig:xsecRatios_epi}
\end{figure}

\begin{figure}[h!]
        \includegraphics[width=0.65\linewidth]{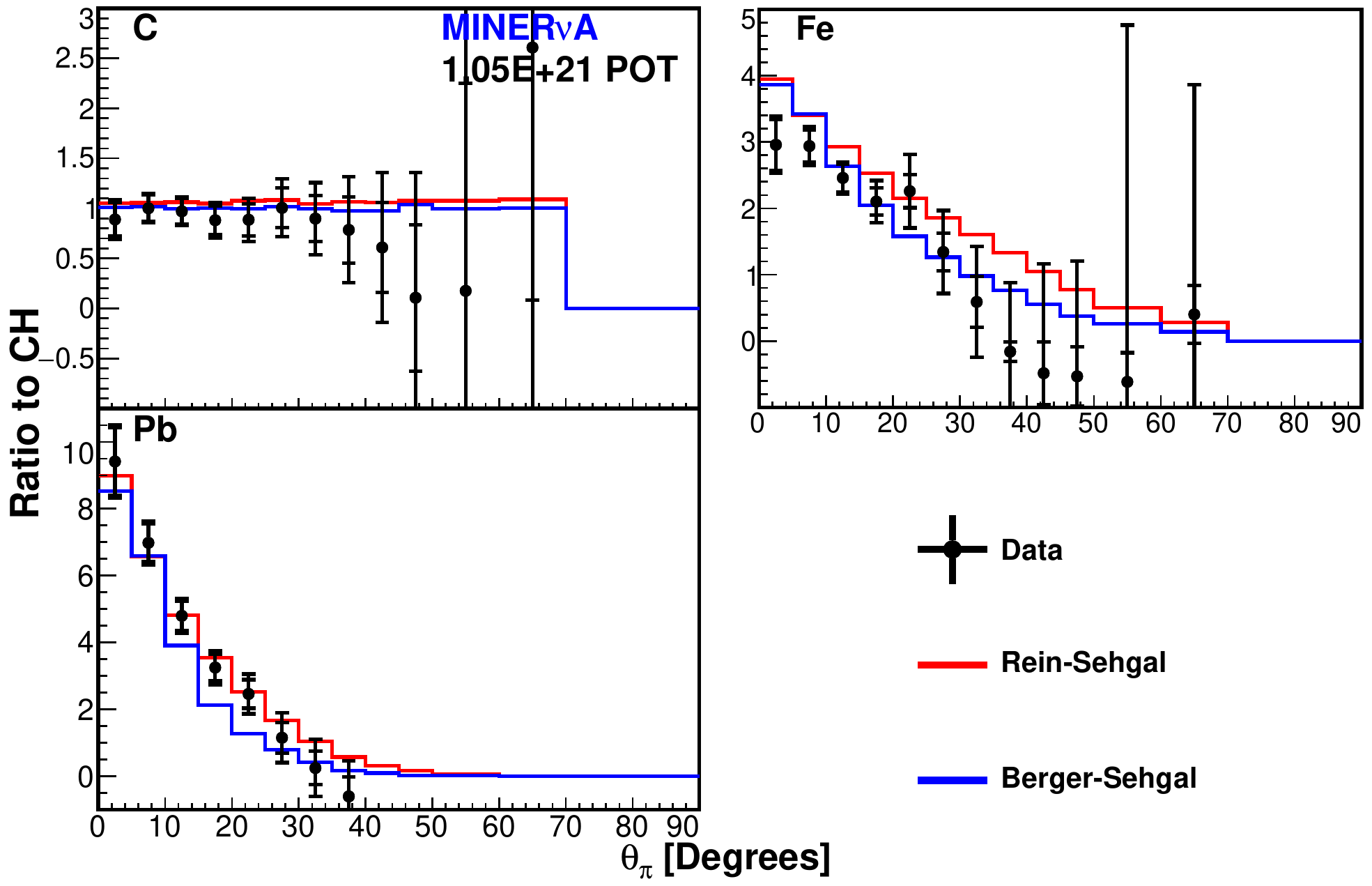}
        \caption{Ratios of the differential cross section as a function of $\theta_{\pi}$. C, Fe and Pb with respect to CH, in reading order. Data is compared to the Rein-Sehgal (red) and Berger-Sehgal (blue) models.}
\label{fig:xsecRatios_thetapi}
\end{figure}

\begin{figure}[h!]
        \includegraphics[width=0.65\linewidth]{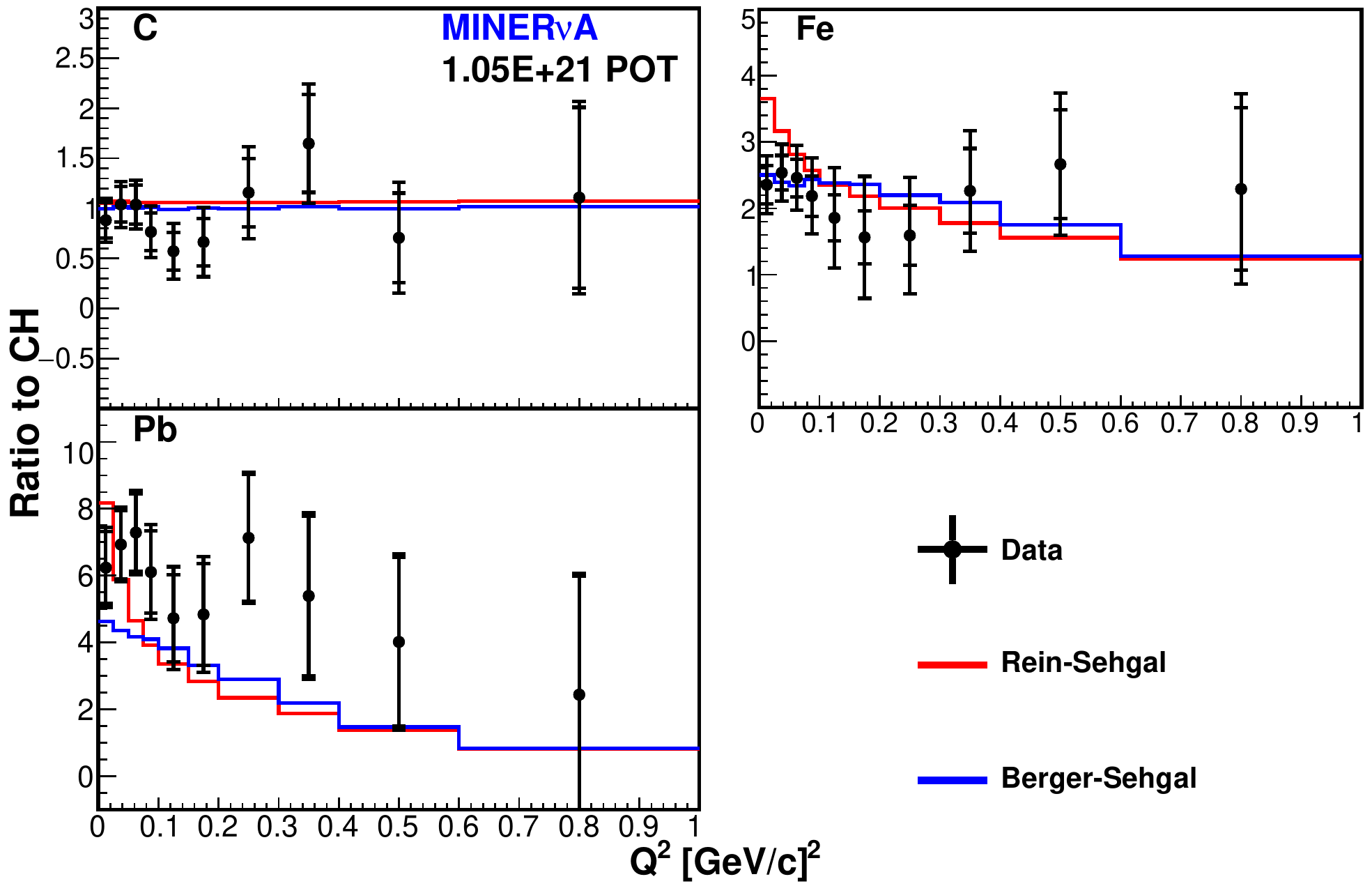}
        \caption{Ratios of the differential cross section as a function of $Q^{2}$. C, Fe and Pb with respect to CH, in reading order. Data is compared to the Rein-Sehgal (red) and Berger-Sehgal (blue) models.}
\label{fig:xsecRatios_q2}
\end{figure}

\begin{figure}[h!]
        \includegraphics[width=0.65\linewidth]{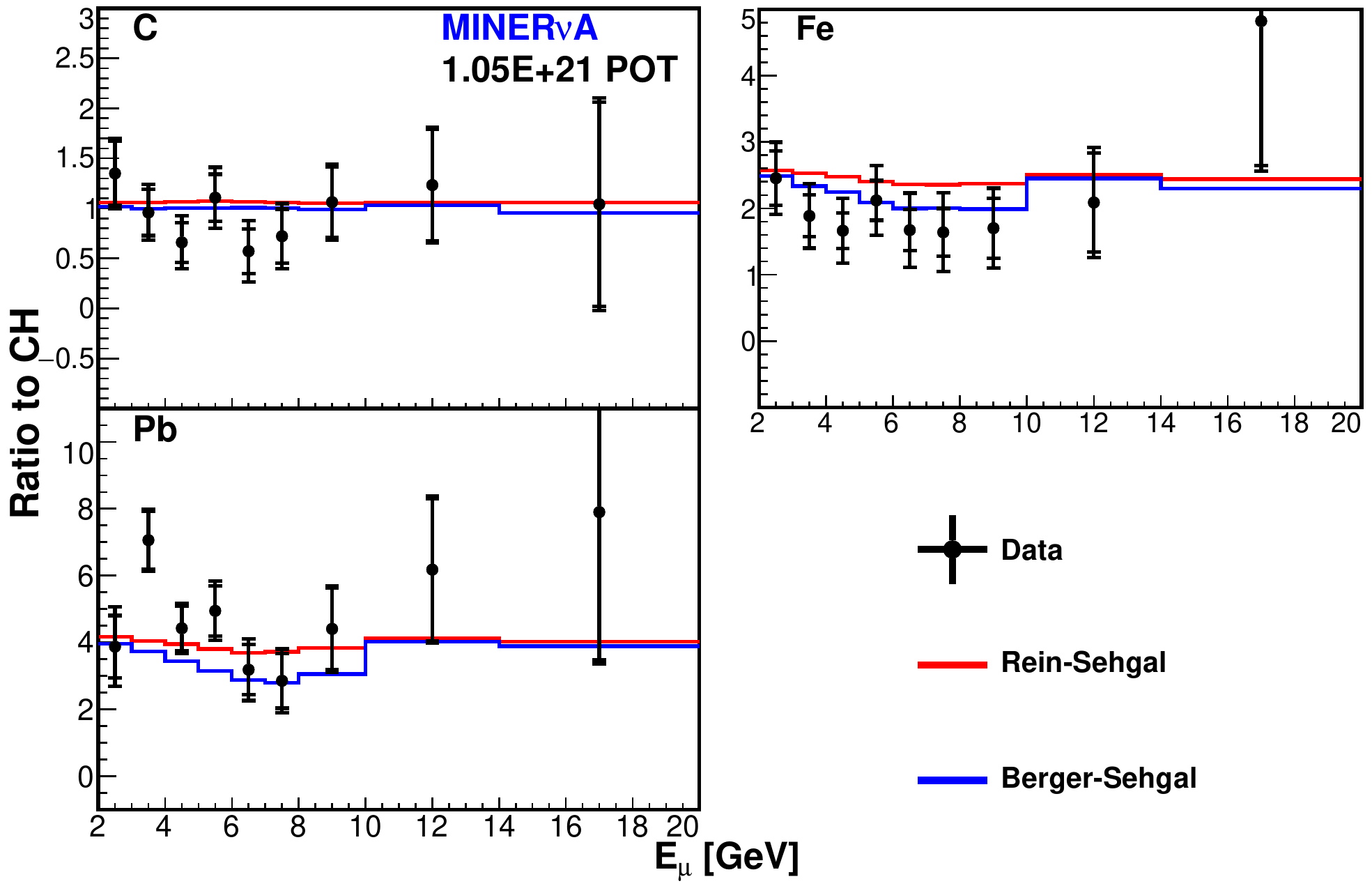}
        \caption{Ratios of the differential cross section as a function of $E_{\mu}$. C, Fe and Pb with respect to CH, in reading order. Data is compared to the Rein-Sehgal (red) and Berger-Sehgal (blue) models.}
\label{fig:xsecRatios_emu}
\end{figure}

\begin{figure}[h!]
        \includegraphics[width=0.65\linewidth]{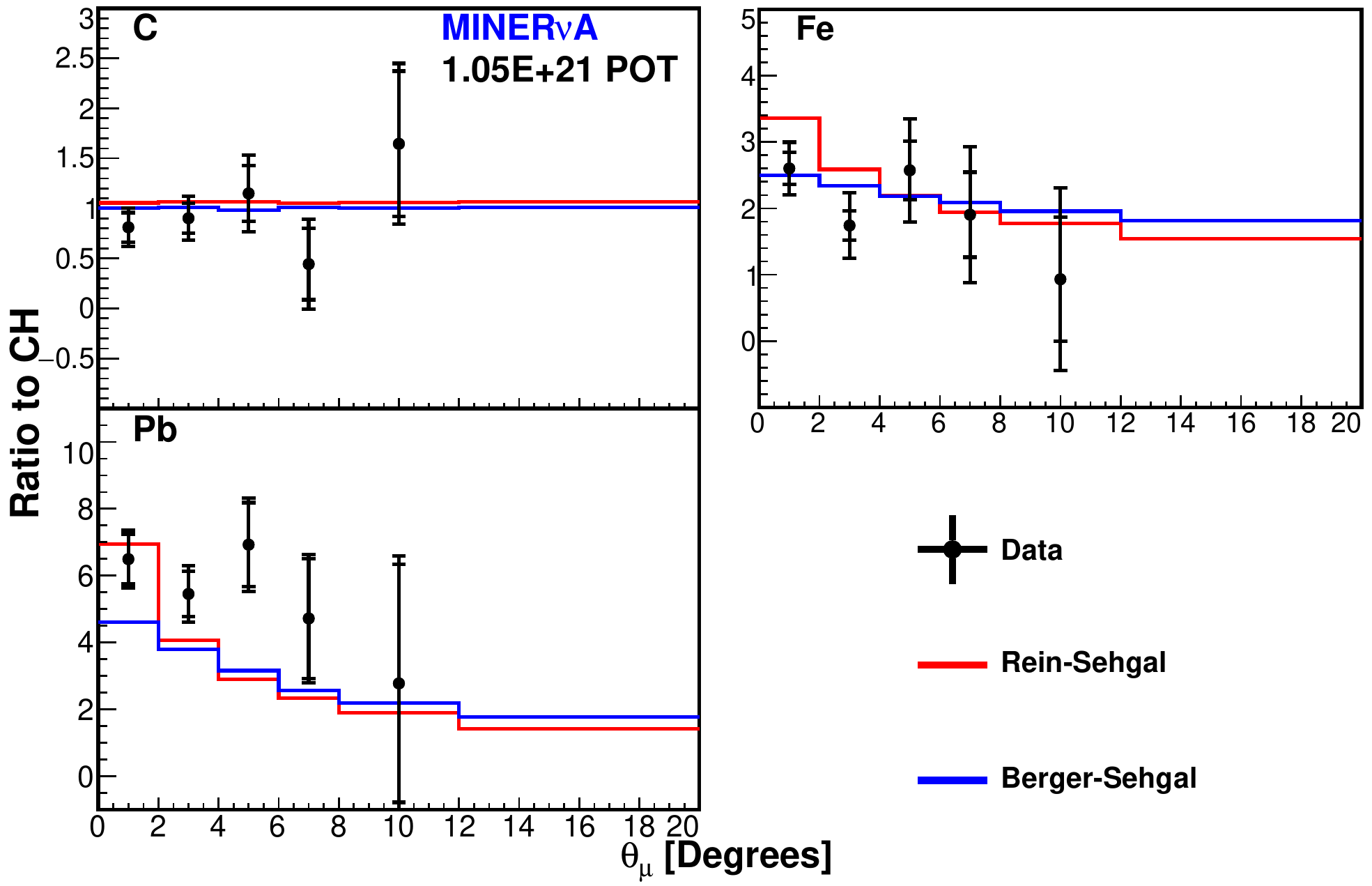}
        \caption{Ratios of the differential cross section as a function of $\theta_{\mu}$. C, Fe and Pb with respect to CH, in reading order. Data is compared to the Rein-Sehgal (red) and Berger-Sehgal (blue) models.}
\label{fig:xsecRatios_thetamu}
\end{figure}
\clearpage
\pagebreak[4]

\section{Cross section tables}
\label{section:xsecTables}

\begin{table}[h!]
  \centering
  \caption{Measured cross section as a function of $E_{\nu}$ on CH, in units of
  $10^{-39}$ $\text{cm}^{2}$/GeV/$^{12}$CH, and the absolute and
  fractional cross section uncertainties.}
\begin{tabular}{c||c||c|c||c|c}
      Bin edges (GeV)  &  $\sigma$  &  Abs. Stat. Unc. & Abs. Tot. Unc.  &  Frac. Stat. Unc.  &  Frac. Tot. Unc. \\
      \hline\hline
        2.0 - 3.0 & 2.377 & 0.254 & 0.586 & 0.107 & 0.246 \\
        3.0 - 4.0 & 3.317 & 0.144 & 0.486 & 0.043 & 0.147 \\
        4.0 - 5.0 & 4.183 & 0.127 & 0.504 & 0.030 & 0.120 \\
        5.0 - 6.0 & 4.661 & 0.114 & 0.478 & 0.024 & 0.103 \\
        6.0 - 7.0 & 4.888 & 0.112 & 0.452 & 0.023 & 0.092 \\
        7.0 - 8.0 & 5.512 & 0.137 & 0.534 & 0.025 & 0.097 \\
       8.0 - 10.0 & 6.836 & 0.209 & 0.785 & 0.031 & 0.115 \\
      10.0 - 14.0 & 10.097 & 0.482 & 1.177 & 0.048 & 0.117 \\
      14.0 - 20.0 & 12.147 & 1.064 & 1.836 & 0.088 & 0.151 \\
    \end{tabular}
  \label{table:EnuXSec_CH}
\end{table}

\begin{table}[h!]
  \centering
  \caption{Statistical covariance matrix of the measured $\sigma(E_{\nu})$ on CH, in units of $10^{-78}$ (cm$^{2}$/GeV/CH)$^{2}$.}
    \begin{tabular}{c||ccccccccc}
      Bin edges (GeV) & 2.0 - 3.0 & 3.0 - 4.0 & 4.0 - 5.0 & 5.0 - 6.0 & 6.0 - 7.0 & 7.0 - 8.0 & 8.0 - 10.0 & 10.0 - 14.0 & 14.0 - 20.0 \\
\hline \hline
        2.0 - 3.0 &  0.0644 &  0.0020 & -0.0020 & -0.0005 & -0.0001 & -0.0001 & -0.0001 & -0.0001 & -0.0001 \\
        3.0 - 4.0 &  0.0020 &  0.0206 &  0.0020 & -0.0014 & -0.0006 & -0.0002 & -0.0001 & -0.0001 & -0.0001 \\
        4.0 - 5.0 & -0.0020 &  0.0020 &  0.0163 &  0.0031 & -0.0016 & -0.0012 & -0.0006 & -0.0002 & -0.0002 \\
        5.0 - 6.0 & -0.0005 & -0.0014 &  0.0031 &  0.0130 &  0.0041 & -0.0016 & -0.0021 & -0.0010 & -0.0003 \\
        6.0 - 7.0 & -0.0001 & -0.0006 & -0.0016 &  0.0041 &  0.0126 &  0.0071 & -0.0019 & -0.0037 & -0.0012 \\
        7.0 - 8.0 & -0.0001 & -0.0002 & -0.0012 & -0.0016 &  0.0071 &  0.0187 &  0.0123 & -0.0061 & -0.0051 \\
       8.0 - 10.0 & -0.0001 & -0.0001 & -0.0006 & -0.0021 & -0.0019 &  0.0123 &  0.0436 &  0.0211 & -0.0159 \\
      10.0 - 14.0 & -0.0001 & -0.0001 & -0.0002 & -0.0010 & -0.0037 & -0.0061 &  0.0211 &  0.2323 &  0.0425 \\
      14.0 - 20.0 & -0.0001 & -0.0001 & -0.0002 & -0.0003 & -0.0012 & -0.0051 & -0.0159 &  0.0425 &  1.1320 \\
    \end{tabular}
  \label{tablecov:CH_Statistical_Enu}
\end{table}

\begin{table}[h!]
  \centering
  \caption{Systematic covariance matrix of the measured $\sigma(E_{\nu})$ on CH, in units of $10^{-78}$ (cm$^{2}$/GeV/CH)$^{2}$.}
    \begin{tabular}{c||ccccccccc}
      Bin edges (GeV) & 2.0 - 3.0 & 3.0 - 4.0 & 4.0 - 5.0 & 5.0 - 6.0 & 6.0 - 7.0 & 7.0 - 8.0 & 8.0 - 10.0 & 10.0 - 14.0 & 14.0 - 20.0 \\
\hline \hline
        2.0 - 3.0 &  0.2790 &  0.2301 &  0.2266 &  0.1924 &  0.1171 &  0.0043 & -0.1085 & -0.1252 &  0.1389 \\
        3.0 - 4.0 &  0.2301 &  0.2160 &  0.2209 &  0.1998 &  0.1455 &  0.0598 & -0.0266 & -0.0267 &  0.2827 \\
        4.0 - 5.0 &  0.2266 &  0.2209 &  0.2374 &  0.2200 &  0.1673 &  0.0824 & -0.0027 &  0.0159 &  0.3826 \\
        5.0 - 6.0 &  0.1924 &  0.1998 &  0.2200 &  0.2158 &  0.1828 &  0.1236 &  0.0625 &  0.0847 &  0.4231 \\
        6.0 - 7.0 &  0.1171 &  0.1455 &  0.1673 &  0.1828 &  0.1913 &  0.1871 &  0.1828 &  0.2154 &  0.4366 \\
        7.0 - 8.0 &  0.0043 &  0.0598 &  0.0824 &  0.1236 &  0.1871 &  0.2667 &  0.3574 &  0.4281 &  0.4455 \\
       8.0 - 10.0 & -0.1085 & -0.0266 & -0.0027 &  0.0625 &  0.1828 &  0.3574 &  0.5728 &  0.7394 &  0.5060 \\
      10.0 - 14.0 & -0.1252 & -0.0267 &  0.0159 &  0.0847 &  0.2154 &  0.4281 &  0.7394 &  1.1522 &  0.9222 \\
      14.0 - 20.0 &  0.1389 &  0.2827 &  0.3826 &  0.4231 &  0.4366 &  0.4455 &  0.5060 &  0.9222 &  2.2386 \\
    \end{tabular}
  \label{tablecov:CH_Systematic_Enu}
\end{table}

\begin{table}[h!]
  \centering
  \caption{Measured cross section as a function of $E_{\nu}$ on C, in units of
  $10^{-39}$ $\text{cm}^{2}$/GeV/$^{12}$C, and the absolute and
  fractional cross section uncertainties.}
    \begin{tabular}{c||c||c|c||c|c}
      Bin edges (GeV)  &  $\sigma$  &  Abs. Stat. Unc. & Abs. Tot. Unc.  &  Frac. Stat. Unc.  &  Frac. Tot. Unc. \\
      \hline\hline
        2.0 - 3.0 & 2.764 & 2.219 & 2.863 & 0.803 & 1.036 \\
        3.0 - 4.0 & 3.712 & 1.252 & 1.687 & 0.337 & 0.454 \\
        4.0 - 5.0 & 5.088 & 1.120 & 1.685 & 0.220 & 0.331 \\
        5.0 - 6.0 & 3.236 & 0.896 & 1.336 & 0.277 & 0.413 \\
        6.0 - 7.0 & 2.848 & 0.835 & 1.297 & 0.293 & 0.455 \\
        7.0 - 8.0 & 5.287 & 1.189 & 1.537 & 0.225 & 0.291 \\
       8.0 - 10.0 & 6.614 & 1.644 & 2.114 & 0.249 & 0.320 \\
      10.0 - 14.0 & 13.743 & 4.093 & 4.476 & 0.298 & 0.326 \\
      14.0 - 20.0 & 6.877 & 6.947 & 7.366 & 1.010 & 1.071 \\
    \end{tabular}
  \label{table:EnuXSec_C}
\end{table}

\begin{table}[h!]
  \centering
  \caption{Statistical covariance matrix of the measured $\sigma(E_{\nu})$ on C, in units of $10^{-78}$ (cm$^{2}$/GeV/C)$^{2}$.}
    \begin{tabular}{c||ccccccccc}
      Bin edges (GeV) & 2.0 - 3.0 & 3.0 - 4.0 & 4.0 - 5.0 & 5.0 - 6.0 & 6.0 - 7.0 & 7.0 - 8.0 & 8.0 - 10.0 & 10.0 - 14.0 & 14.0 - 20.0 \\
\hline \hline
        2.0 - 3.0 &  4.9248 &  0.1154 & -0.1384 & -0.0182 & -0.0012 & -0.0035 &  0.0010 & -0.0031 & -0.0023 \\
        3.0 - 4.0 &  0.1154 &  1.5675 &  0.0466 & -0.0840 & -0.0237 & -0.0049 & -0.0049 & -0.0056 &  0.0007 \\
        4.0 - 5.0 & -0.1384 &  0.0466 &  1.2547 &  0.1217 & -0.0881 & -0.0675 & -0.0227 & -0.0139 & -0.0034 \\
        5.0 - 6.0 & -0.0182 & -0.0840 &  0.1217 &  0.8025 &  0.2216 & -0.1081 & -0.0935 & -0.0354 & -0.0106 \\
        6.0 - 7.0 & -0.0012 & -0.0237 & -0.0881 &  0.2216 &  0.6965 &  0.3806 & -0.1354 & -0.1647 & -0.0150 \\
        7.0 - 8.0 & -0.0035 & -0.0049 & -0.0675 & -0.1081 &  0.3806 &  1.4138 &  0.5926 & -0.5660 & -0.1693 \\
       8.0 - 10.0 &  0.0010 & -0.0049 & -0.0227 & -0.0935 & -0.1354 &  0.5926 &  2.7026 &  0.7314 & -0.5316 \\
      10.0 - 14.0 & -0.0031 & -0.0056 & -0.0139 & -0.0354 & -0.1647 & -0.5660 &  0.7314 & 16.7528 &  1.5120 \\
      14.0 - 20.0 & -0.0023 &  0.0007 & -0.0034 & -0.0106 & -0.0150 & -0.1693 & -0.5316 &  1.5120 & 48.2633 \\
    \end{tabular}
  \label{tablecov:C_Statistical_Enu}
\end{table}

\begin{table}[h!]
  \centering
  \caption{Systematic covariance matrix of the measured $\sigma(E_{\nu})$ on C, in units of $10^{-78}$ (cm$^{2}$/GeV/C)$^{2}$.}
    \begin{tabular}{c||ccccccccc}
      Bin edges (GeV) & 2.0 - 3.0 & 3.0 - 4.0 & 4.0 - 5.0 & 5.0 - 6.0 & 6.0 - 7.0 & 7.0 - 8.0 & 8.0 - 10.0 & 10.0 - 14.0 & 14.0 - 20.0 \\
\hline \hline
        2.0 - 3.0 &  3.2724 &  1.7230 &  1.8016 &  0.9754 &  0.9961 &  0.7259 & -0.9863 & -0.5744 &  0.7490 \\
        3.0 - 4.0 &  1.7230 &  1.2776 &  1.3792 &  0.9590 &  0.9541 &  0.7800 & -0.1119 &  0.1359 &  1.3682 \\
        4.0 - 5.0 &  1.8016 &  1.3792 &  1.5859 &  1.1263 &  1.1064 &  0.9143 & -0.0131 &  0.3331 &  1.7378 \\
        5.0 - 6.0 &  0.9754 &  0.9590 &  1.1263 &  0.9813 &  0.9464 &  0.8497 &  0.4505 &  0.6405 &  1.8201 \\
        6.0 - 7.0 &  0.9961 &  0.9541 &  1.1064 &  0.9464 &  0.9860 &  0.9121 &  0.4968 &  0.6462 &  1.6526 \\
        7.0 - 8.0 &  0.7259 &  0.7800 &  0.9143 &  0.8497 &  0.9121 &  0.9484 &  0.7463 &  0.9362 &  1.5766 \\
       8.0 - 10.0 & -0.9863 & -0.1119 & -0.0131 &  0.4505 &  0.4968 &  0.7463 &  1.7679 &  1.9288 &  1.7080 \\
      10.0 - 14.0 & -0.5744 &  0.1359 &  0.3331 &  0.6405 &  0.6462 &  0.9362 &  1.9288 &  3.2803 &  2.6492 \\
      14.0 - 20.0 &  0.7490 &  1.3682 &  1.7378 &  1.8201 &  1.6526 &  1.5766 &  1.7080 &  2.6492 &  5.9903 \\
    \end{tabular}
  \label{tablecov:C_Systematic_Enu}
\end{table}

\begin{table}[h!]
  \centering
  \caption{Measured cross section as a function of $E_{\nu}$ on Fe, in units of
  $10^{-39}$ $\text{cm}^{2}$/GeV/$^{56}$Fe, and the absolute and
  fractional cross section uncertainties.}
    \begin{tabular}{c||c||c|c||c|c}
      Bin edges (GeV)  &  $\sigma$  &  Abs. Stat. Unc. & Abs. Tot. Unc.  &  Frac. Stat. Unc.  &  Frac. Tot. Unc. \\
      \hline\hline
        2.0 - 3.0 & 3.645 & 2.600 & 4.099 & 0.713 & 1.125 \\
        3.0 - 4.0 & 7.275 & 1.651 & 3.116 & 0.227 & 0.428 \\
        4.0 - 5.0 & 6.119 & 1.305 & 2.862 & 0.213 & 0.468 \\
        5.0 - 6.0 & 7.976 & 1.178 & 2.914 & 0.148 & 0.365 \\
        6.0 - 7.0 & 9.672 & 1.211 & 2.726 & 0.125 & 0.282 \\
        7.0 - 8.0 & 11.415 & 1.424 & 2.895 & 0.125 & 0.254 \\
       8.0 - 10.0 & 15.748 & 2.205 & 3.734 & 0.140 & 0.237 \\
      10.0 - 14.0 & 24.823 & 5.127 & 6.448 & 0.207 & 0.260 \\
      14.0 - 20.0 & 29.653 & 11.370 & 13.041 & 0.383 & 0.440 \\
    \end{tabular}
  \label{table:EnuXSec_Fe}
\end{table}

\begin{table}[h!]
  \centering
  \caption{Statistical covariance matrix of the measured $\sigma(E_{\nu})$ on Fe, in units of $10^{-78}$ (cm$^{2}$/GeV/Fe)$^{2}$.}
    \begin{tabular}{c||ccccccccc}
      Bin edges (GeV) & 2.0 - 3.0 & 3.0 - 4.0 & 4.0 - 5.0 & 5.0 - 6.0 & 6.0 - 7.0 & 7.0 - 8.0 & 8.0 - 10.0 & 10.0 - 14.0 & 14.0 - 20.0 \\
\hline \hline
        2.0 - 3.0 &  6.7618 &  0.2772 & -0.1783 & -0.0454 & -0.0060 & -0.0041 & -0.0047 & -0.0097 & -0.0004 \\
        3.0 - 4.0 &  0.2772 &  2.7247 &  0.1849 & -0.1708 & -0.0691 & -0.0199 & -0.0121 & -0.0162 & -0.0188 \\
        4.0 - 5.0 & -0.1783 &  0.1849 &  1.7034 &  0.3364 & -0.1621 & -0.1079 & -0.0466 & -0.0211 & -0.0227 \\
        5.0 - 6.0 & -0.0454 & -0.1708 &  0.3364 &  1.3873 &  0.4109 & -0.1928 & -0.2062 & -0.0926 & -0.0349 \\
        6.0 - 7.0 & -0.0060 & -0.0691 & -0.1621 &  0.4109 &  1.4674 &  0.7105 & -0.2779 & -0.3884 & -0.1054 \\
        7.0 - 8.0 & -0.0041 & -0.0199 & -0.1079 & -0.1928 &  0.7105 &  2.0276 &  1.1923 & -0.7401 & -0.4978 \\
       8.0 - 10.0 & -0.0047 & -0.0121 & -0.0466 & -0.2062 & -0.2779 &  1.1923 &  4.8623 &  2.0433 & -1.7170 \\
      10.0 - 14.0 & -0.0097 & -0.0162 & -0.0211 & -0.0926 & -0.3884 & -0.7401 &  2.0433 & 26.2871 &  3.7779 \\
      14.0 - 20.0 & -0.0004 & -0.0188 & -0.0227 & -0.0349 & -0.1054 & -0.4978 & -1.7170 &  3.7779 & 129.2868 \\
    \end{tabular}
  \label{tablecov:Fe_Statistical_Enu}
\end{table}

\begin{table}[h!]
  \centering
  \caption{Systematic covariance matrix of the measured $\sigma(E_{\nu})$ on Fe, in units of $10^{-78}$ (cm$^{2}$/GeV/Fe)$^{2}$.}
    \begin{tabular}{c||ccccccccc}
      Bin edges (GeV) & 2.0 - 3.0 & 3.0 - 4.0 & 4.0 - 5.0 & 5.0 - 6.0 & 6.0 - 7.0 & 7.0 - 8.0 & 8.0 - 10.0 & 10.0 - 14.0 & 14.0 - 20.0 \\
\hline \hline
        2.0 - 3.0 & 10.0401 &  7.5886 &  6.5583 &  6.5540 &  4.1295 &  2.1339 &  0.2314 &  1.3701 &  4.8032 \\
        3.0 - 4.0 &  7.5886 &  6.9875 &  6.4540 &  6.6337 &  5.0889 &  3.7535 &  2.4344 &  3.5779 &  8.9142 \\
        4.0 - 5.0 &  6.5583 &  6.4540 &  6.4872 &  6.7361 &  5.6069 &  4.5848 &  3.6393 &  4.8048 & 10.8173 \\
        5.0 - 6.0 &  6.5540 &  6.6337 &  6.7361 &  7.1030 &  6.0831 &  5.1319 &  4.2520 &  5.4613 & 11.9276 \\
        6.0 - 7.0 &  4.1295 &  5.0889 &  5.6069 &  6.0831 &  5.9642 &  5.7731 &  5.6431 &  6.5948 & 12.3660 \\
        7.0 - 8.0 &  2.1339 &  3.7535 &  4.5848 &  5.1319 &  5.7731 &  6.3506 &  7.0722 &  7.9518 & 12.3611 \\
       8.0 - 10.0 &  0.2314 &  2.4344 &  3.6393 &  4.2520 &  5.6431 &  7.0722 &  9.0818 & 10.6140 & 12.8714 \\
      10.0 - 14.0 &  1.3701 &  3.5779 &  4.8048 &  5.4613 &  6.5948 &  7.9518 & 10.6140 & 15.2864 & 18.2801 \\
      14.0 - 20.0 &  4.8032 &  8.9142 & 10.8173 & 11.9276 & 12.3660 & 12.3611 & 12.8714 & 18.2801 & 40.7697 \\
    \end{tabular}
  \label{tablecov:Fe_Systematic_Enu}
\end{table}

\begin{table}[h!]
  \centering
  \caption{Measured cross section as a function of $E_{\nu}$ on Pb, in units of
  $10^{-39}$ $\text{cm}^{2}$/GeV/$^{207}$Pb, and the absolute and
  fractional cross section uncertainties.}
    \begin{tabular}{c||c||c|c||c|c}
      Bin edges (GeV)  &  $\sigma$  &  Abs. Stat. Unc. & Abs. Tot. Unc.  &  Frac. Stat. Unc.  &  Frac. Tot. Unc. \\
      \hline\hline
        2.0 - 3.0 & -2.085 & 4.016 & 5.184 & -1.926 & -2.487 \\
        3.0 - 4.0 & 13.667 & 3.134 & 4.579 & 0.229 & 0.335 \\
        4.0 - 5.0 & 19.545 & 2.777 & 4.373 & 0.142 & 0.224 \\
        5.0 - 6.0 & 23.350 & 2.609 & 4.031 & 0.112 & 0.173 \\
        6.0 - 7.0 & 24.214 & 2.670 & 3.819 & 0.110 & 0.158 \\
        7.0 - 8.0 & 25.476 & 3.191 & 4.425 & 0.125 & 0.174 \\
       8.0 - 10.0 & 33.061 & 4.730 & 6.124 & 0.143 & 0.185 \\
      10.0 - 14.0 & 63.446 & 11.662 & 13.263 & 0.184 & 0.209 \\
      14.0 - 20.0 & 82.615 & 28.103 & 29.643 & 0.340 & 0.359 \\
    \end{tabular}
  \label{table:EnuXSec_Pb}
\end{table}

\begin{table}[h!]
  \centering
  \caption{Statistical covariance matrix of the measured $\sigma(E_{\nu})$ on Pb, in units of $10^{-78}$ (cm$^{2}$/GeV/Pb)$^{2}$.}
    \begin{tabular}{c||ccccccccc}
      Bin edges (GeV) & 2.0 - 3.0 & 3.0 - 4.0 & 4.0 - 5.0 & 5.0 - 6.0 & 6.0 - 7.0 & 7.0 - 8.0 & 8.0 - 10.0 & 10.0 - 14.0 & 14.0 - 20.0 \\
\hline \hline
        2.0 - 3.0 & 16.1250 &  3.2152 &  0.4663 &  0.0804 &  0.0356 &  0.0142 &  0.0150 &  0.0006 &  0.0251 \\
        3.0 - 4.0 &  3.2152 &  9.8220 &  3.0035 &  0.5524 &  0.1496 &  0.0768 &  0.0431 &  0.0582 &  0.1578 \\
        4.0 - 5.0 &  0.4663 &  3.0035 &  7.7132 &  3.3863 &  0.8484 &  0.2482 &  0.1378 &  0.1873 &  0.1200 \\
        5.0 - 6.0 &  0.0804 &  0.5524 &  3.3863 &  6.8069 &  3.7850 &  1.3710 &  0.5196 &  0.2818 &  0.1661 \\
        6.0 - 7.0 &  0.0356 &  0.1496 &  0.8484 &  3.7850 &  7.1306 &  5.1985 &  2.3620 &  1.0254 &  0.5773 \\
        7.0 - 8.0 &  0.0142 &  0.0768 &  0.2482 &  1.3710 &  5.1985 & 10.1811 &  9.3051 &  4.2276 &  0.9908 \\
       8.0 - 10.0 &  0.0150 &  0.0431 &  0.1378 &  0.5196 &  2.3620 &  9.3051 & 22.3683 & 24.5201 &  8.7950 \\
      10.0 - 14.0 &  0.0006 &  0.0582 &  0.1873 &  0.2818 &  1.0254 &  4.2276 & 24.5201 & 136.0103 & 107.0414 \\
      14.0 - 20.0 &  0.0251 &  0.1578 &  0.1200 &  0.1661 &  0.5773 &  0.9908 &  8.7950 & 107.0414 & 789.7512 \\
    \end{tabular}
  \label{tablecov:Pb_Statistical_Enu}
\end{table}

\begin{table}[h!]
  \centering
  \caption{Systematic covariance matrix of the measured $\sigma(E_{\nu})$ on Pb, in units of $10^{-78}$ (cm$^{2}$/GeV/Pb)$^{2}$.}
    \begin{tabular}{c||ccccccccc}
      Bin edges (GeV) & 2.0 - 3.0 & 3.0 - 4.0 & 4.0 - 5.0 & 5.0 - 6.0 & 6.0 - 7.0 & 7.0 - 8.0 & 8.0 - 10.0 & 10.0 - 14.0 & 14.0 - 20.0 \\
\hline \hline
        2.0 - 3.0 & 10.7501 & 10.2017 &  9.4881 &  7.6136 &  4.6177 &  1.5630 & -0.6901 & -1.1925 &  2.2098 \\
        3.0 - 4.0 & 10.2017 & 11.1466 & 10.9766 &  9.1403 &  5.5968 &  1.9597 & -0.7532 & -0.4484 &  5.2480 \\
        4.0 - 5.0 &  9.4881 & 10.9766 & 11.4078 &  9.9918 &  6.7780 &  3.4311 &  0.8750 &  2.1669 &  9.8261 \\
        5.0 - 6.0 &  7.6136 &  9.1403 &  9.9918 &  9.4440 &  7.3828 &  5.1562 &  3.4300 &  5.5130 & 13.2359 \\
        6.0 - 7.0 &  4.6177 &  5.5968 &  6.7780 &  7.3828 &  7.4514 &  7.2947 &  7.1126 & 10.0302 & 15.8811 \\
        7.0 - 8.0 &  1.5630 &  1.9597 &  3.4311 &  5.1562 &  7.2947 &  9.3996 & 11.1009 & 15.2224 & 18.2495 \\
       8.0 - 10.0 & -0.6901 & -0.7532 &  0.8750 &  3.4300 &  7.1126 & 11.1009 & 15.1331 & 22.2243 & 22.8958 \\
      10.0 - 14.0 & -1.1925 & -0.4484 &  2.1669 &  5.5130 & 10.0302 & 15.2224 & 22.2243 & 39.9002 & 45.8082 \\
      14.0 - 20.0 &  2.2098 &  5.2480 &  9.8261 & 13.2359 & 15.8811 & 18.2495 & 22.8958 & 45.8082 & 88.9614 \\
    \end{tabular}
  \label{tablecov:Pb_Systematic_Enu}
\end{table}

\begin{table}[h!]
  \centering
  \caption{Measured cross section as a function of $E_{\pi}$ on CH, in units of
  $10^{-39}$ $\text{cm}^{2}$/GeV/$^{12}$CH, and the absolute and
  fractional cross section uncertainties.}
    \begin{tabular}{c||c||c|c||c|c}
      Bin edges (GeV)  &  $d\sigma/dE_{\pi}$  &  Abs. Stat. Unc. & Abs. Tot. Unc.  &  Frac. Stat. Unc.  &  Frac. Tot. Unc. \\
      \hline\hline
        0 - 0.25 & 0.708 & 0.053 & 0.278 & 0.074 & 0.392 \\
      0.25 - 0.5 & 3.440 & 0.107 & 0.563 & 0.031 & 0.164 \\
      0.5 - 0.75 & 1.725 & 0.063 & 0.235 & 0.037 & 0.136 \\
      0.75 - 1.0 & 1.974 & 0.055 & 0.197 & 0.028 & 0.100 \\
       1.0 - 1.5 & 1.629 & 0.043 & 0.168 & 0.026 & 0.103 \\
       1.5 - 2.0 & 1.246 & 0.035 & 0.137 & 0.028 & 0.110 \\
       2.0 - 3.0 & 0.880 & 0.026 & 0.086 & 0.030 & 0.098 \\
       3.0 - 6.0 & 0.299 & 0.012 & 0.041 & 0.040 & 0.137 \\
    \end{tabular}
  \label{table:EpiXSec_CH}
\end{table}

\begin{table}[h!]
  \centering
  \caption{Statistical covariance matrix of the measured $d\sigma/dE_{\pi}$ on CH, in units of $10^{-80}$ (cm$^{2}$/GeV/CH)$^{2}$.}
    \begin{tabular}{c||cccccccc}
      Bin edges (GeV) & 0 - 0.25 & 0.25 - 0.5 & 0.5 - 0.75 & 0.75 - 1.0 & 1.0 - 1.5 & 1.5 - 2.0 & 2.0 - 3.0 & 3.0 - 6.0 \\
\hline \hline
        0 - 0.25 &  0.2760 & -0.0408 & -0.0353 & -0.0106 &  0.0010 &  0.0018 &  0.0005 &  0.0000 \\
      0.25 - 0.5 & -0.0408 &  1.1345 &  0.0136 & -0.0975 & -0.0497 & -0.0050 &  0.0030 &  0.0006 \\
      0.5 - 0.75 & -0.0353 &  0.0136 &  0.4026 &  0.1914 & -0.0234 & -0.0415 & -0.0085 &  0.0008 \\
      0.75 - 1.0 & -0.0106 & -0.0975 &  0.1914 &  0.3024 &  0.1234 & -0.0285 & -0.0261 & -0.0008 \\
       1.0 - 1.5 &  0.0010 & -0.0497 & -0.0234 &  0.1234 &  0.1823 &  0.0665 & -0.0158 & -0.0056 \\
       1.5 - 2.0 &  0.0018 & -0.0050 & -0.0415 & -0.0285 &  0.0665 &  0.1259 &  0.0451 & -0.0097 \\
       2.0 - 3.0 &  0.0005 &  0.0030 & -0.0085 & -0.0261 & -0.0158 &  0.0451 &  0.0685 &  0.0037 \\
       3.0 - 6.0 &  0.0000 &  0.0006 &  0.0008 & -0.0008 & -0.0056 & -0.0097 &  0.0037 &  0.0140 \\
    \end{tabular}
  \label{tablecov:CH_Statistical_Epi}
\end{table}

\begin{table}[h!]
  \centering
  \caption{Systematic covariance matrix of the measured $d\sigma/dE_{\pi}$ on CH, in units of $10^{-80}$ (cm$^{2}$/GeV/CH)$^{2}$.}
    \begin{tabular}{c||cccccccc}
      Bin edges (GeV) & 0 - 0.25 & 0.25 - 0.5 & 0.5 - 0.75 & 0.75 - 1.0 & 1.0 - 1.5 & 1.5 - 2.0 & 2.0 - 3.0 & 3.0 - 6.0 \\
\hline \hline
        0 - 0.25 &  7.4352 &  6.6553 & -2.1796 &  0.0648 &  1.7209 &  1.7141 &  0.3655 & -0.6048 \\
      0.25 - 0.5 &  6.6553 & 30.5695 &  5.9236 &  3.6695 &  2.3474 &  1.7564 &  1.0188 &  0.0588 \\
      0.5 - 0.75 & -2.1796 &  5.9236 &  5.1158 &  3.3984 &  1.3836 &  0.6822 &  0.9287 &  0.6777 \\
      0.75 - 1.0 &  0.0648 &  3.6695 &  3.3984 &  3.5701 &  2.6030 &  1.8554 &  1.3722 &  0.5266 \\
       1.0 - 1.5 &  1.7209 &  2.3474 &  1.3836 &  2.6030 &  2.6241 &  2.0963 &  1.2584 &  0.2569 \\
       1.5 - 2.0 &  1.7141 &  1.7564 &  0.6822 &  1.8554 &  2.0963 &  1.7386 &  0.9981 &  0.1575 \\
       2.0 - 3.0 &  0.3655 &  1.0188 &  0.9287 &  1.3722 &  1.2584 &  0.9981 &  0.6778 &  0.2097 \\
       3.0 - 6.0 & -0.6048 &  0.0588 &  0.6777 &  0.5266 &  0.2569 &  0.1575 &  0.2097 &  0.1529 \\
    \end{tabular}
  \label{tablecov:CH_Systematic_Epi}
\end{table}

\begin{table}[h!]
  \centering
  \caption{Measured cross section as a function of $E_{\pi}$ on C, in units of
  $10^{-39}$ $\text{cm}^{2}$/GeV/$^{12}$C, and the absolute and
  fractional cross section uncertainties.}
    \begin{tabular}{c||c||c|c||c|c}
      Bin edges (GeV)  &  $d\sigma/dE_{\pi}$  &  Abs. Stat. Unc. & Abs. Tot. Unc.  &  Frac. Stat. Unc.  &  Frac. Tot. Unc. \\
      \hline\hline
        0 - 0.25 & 0.308 & 0.210 & 0.350 & 0.682 & 1.135 \\
      0.25 - 0.5 & 2.893 & 0.698 & 1.682 & 0.241 & 0.581 \\
      0.5 - 0.75 & 1.636 & 0.373 & 0.649 & 0.228 & 0.397 \\
      0.75 - 1.0 & 1.929 & 0.353 & 0.570 & 0.183 & 0.295 \\
       1.0 - 1.5 & 1.587 & 0.276 & 0.413 & 0.174 & 0.260 \\
       1.5 - 2.0 & 1.334 & 0.233 & 0.287 & 0.174 & 0.215 \\
       2.0 - 3.0 & 0.913 & 0.171 & 0.192 & 0.188 & 0.210 \\
       3.0 - 6.0 & 0.251 & 0.080 & 0.100 & 0.317 & 0.399 \\
    \end{tabular}
  \label{table:EpiXSec_C}
\end{table}

\begin{table}[h!]
  \centering
  \caption{Statistical covariance matrix of the measured $d\sigma/dE_{\pi}$ on C, in units of $10^{-80}$ (cm$^{2}$/GeV/C)$^{2}$.}
    \begin{tabular}{c||cccccccc}
      Bin edges (GeV) & 0 - 0.25 & 0.25 - 0.5 & 0.5 - 0.75 & 0.75 - 1.0 & 1.0 - 1.5 & 1.5 - 2.0 & 2.0 - 3.0 & 3.0 - 6.0 \\
\hline \hline
        0 - 0.25 &  4.4170 &  5.7037 & -0.7952 & -0.6720 & -0.2486 & -0.0324 &  0.0016 & -0.0007 \\
      0.25 - 0.5 &  5.7037 & 48.6925 &  5.3324 & -1.6066 & -2.1061 & -0.8416 & -0.1604 & -0.0071 \\
      0.5 - 0.75 & -0.7952 &  5.3324 & 13.9190 &  8.5817 &  0.9126 & -1.1168 & -0.4201 & -0.0343 \\
      0.75 - 1.0 & -0.6720 & -1.6066 &  8.5817 & 12.4958 &  6.3305 & -0.1191 & -0.7724 & -0.0890 \\
       1.0 - 1.5 & -0.2486 & -2.1061 &  0.9126 &  6.3305 &  7.6198 &  3.1542 & -0.2496 & -0.1697 \\
       1.5 - 2.0 & -0.0324 & -0.8416 & -1.1168 & -0.1191 &  3.1542 &  5.4208 &  2.1045 & -0.2151 \\
       2.0 - 3.0 &  0.0016 & -0.1604 & -0.4201 & -0.7724 & -0.2496 &  2.1045 &  2.9358 &  0.3287 \\
       3.0 - 6.0 & -0.0007 & -0.0071 & -0.0343 & -0.0890 & -0.1697 & -0.2151 &  0.3287 &  0.6341 \\
    \end{tabular}
  \label{tablecov:C_Statistical_Epi}
\end{table}

\begin{table}[h!]
  \centering
  \caption{Systematic covariance matrix of the measured $d\sigma/dE_{\pi}$ on C, in units of $10^{-80}$ (cm$^{2}$/GeV/C)$^{2}$.}
    \begin{tabular}{c||cccccccc}
      Bin edges (GeV) & 0 - 0.25 & 0.25 - 0.5 & 0.5 - 0.75 & 0.75 - 1.0 & 1.0 - 1.5 & 1.5 - 2.0 & 2.0 - 3.0 & 3.0 - 6.0 \\
\hline \hline
        0 - 0.25 &  7.8336 & 37.0751 &  6.0245 &  2.0742 &  1.0786 &  0.6509 & -0.8443 & -1.0527 \\
      0.25 - 0.5 & 37.0751 & 234.2113 & 60.8254 & 33.8527 & 16.9079 &  6.5851 & -2.7988 & -4.4001 \\
      0.5 - 0.75 &  6.0245 & 60.8254 & 28.1731 & 21.6607 & 12.0346 &  4.7428 &  0.9579 & -0.2510 \\
      0.75 - 1.0 &  2.0742 & 33.8527 & 21.6607 & 20.0101 & 12.8231 &  5.8221 &  1.5520 & -0.0197 \\
       1.0 - 1.5 &  1.0786 & 16.9079 & 12.0346 & 12.8231 &  9.4153 &  4.8126 &  1.0894 & -0.2466 \\
       1.5 - 2.0 &  0.6509 &  6.5851 &  4.7428 &  5.8221 &  4.8126 &  2.8133 &  0.7237 & -0.1653 \\
       2.0 - 3.0 & -0.8443 & -2.7988 &  0.9579 &  1.5520 &  1.0894 &  0.7237 &  0.7465 &  0.3833 \\
       3.0 - 6.0 & -1.0527 & -4.4001 & -0.2510 & -0.0197 & -0.2466 & -0.1653 &  0.3833 &  0.3703 \\
    \end{tabular}
  \label{tablecov:C_Systematic_Epi}
\end{table}

\begin{table}[h!]
  \centering
  \caption{Measured cross section as a function of $E_{\pi}$ on Fe, in units of
  $10^{-39}$ $\text{cm}^{2}$/GeV/$^{56}$Fe, and the absolute and
  fractional cross section uncertainties.}
    \begin{tabular}{c||c||c|c||c|c}
      Bin edges (GeV)  &  $d\sigma/dE_{\pi}$  &  Abs. Stat. Unc. & Abs. Tot. Unc.  &  Frac. Stat. Unc.  &  Frac. Tot. Unc. \\
      \hline\hline
        0 - 0.25 & 1.984 & 0.613 & 1.262 & 0.309 & 0.636 \\
      0.25 - 0.5 & 2.804 & 0.880 & 3.226 & 0.314 & 1.150 \\
      0.5 - 0.75 & 2.318 & 0.521 & 1.419 & 0.225 & 0.612 \\
      0.75 - 1.0 & 3.641 & 0.482 & 0.978 & 0.132 & 0.269 \\
       1.0 - 1.5 & 4.654 & 0.441 & 0.856 & 0.095 & 0.184 \\
       1.5 - 2.0 & 3.695 & 0.391 & 0.509 & 0.106 & 0.138 \\
       2.0 - 3.0 & 2.176 & 0.275 & 0.327 & 0.126 & 0.150 \\
       3.0 - 6.0 & 0.986 & 0.156 & 0.216 & 0.158 & 0.219 \\
    \end{tabular}
  \label{table:EpiXSec_Fe}
\end{table}

\begin{table}[h!]
  \centering
  \caption{Statistical covariance matrix of the measured $d\sigma/dE_{\pi}$ on Fe, in units of $10^{-80}$ (cm$^{2}$/GeV/Fe)$^{2}$.}
    \begin{tabular}{c||cccccccc}
      Bin edges (GeV) & 0 - 0.25 & 0.25 - 0.5 & 0.5 - 0.75 & 0.75 - 1.0 & 1.0 - 1.5 & 1.5 - 2.0 & 2.0 - 3.0 & 3.0 - 6.0 \\
\hline \hline
        0 - 0.25 & 37.5947 &  2.3571 & -7.7105 & -4.3877 & -0.8628 &  0.4330 &  0.1895 &  0.0066 \\
      0.25 - 0.5 &  2.3571 & 77.4070 & 12.8532 & -3.0434 & -5.4244 & -1.5835 &  0.0508 &  0.0642 \\
      0.5 - 0.75 & -7.7105 & 12.8532 & 27.0992 & 15.7089 & -0.5231 & -3.5119 & -0.7574 &  0.0805 \\
      0.75 - 1.0 & -4.3877 & -3.0434 & 15.7089 & 23.2364 & 11.3064 & -3.7754 & -2.3192 &  0.0062 \\
       1.0 - 1.5 & -0.8628 & -5.4244 & -0.5231 & 11.3064 & 19.4475 &  6.2474 & -2.1927 & -0.6301 \\
       1.5 - 2.0 &  0.4330 & -1.5835 & -3.5119 & -3.7754 &  6.2474 & 15.2930 &  5.3661 & -1.3946 \\
       2.0 - 3.0 &  0.1895 &  0.0508 & -0.7574 & -2.3192 & -2.1927 &  5.3661 &  7.5669 &  0.4711 \\
       3.0 - 6.0 &  0.0066 &  0.0642 &  0.0805 &  0.0062 & -0.6301 & -1.3946 &  0.4711 &  2.4289 \\
    \end{tabular}
  \label{tablecov:Fe_Statistical_Epi}
\end{table}

\begin{table}[h!]
  \centering
  \caption{Systematic covariance matrix of the measured $d\sigma/dE_{\pi}$ on Fe, in units of $10^{-80}$ (cm$^{2}$/GeV/Fe)$^{2}$.}
    \begin{tabular}{c||cccccccc}
      Bin edges (GeV) & 0 - 0.25 & 0.25 - 0.5 & 0.5 - 0.75 & 0.75 - 1.0 & 1.0 - 1.5 & 1.5 - 2.0 & 2.0 - 3.0 & 3.0 - 6.0 \\
\hline \hline
        0 - 0.25 & 121.5712 & 123.3174 & 22.8534 & 49.7716 & 55.1985 & 14.8221 &  2.8513 & -8.7838 \\
      0.25 - 0.5 & 123.3174 & 963.1823 & 379.7737 & 175.6837 & 40.7910 & 20.6318 & 25.5564 & 14.6491 \\
      0.5 - 0.75 & 22.8534 & 379.7737 & 174.1696 & 84.4061 & 20.1848 & 12.5757 & 13.3942 &  9.4542 \\
      0.75 - 1.0 & 49.7716 & 175.6837 & 84.4061 & 72.3703 & 49.3928 & 19.4188 &  9.2768 &  0.8456 \\
       1.0 - 1.5 & 55.1985 & 40.7910 & 20.1848 & 49.3928 & 53.7995 & 20.1270 &  6.1711 & -3.1294 \\
       1.5 - 2.0 & 14.8221 & 20.6318 & 12.5757 & 19.4188 & 20.1270 & 10.6391 &  4.8107 &  0.9558 \\
       2.0 - 3.0 &  2.8513 & 25.5564 & 13.3942 &  9.2768 &  6.1711 &  4.8107 &  3.1322 &  1.6987 \\
       3.0 - 6.0 & -8.7838 & 14.6491 &  9.4542 &  0.8456 & -3.1294 &  0.9558 &  1.6987 &  2.2431 \\
    \end{tabular}
  \label{tablecov:Fe_Systematic_Epi}
\end{table}

\begin{table}[h!]
  \centering
  \caption{Measured cross section as a function of $E_{\pi}$ on Pb, in units of
  $10^{-39}$ $\text{cm}^{2}$/GeV/$^{207}$Pb, and the absolute and
  fractional cross section uncertainties.}
    \begin{tabular}{c||c||c|c||c|c}
      Bin edges (GeV)  &  $d\sigma/dE_{\pi}$  &  Abs. Stat. Unc. & Abs. Tot. Unc.  &  Frac. Stat. Unc.  &  Frac. Tot. Unc. \\
      \hline\hline
        0 - 0.25 & 1.111 & 0.839 & 1.522 & 0.755 & 1.370 \\
      0.25 - 0.5 & 6.546 & 1.888 & 3.964 & 0.288 & 0.606 \\
      0.5 - 0.75 & 9.105 & 1.334 & 2.026 & 0.147 & 0.223 \\
      0.75 - 1.0 & 8.854 & 1.035 & 1.302 & 0.117 & 0.147 \\
       1.0 - 1.5 & 9.703 & 0.988 & 1.346 & 0.102 & 0.139 \\
       1.5 - 2.0 & 7.977 & 0.856 & 1.059 & 0.107 & 0.133 \\
       2.0 - 3.0 & 5.517 & 0.657 & 0.746 & 0.119 & 0.135 \\
       3.0 - 6.0 & 2.379 & 0.371 & 0.492 & 0.156 & 0.207 \\
    \end{tabular}
  \label{table:EpiXSec_Pb}
\end{table}

\begin{table}[h!]
  \centering
  \caption{Statistical covariance matrix of the measured $d\sigma/dE_{\pi}$ on Pb, in units of $10^{-80}$ (cm$^{2}$/GeV/Pb)$^{2}$.}
    \begin{tabular}{c||cccccccc}
      Bin edges (GeV) & 0 - 0.25 & 0.25 - 0.5 & 0.5 - 0.75 & 0.75 - 1.0 & 1.0 - 1.5 & 1.5 - 2.0 & 2.0 - 3.0 & 3.0 - 6.0 \\
\hline \hline
        0 - 0.25 & 70.3302 & 57.2517 & -4.6570 & -7.4049 & -3.6718 & -0.9772 & -0.1680 & -0.0341 \\
      0.25 - 0.5 & 57.2517 & 356.6308 & 91.8816 & -8.8042 & -21.3034 & -9.6179 & -2.4860 & -0.2734 \\
      0.5 - 0.75 & -4.6570 & 91.8816 & 177.9462 & 82.7665 &  5.7367 & -14.9631 & -5.4753 & -0.5958 \\
      0.75 - 1.0 & -7.4049 & -8.8042 & 82.7665 & 107.2260 & 60.9311 & -2.0950 & -8.5282 & -1.4282 \\
       1.0 - 1.5 & -3.6718 & -21.3034 &  5.7367 & 60.9311 & 97.6004 & 45.5293 & -1.4870 & -3.7291 \\
       1.5 - 2.0 & -0.9772 & -9.6179 & -14.9631 & -2.0950 & 45.5293 & 73.2230 & 29.1632 & -3.6135 \\
       2.0 - 3.0 & -0.1680 & -2.4860 & -5.4753 & -8.5282 & -1.4870 & 29.1632 & 43.1505 &  6.5841 \\
       3.0 - 6.0 & -0.0341 & -0.2734 & -0.5958 & -1.4282 & -3.7291 & -3.6135 &  6.5841 & 13.7705 \\
    \end{tabular}
  \label{tablecov:Pb_Statistical_Epi}
\end{table}

\begin{table}[h!]
  \centering
  \caption{Systematic covariance matrix of the measured $d\sigma/dE_{\pi}$ on Pb, in units of $10^{-80}$ (cm$^{2}$/GeV/Pb)$^{2}$.}
    \begin{tabular}{c||cccccccc}
      Bin edges (GeV) & 0 - 0.25 & 0.25 - 0.5 & 0.5 - 0.75 & 0.75 - 1.0 & 1.0 - 1.5 & 1.5 - 2.0 & 2.0 - 3.0 & 3.0 - 6.0 \\
\hline \hline
        0 - 0.25 & 161.3889 & 274.5015 & 79.0755 & 57.1965 & 73.5423 & 44.9724 & 13.7603 & -24.4509 \\
      0.25 - 0.5 & 274.5015 & 1214.8783 & 472.2045 & 143.9961 & 60.3931 & 32.5159 & 27.5435 & -3.9565 \\
      0.5 - 0.75 & 79.0755 & 472.2045 & 232.4696 & 84.5246 & 34.1260 & 21.7769 & 23.5233 & 12.4378 \\
      0.75 - 1.0 & 57.1965 & 143.9961 & 84.5246 & 62.3161 & 60.4339 & 40.0911 & 22.1608 & -1.4281 \\
       1.0 - 1.5 & 73.5423 & 60.3931 & 34.1260 & 60.4339 & 83.5409 & 55.5527 & 23.6323 & -11.9043 \\
       1.5 - 2.0 & 44.9724 & 32.5159 & 21.7769 & 40.0911 & 55.5527 & 38.8312 & 18.4448 & -5.2071 \\
       2.0 - 3.0 & 13.7603 & 27.5435 & 23.5233 & 22.1608 & 23.6323 & 18.4448 & 12.4749 &  3.1934 \\
       3.0 - 6.0 & -24.4509 & -3.9565 & 12.4378 & -1.4281 & -11.9043 & -5.2071 &  3.1934 & 10.3920 \\
    \end{tabular}
  \label{tablecov:Pb_Systematic_Epi}
\end{table}

\begin{table}[h!]
  \centering
  \caption{Measured cross section as a function of $\theta_{\pi}$ on CH, in units of
  $10^{-41}$ $\text{cm}^{2}$/Degrees/$^{12}$CH, and the absolute and
  fractional cross section uncertainties.}
    \begin{tabular}{c||c||c|c||c|c}
      Bin edges (Degrees)  &  $d\sigma/d\theta_{\pi}$  &  Abs. Stat. Unc. & Abs. Tot. Unc.  &  Frac. Stat. Unc.  &  Frac. Tot. Unc. \\
      \hline\hline
        0 - 5 & 15.847 & 0.388 & 1.695 & 0.024 & 0.107 \\
       5 - 10 & 24.663 & 0.430 & 2.273 & 0.017 & 0.092 \\
      10 - 15 & 19.800 & 0.342 & 1.693 & 0.017 & 0.086 \\
      15 - 20 & 14.322 & 0.275 & 1.244 & 0.019 & 0.087 \\
      20 - 25 & 9.965 & 0.235 & 0.897 & 0.024 & 0.090 \\
      25 - 30 & 6.918 & 0.179 & 0.680 & 0.026 & 0.098 \\
      30 - 35 & 4.723 & 0.148 & 0.552 & 0.031 & 0.117 \\
      35 - 40 & 2.997 & 0.128 & 0.476 & 0.043 & 0.159 \\
      40 - 45 & 1.872 & 0.112 & 0.354 & 0.060 & 0.189 \\
      45 - 50 & 0.997 & 0.099 & 0.326 & 0.100 & 0.327 \\
      50 - 60 & 0.341 & 0.084 & 0.318 & 0.247 & 0.932 \\ 
      60 - 70 & -0.152 & 0.084 & 0.365 & -0.553 & -2.405 \\
    \end{tabular}
  \label{table:ThetaPiXSec_CH}
\end{table}

\begin{table}[h!]
  \centering
  \caption{Measured cross section as a function of $\theta_{\pi}$ on C, in units of
  $10^{-41}$ $\text{cm}^{2}$/Degrees/$^{12}$C, and the absolute and
  fractional cross section uncertainties.}
    \begin{tabular}{c||c||c|c||c|c}
      Bin edges (Degrees)  &  $d\sigma/d\theta_{\pi}$  &  Abs. Stat. Unc. & Abs. Tot. Unc.  &  Frac. Stat. Unc.  &  Frac. Tot. Unc. \\
      \hline\hline
        0 - 5 & 14.113 & 2.644 & 3.735 & 0.187 & 0.265 \\
       5 - 10 & 24.197 & 3.166 & 4.343 & 0.131 & 0.180 \\
      10 - 15 & 18.827 & 2.513 & 3.271 & 0.133 & 0.174 \\
      15 - 20 & 12.399 & 1.961 & 2.803 & 0.158 & 0.226 \\
      20 - 25 & 8.693 & 1.557 & 2.405 & 0.179 & 0.277 \\
      25 - 30 & 6.948 & 1.348 & 2.255 & 0.194 & 0.325 \\
      30 - 35 & 4.271 & 1.092 & 1.868 & 0.256 & 0.437 \\
      35 - 40 & 2.289 & 0.957 & 1.644 & 0.418 & 0.718 \\
      40 - 45 & 1.120 & 0.827 & 1.426 & 0.738 & 1.273 \\
      45 - 50 & 0.107 & 0.736 & 1.263 & 6.859 & 11.779 \\
      50 - 60 & 0.056 & 0.658 & 1.116 & 11.818 & 20.040 \\
      60 - 70 & -0.635 & 0.562 & 0.851 & -0.886 & -1.339 \\
    \end{tabular}
  \label{table:ThetaPiXSec_C}
\end{table}

\begin{table}[h!]
  \centering
  \caption{Measured cross section as a function of $\theta_{\pi}$ on Fe, in units of
  $10^{-41}$ $\text{cm}^{2}$/Degrees/$^{56}$Fe, and the absolute and
  fractional cross section uncertainties.}
    \begin{tabular}{c||c||c|c||c|c}
      Bin edges (Degrees)  &  $d\sigma/d\theta_{\pi}$  &  Abs. Stat. Unc. & Abs. Tot. Unc.  &  Frac. Stat. Unc.  &  Frac. Tot. Unc. \\
      \hline\hline
        0 - 5 & 47.141 & 5.930 & 7.924 & 0.126 & 0.168 \\
       5 - 10 & 74.730 & 6.020 & 9.424 & 0.081 & 0.126 \\
      10 - 15 & 50.799 & 4.233 & 6.204 & 0.083 & 0.122 \\
      15 - 20 & 31.157 & 2.965 & 5.762 & 0.095 & 0.185 \\
      20 - 25 & 23.392 & 2.500 & 6.476 & 0.107 & 0.277 \\
      25 - 30 & 9.536 & 2.017 & 4.787 & 0.212 & 0.502 \\
      30 - 35 & 2.794 & 1.815 & 4.108 & 0.650 & 1.470 \\
      35 - 40 & -0.479 & 0.448 & 3.138 & -0.934 & -6.547 \\
      40 - 45 & -0.905 & 0.880 & 2.991 & -0.972 & -3.304 \\
      45 - 50 & -0.520 & 0.433 & 1.466 & -0.832 & -2.821 \\
      50 - 60 & -0.208 & 0.132 & 0.474 & -0.637 & -2.280 \\
      60 - 70 & -0.052 & 0.035 & 0.116 & -0.673 & -2.242 \\
    \end{tabular}
  \label{table:ThetaPiXSec_Fe}
\end{table}

\begin{table}[h!]
  \centering
  \caption{Measured cross section as a function of $\theta_{\pi}$ on Pb, in units of
  $10^{-41}$ $\text{cm}^{2}$/Degrees/$^{207}$Pb, and the absolute and
  fractional cross section uncertainties.}
    \begin{tabular}{c||c||c|c||c|c}
      Bin edges (Degrees)  &  $d\sigma/d\theta_{\pi}$  &  Abs. Stat. Unc. & Abs. Tot. Unc.  &  Frac. Stat. Unc.  &  Frac. Tot. Unc. \\
      \hline\hline
        0 - 5 & 149.597 & 15.574 & 20.378 & 0.104 & 0.136 \\
       5 - 10 & 174.018 & 13.720 & 18.445 & 0.079 & 0.106 \\
      10 - 15 & 95.970 & 8.498 & 11.789 & 0.089 & 0.123 \\
      15 - 20 & 46.906 & 5.578 & 8.244 & 0.119 & 0.176 \\
      20 - 25 & 24.608 & 4.248 & 6.437 & 0.173 & 0.262 \\
      25 - 30 & 7.935 & 3.163 & 5.315 & 0.399 & 0.670 \\
      30 - 35 & 1.138 & 2.361 & 4.080 & 2.075 & 3.586 \\
      35 - 40 & -1.836 & 1.742 & 3.128 & -0.949 & -1.703 \\
      40 - 45 & -2.219 & 1.129 & 2.043 & -0.509 & -0.921 \\
      45 - 50 & -2.875 & 1.184 & 1.789 & -0.412 & -0.622 \\
      50 - 60 & -1.287 & 0.685 & 0.859 & -0.532 & -0.667 \\
      60 - 70 & 0.000 & 0.000 & 0.000 & --- & --- \\
    \end{tabular}
  \label{table:ThetaPiXSec_Pb}
\end{table}

\begin{turnpage}
 \begin{table}[h!]
  \centering
  \caption{Statistical covariance matrix of the measured $d\sigma/d\theta_{\pi}$ on CH, in units of $10^{-82}$ (cm$^{2}$/Degrees/CH)$^{2}$.}
    \begin{tabular}{c||cccccccccccc}
      Bin edges (Degrees) & 0 - 5 & 5 - 10 & 10 - 15 & 15 - 20 & 20 - 25 & 25 - 30 & 30 - 35 & 35 - 40 & 40 - 45 & 45 - 50 & 50 - 60 & 60 - 70 \\
\hline \hline
        0 - 5 &  0.15022 &  0.06762 &  0.01108 &  0.00271 &  0.00119 &  0.00061 &  0.00037 &  0.00023 &  0.00017 &  0.00012 &  0.00009 &  0.00006 \\
       5 - 10 &  0.06762 &  0.18463 &  0.05092 &  0.01023 &  0.00352 &  0.00167 &  0.00097 &  0.00060 &  0.00038 &  0.00024 &  0.00015 &  0.00008 \\
      10 - 15 &  0.01108 &  0.05092 &  0.11678 &  0.03526 &  0.00842 &  0.00302 &  0.00156 &  0.00094 &  0.00059 &  0.00038 &  0.00024 &  0.00012 \\
      15 - 20 &  0.00271 &  0.01023 &  0.03526 &  0.07563 &  0.02695 &  0.00641 &  0.00246 &  0.00136 &  0.00085 &  0.00056 &  0.00033 &  0.00017 \\
      20 - 25 &  0.00119 &  0.00352 &  0.00842 &  0.02695 &  0.05540 &  0.01816 &  0.00480 &  0.00203 &  0.00113 &  0.00073 &  0.00044 &  0.00022 \\
      25 - 30 &  0.00061 &  0.00167 &  0.00302 &  0.00641 &  0.01816 &  0.03206 &  0.01243 &  0.00372 &  0.00161 &  0.00095 &  0.00054 &  0.00028 \\
      30 - 35 &  0.00037 &  0.00097 &  0.00156 &  0.00246 &  0.00480 &  0.01243 &  0.02188 &  0.00964 &  0.00296 &  0.00127 &  0.00067 &  0.00033 \\
      35 - 40 &  0.00023 &  0.00060 &  0.00094 &  0.00136 &  0.00203 &  0.00372 &  0.00964 &  0.01626 &  0.00755 &  0.00241 &  0.00088 &  0.00039 \\
      40 - 45 &  0.00017 &  0.00038 &  0.00059 &  0.00085 &  0.00113 &  0.00161 &  0.00296 &  0.00755 &  0.01247 &  0.00617 &  0.00155 &  0.00050 \\
      45 - 50 &  0.00012 &  0.00024 &  0.00038 &  0.00056 &  0.00073 &  0.00095 &  0.00127 &  0.00241 &  0.00617 &  0.00990 &  0.00405 &  0.00087 \\
      50 - 60 &  0.00009 &  0.00015 &  0.00024 &  0.00033 &  0.00044 &  0.00054 &  0.00067 &  0.00088 &  0.00155 &  0.00405 &  0.00711 &  0.00253 \\
      60 - 70 &  0.00006 &  0.00008 &  0.00012 &  0.00017 &  0.00022 &  0.00028 &  0.00033 &  0.00039 &  0.00050 &  0.00087 &  0.00253 &  0.00703 \\
    \end{tabular}
  \label{tablecov:CH_Statistical_ThetaPi}
  \end{table}
\end{turnpage}

\begin{turnpage}
 \begin{table}[h!]
  \centering
  \caption{Systematic covariance matrix of the measured $d\sigma/d\theta_{\pi}$ on CH, in units of $10^{-82}$ (cm$^{2}$/Degrees/CH)$^{2}$.}
    \begin{tabular}{c||cccccccccccc}
      Bin edges (Degrees) & 0 - 5 & 5 - 10 & 10 - 15 & 15 - 20 & 20 - 25 & 25 - 30 & 30 - 35 & 35 - 40 & 40 - 45 & 45 - 50 & 50 - 60 & 60 - 70 \\
\hline \hline
        0 - 5 &  2.72379 &  3.64432 &  2.59066 &  1.71175 &  1.04589 &  0.68129 &  0.45271 &  0.35331 &  0.18791 &  0.11965 &  0.10532 &  0.19777 \\
       5 - 10 &  3.64432 &  4.98285 &  3.62489 &  2.45822 &  1.53001 &  0.99810 &  0.64144 &  0.46551 &  0.23676 &  0.13264 &  0.10081 &  0.20472 \\
      10 - 15 &  2.59066 &  3.62489 &  2.74969 &  1.95433 &  1.27232 &  0.85674 &  0.55764 &  0.39361 &  0.19909 &  0.10231 &  0.05973 &  0.12776 \\
      15 - 20 &  1.71175 &  2.45822 &  1.95433 &  1.47295 &  1.01639 &  0.71259 &  0.47978 &  0.33507 &  0.17657 &  0.08817 &  0.03451 &  0.06223 \\
      20 - 25 &  1.04589 &  1.53001 &  1.27232 &  1.01639 &  0.74909 &  0.55221 &  0.39722 &  0.28684 &  0.16337 &  0.08799 &  0.03185 &  0.03333 \\
      25 - 30 &  0.68129 &  0.99810 &  0.85674 &  0.71259 &  0.55221 &  0.42986 &  0.33101 &  0.25180 &  0.15462 &  0.09421 &  0.04326 &  0.03802 \\
      30 - 35 &  0.45271 &  0.64144 &  0.55764 &  0.47978 &  0.39722 &  0.33101 &  0.28272 &  0.23342 &  0.15472 &  0.10584 &  0.06055 &  0.05348 \\
      35 - 40 &  0.35331 &  0.46551 &  0.39361 &  0.33507 &  0.28684 &  0.25180 &  0.23342 &  0.21071 &  0.14610 &  0.11014 &  0.07672 &  0.07945 \\
      40 - 45 &  0.18791 &  0.23676 &  0.19909 &  0.17657 &  0.16337 &  0.15462 &  0.15472 &  0.14610 &  0.11302 &  0.09637 &  0.07528 &  0.06780 \\
      45 - 50 &  0.11965 &  0.13264 &  0.10231 &  0.08817 &  0.08799 &  0.09421 &  0.10584 &  0.11014 &  0.09637 &  0.09617 &  0.08786 &  0.08008 \\
      50 - 60 &  0.10532 &  0.10081 &  0.05973 &  0.03451 &  0.03185 &  0.04326 &  0.06055 &  0.07672 &  0.07528 &  0.08786 &  0.09406 &  0.09559 \\
      60 - 70 &  0.19777 &  0.20472 &  0.12776 &  0.06223 &  0.03333 &  0.03802 &  0.05348 &  0.07945 &  0.06780 &  0.08008 &  0.09559 &  0.12606 \\
    \end{tabular}
  \label{tablecov:CH_Systematic_ThetaPi}
  \end{table}
\end{turnpage}

\begin{turnpage}
 \begin{table}[h!]
  \centering
  \caption{Statistical covariance matrix of the measured $d\sigma/d\theta_{\pi}$ on C, in units of $10^{-82}$ (cm$^{2}$/Degrees/C)$^{2}$.}
    \begin{tabular}{c||cccccccccccc}
      Bin edges (Degrees) & 0 - 5 & 5 - 10 & 10 - 15 & 15 - 20 & 20 - 25 & 25 - 30 & 30 - 35 & 35 - 40 & 40 - 45 & 45 - 50 & 50 - 60 & 60 - 70 \\
\hline \hline
        0 - 5 &  6.98980 &  3.84301 &  0.67443 &  0.14305 &  0.08112 &  0.03548 &  0.02212 &  0.01966 &  0.01251 &  0.00793 &  0.00849 &  0.01550 \\
       5 - 10 &  3.84301 & 10.02638 &  3.10618 &  0.64788 &  0.25251 &  0.13271 &  0.07565 &  0.04025 &  0.02159 &  0.01995 &  0.02230 &  0.01501 \\
      10 - 15 &  0.67443 &  3.10618 &  6.31476 &  2.02052 &  0.53587 &  0.21131 &  0.12503 &  0.06753 &  0.04492 &  0.03438 &  0.02526 &  0.01268 \\
      15 - 20 &  0.14305 &  0.64788 &  2.02052 &  3.84585 &  1.37971 &  0.41266 &  0.17748 &  0.09676 &  0.06915 &  0.03837 &  0.02314 &  0.01246 \\
      20 - 25 &  0.08112 &  0.25251 &  0.53587 &  1.37971 &  2.42569 &  1.05707 &  0.34085 &  0.15226 &  0.09125 &  0.05684 &  0.03786 &  0.02042 \\
      25 - 30 &  0.03548 &  0.13271 &  0.21131 &  0.41266 &  1.05707 &  1.81715 &  0.78115 &  0.29038 &  0.14797 &  0.09071 &  0.06102 &  0.03514 \\
      30 - 35 &  0.02212 &  0.07565 &  0.12503 &  0.17748 &  0.34085 &  0.78115 &  1.19178 &  0.58065 &  0.22922 &  0.11877 &  0.07560 &  0.03664 \\
      35 - 40 &  0.01966 &  0.04025 &  0.06753 &  0.09676 &  0.15226 &  0.29038 &  0.58065 &  0.91606 &  0.49124 &  0.19236 &  0.09862 &  0.05227 \\
      40 - 45 &  0.01251 &  0.02159 &  0.04492 &  0.06915 &  0.09125 &  0.14797 &  0.22922 &  0.49124 &  0.68318 &  0.37083 &  0.13331 &  0.04651 \\
      45 - 50 &  0.00793 &  0.01995 &  0.03438 &  0.03837 &  0.05684 &  0.09071 &  0.11877 &  0.19236 &  0.37083 &  0.54116 &  0.29857 &  0.08034 \\
      50 - 60 &  0.00849 &  0.02230 &  0.02526 &  0.02314 &  0.03786 &  0.06102 &  0.07560 &  0.09862 &  0.13331 &  0.29857 &  0.43341 &  0.18068 \\
      60 - 70 &  0.01550 &  0.01501 &  0.01268 &  0.01246 &  0.02042 &  0.03514 &  0.03664 &  0.05227 &  0.04651 &  0.08034 &  0.18068 &  0.31627 \\
    \end{tabular}
  \label{tablecov:C_Statistical_ThetaPi}
  \end{table}
\end{turnpage}

\begin{turnpage}
 \begin{table}[h!]
  \centering
  \caption{Systematic covariance matrix of the measured $d\sigma/d\theta_{\pi}$ on C, in units of $10^{-82}$ (cm$^{2}$/Degrees/C)$^{2}$.}
    \begin{tabular}{c||cccccccccccc}
      Bin edges (Degrees) & 0 - 5 & 5 - 10 & 10 - 15 & 15 - 20 & 20 - 25 & 25 - 30 & 30 - 35 & 35 - 40 & 40 - 45 & 45 - 50 & 50 - 60 & 60 - 70 \\
\hline \hline
        0 - 5 &  6.95677 &  7.26890 &  3.07948 &  2.04259 &  1.53132 &  1.41817 &  1.17586 &  1.20298 &  1.23229 &  0.79065 &  0.49969 &  0.38967 \\
       5 - 10 &  7.26890 &  8.83938 &  4.94581 &  3.66890 &  2.89344 &  2.76229 &  2.16670 &  1.95206 &  1.77766 &  1.23827 &  0.79736 &  0.44907 \\
      10 - 15 &  3.07948 &  4.94581 &  4.38309 &  3.91850 &  3.34185 &  3.20779 &  2.52158 &  2.07264 &  1.62714 &  1.22479 &  0.83741 &  0.37920 \\
      15 - 20 &  2.04259 &  3.66890 &  3.91850 &  4.00877 &  3.60273 &  3.45696 &  2.76270 &  2.26945 &  1.74443 &  1.30781 &  0.86964 &  0.38599 \\
      20 - 25 &  1.53132 &  2.89344 &  3.34185 &  3.60273 &  3.35998 &  3.26355 &  2.65265 &  2.21307 &  1.72323 &  1.30426 &  0.86581 &  0.38884 \\
      25 - 30 &  1.41817 &  2.76229 &  3.20779 &  3.45696 &  3.26355 &  3.26608 &  2.69645 &  2.27838 &  1.81686 &  1.41218 &  0.97143 &  0.45127 \\
      30 - 35 &  1.17586 &  2.16670 &  2.52158 &  2.76270 &  2.65265 &  2.69645 &  2.29635 &  1.98365 &  1.61540 &  1.28512 &  0.90824 &  0.45110 \\
      35 - 40 &  1.20298 &  1.95206 &  2.07264 &  2.26945 &  2.21307 &  2.27838 &  1.98365 &  1.78638 &  1.50990 &  1.23392 &  0.90095 &  0.48104 \\
      40 - 45 &  1.23229 &  1.77766 &  1.62714 &  1.74443 &  1.72323 &  1.81686 &  1.61540 &  1.50990 &  1.35026 &  1.14470 &  0.87503 &  0.50487 \\
      45 - 50 &  0.79065 &  1.23827 &  1.22479 &  1.30781 &  1.30426 &  1.41218 &  1.28512 &  1.23392 &  1.14470 &  1.05499 &  0.87711 &  0.53907 \\
      50 - 60 &  0.49969 &  0.79736 &  0.83741 &  0.86964 &  0.86581 &  0.97143 &  0.90824 &  0.90095 &  0.87503 &  0.87711 &  0.81284 &  0.54262 \\
      60 - 70 &  0.38967 &  0.44907 &  0.37920 &  0.38599 &  0.38884 &  0.45127 &  0.45110 &  0.48104 &  0.50487 &  0.53907 &  0.54262 &  0.40731 \\
    \end{tabular}
  \label{tablecov:C_Systematic_ThetaPi}
  \end{table}
\end{turnpage}

\begin{turnpage}
 \begin{table}[h!]
  \centering
  \caption{Statistical covariance matrix of the measured $d\sigma/d\theta_{\pi}$ on Fe, in units of $10^{-82}$ (cm$^{2}$/Degrees/Fe)$^{2}$.}
    \begin{tabular}{c||cccccccccccc}
      Bin edges (Degrees) & 0 - 5 & 5 - 10 & 10 - 15 & 15 - 20 & 20 - 25 & 25 - 30 & 30 - 35 & 35 - 40 & 40 - 45 & 45 - 50 & 50 - 60 & 60 - 70 \\
\hline \hline
        0 - 5 & 35.16603 &  6.89135 & -2.49103 & -0.97904 & -0.39292 & -0.07174 &  0.07874 & -0.04683 & -0.09115 & -0.04485 & -0.01244 & -0.00029 \\
       5 - 10 &  6.89135 & 36.24242 &  5.05513 & -1.57726 & -1.12329 & -0.15973 &  0.25032 & -0.11827 & -0.23437 & -0.11345 & -0.03396 & -0.01652 \\
      10 - 15 & -2.49103 &  5.05513 & 17.91881 &  3.80610 & -0.73742 & -0.30101 &  0.23834 & -0.11640 & -0.23119 & -0.10911 & -0.03177 & -0.00721 \\
      15 - 20 & -0.97904 & -1.57726 &  3.80610 &  8.78915 &  2.82792 &  0.20231 &  0.37448 & -0.13168 & -0.26310 & -0.12414 & -0.03464 & -0.01014 \\
      20 - 25 & -0.39292 & -1.12329 & -0.73742 &  2.82792 &  6.25186 &  2.54411 &  1.19428 & -0.22547 & -0.45027 & -0.22320 & -0.06514 & -0.01776 \\
      25 - 30 & -0.07174 & -0.15973 & -0.30101 &  0.20231 &  2.54411 &  4.06903 &  2.79543 & -0.40654 & -0.79125 & -0.39381 & -0.12556 & -0.02911 \\
      30 - 35 &  0.07874 &  0.25032 &  0.23834 &  0.37448 &  1.19428 &  2.79543 &  3.29282 & -0.61462 & -1.19251 & -0.58760 & -0.18622 & -0.04427 \\
      35 - 40 & -0.04683 & -0.11827 & -0.11640 & -0.13168 & -0.22547 & -0.40654 & -0.61462 &  0.20046 &  0.39392 &  0.19351 &  0.05906 &  0.01524 \\
      40 - 45 & -0.09115 & -0.23437 & -0.23119 & -0.26310 & -0.45027 & -0.79125 & -1.19251 &  0.39392 &  0.77485 &  0.38049 &  0.11601 &  0.02999 \\
      45 - 50 & -0.04485 & -0.11345 & -0.10911 & -0.12414 & -0.22320 & -0.39381 & -0.58760 &  0.19351 &  0.38049 &  0.18710 &  0.05700 &  0.01473 \\
      50 - 60 & -0.01244 & -0.03396 & -0.03177 & -0.03464 & -0.06514 & -0.12556 & -0.18622 &  0.05906 &  0.11601 &  0.05700 &  0.01749 &  0.00448 \\
      60 - 70 & -0.00029 & -0.01652 & -0.00721 & -0.01014 & -0.01776 & -0.02911 & -0.04427 &  0.01524 &  0.02999 &  0.01473 &  0.00448 &  0.00121 \\
    \end{tabular}
  \label{tablecov:Fe_Statistical_ThetaPi}
  \end{table}
\end{turnpage}

\begin{turnpage}
 \begin{table}[h!]
  \centering
  \caption{Systematic covariance matrix of the measured $d\sigma/d\theta_{\pi}$ on Fe, in units of $10^{-82}$ (cm$^{2}$/Degrees/Fe)$^{2}$.}
    \begin{tabular}{c||cccccccccccc}
      Bin edges (Degrees) & 0 - 5 & 5 - 10 & 10 - 15 & 15 - 20 & 20 - 25 & 25 - 30 & 30 - 35 & 35 - 40 & 40 - 45 & 45 - 50 & 50 - 60 & 60 - 70 \\
\hline \hline
        0 - 5 & 27.61667 & 32.96602 & 15.68630 & 12.11356 & 11.05214 &  5.72072 &  4.18748 &  0.38531 & -1.08606 & -0.71713 & -0.18880 & -0.05342 \\
       5 - 10 & 32.96602 & 52.57652 & 27.75217 & 28.55298 & 31.39175 & 18.19445 & 14.05411 &  2.32195 & -2.87945 & -2.16256 & -0.56650 & -0.17188 \\
      10 - 15 & 15.68630 & 27.75217 & 20.57183 & 19.94198 & 19.96012 & 11.64074 &  7.37233 & -1.23172 & -3.59124 & -2.06450 & -0.63074 & -0.16455 \\
      15 - 20 & 12.11356 & 28.55298 & 19.94198 & 24.41427 & 28.29986 & 18.43032 & 13.58408 & -0.16180 & -4.98354 & -3.07918 & -0.91758 & -0.24900 \\
      20 - 25 & 11.05214 & 31.39175 & 19.96012 & 28.29986 & 35.69179 & 24.51090 & 19.23526 &  0.29319 & -6.76264 & -4.24555 & -1.25867 & -0.34549 \\
      25 - 30 &  5.72072 & 18.19445 & 11.64074 & 18.43032 & 24.51090 & 18.84836 & 15.42946 &  0.39506 & -5.44913 & -3.38528 & -1.03853 & -0.28161 \\
      30 - 35 &  4.18748 & 14.05411 &  7.37233 & 13.58408 & 19.23526 & 15.42946 & 13.58248 &  2.36798 & -3.08583 & -2.20023 & -0.65940 & -0.18982 \\
      35 - 40 &  0.38531 &  2.32195 & -1.23172 & -0.16180 &  0.29319 &  0.39506 &  2.36798 &  9.64452 &  7.75041 &  3.45279 &  1.14582 &  0.26172 \\
      40 - 45 & -1.08606 & -2.87945 & -3.59124 & -4.98354 & -6.76264 & -5.44913 & -3.08583 &  7.75041 &  8.17273 &  3.96125 &  1.28993 &  0.30995 \\
      45 - 50 & -0.71713 & -2.16256 & -2.06450 & -3.07918 & -4.24555 & -3.38528 & -2.20023 &  3.45279 &  3.96125 &  1.96260 &  0.63515 &  0.15457 \\
      50 - 60 & -0.18880 & -0.56650 & -0.63074 & -0.91758 & -1.25867 & -1.03853 & -0.65940 &  1.14582 &  1.28993 &  0.63515 &  0.20672 &  0.05009 \\
      60 - 70 & -0.05342 & -0.17188 & -0.16455 & -0.24900 & -0.34549 & -0.28161 & -0.18982 &  0.26172 &  0.30995 &  0.15457 &  0.05009 &  0.01224 \\
    \end{tabular}
  \label{tablecov:Fe_Systematic_ThetaPi}
  \end{table}
\end{turnpage}

\begin{turnpage}
 \begin{table}[h!]
  \centering
  \caption{Statistical covariance matrix of the measured $d\sigma/d\theta_{\pi}$ on Pb, in units of $10^{-82}$ (cm$^{2}$/Degrees/Pb)$^{2}$.}
    \begin{tabular}{c||cccccccccccc}
      Bin edges (Degrees) & 0 - 5 & 5 - 10 & 10 - 15 & 15 - 20 & 20 - 25 & 25 - 30 & 30 - 35 & 35 - 40 & 40 - 45 & 45 - 50 & 50 - 60 & 60 - 70 \\
\hline \hline
        0 - 5 & 242.54950 & 109.72048 & 19.83020 &  4.82077 &  1.85123 &  1.05464 &  0.49452 &  0.31882 &  0.11771 &  0.03703 &  0.21595 &  0.00000 \\
       5 - 10 & 109.72048 & 188.24631 & 60.02301 & 14.86457 &  5.48972 &  2.64400 &  1.41532 &  0.82293 &  0.42083 &  0.25489 &  0.15815 &  0.00000 \\
      10 - 15 & 19.83020 & 60.02301 & 72.21860 & 28.70615 & 10.51970 &  4.74159 &  2.50820 &  1.48138 &  0.93924 &  0.80084 &  0.09803 &  0.00000 \\
      15 - 20 &  4.82077 & 14.86457 & 28.70615 & 31.11766 & 17.08810 &  7.38023 &  3.65523 &  2.12210 &  1.19593 &  0.78367 &  0.10650 &  0.00000 \\
      20 - 25 &  1.85123 &  5.48972 & 10.51970 & 17.08810 & 18.04922 & 10.45831 &  5.19757 &  2.58077 &  1.33066 &  0.81171 &  0.22366 &  0.00000 \\
      25 - 30 &  1.05464 &  2.64400 &  4.74159 &  7.38023 & 10.45831 & 10.00160 &  6.38026 &  3.39902 &  1.73989 &  1.04106 &  0.28529 &  0.00000 \\
      30 - 35 &  0.49452 &  1.41532 &  2.50820 &  3.65523 &  5.19757 &  6.38026 &  5.57272 &  3.56192 &  1.77088 &  0.88699 &  0.17444 &  0.00000 \\
\end{tabular}
  \label{tablecov:Pb_Statistical_ThetaPi}
  \end{table}
\end{turnpage}

\begin{turnpage}
 \begin{table}[h!]
  \centering
  \caption{Systematic covariance matrix of the measured $d\sigma/d\theta_{\pi}$ on Pb, in units of $10^{-82}$ (cm$^{2}$/Degrees/Pb)$^{2}$.}
    \begin{tabular}{c||cccccccccccc}
      Bin edges (Degrees) & 0 - 5 & 5 - 10 & 10 - 15 & 15 - 20 & 20 - 25 & 25 - 30 & 30 - 35 & 35 - 40 & 40 - 45 & 45 - 50 & 50 - 60 & 60 - 70 \\
\hline \hline
        0 - 5 & 172.70143 & 152.55724 & 90.76358 & 54.48454 & 34.06493 & 22.48966 & 14.19810 &  9.17525 &  5.16481 &  3.82080 &  0.86722 &  0.00000 \\
       5 - 10 & 152.55724 & 151.95794 & 95.08965 & 58.76473 & 37.66026 & 22.83444 & 13.80425 &  7.72315 &  3.44638 &  2.57082 &  0.64773 &  0.00000 \\
      10 - 15 & 90.76358 & 95.08965 & 66.75954 & 46.36148 & 32.24179 & 21.21012 & 13.81165 &  8.09449 &  3.82935 &  2.77460 &  0.79811 &  0.00000 \\
      15 - 20 & 54.48454 & 58.76473 & 46.36148 & 36.85222 & 28.22724 & 21.07642 & 14.78800 &  9.46230 &  4.99360 &  3.47652 &  1.08150 &  0.00000 \\
      20 - 25 & 34.06493 & 37.66026 & 32.24179 & 28.22724 & 23.39157 & 19.28490 & 14.15062 &  9.68684 &  5.47718 &  3.86973 &  1.25901 &  0.00000 \\
      25 - 30 & 22.48966 & 22.83444 & 21.21012 & 21.07642 & 19.28490 & 18.24730 & 14.07490 & 10.47337 &  6.39302 &  4.58464 &  1.52507 &  0.00000 \\
      30 - 35 & 14.19810 & 13.80425 & 13.81165 & 14.78800 & 14.15062 & 14.07490 & 11.07761 &  8.43812 &  5.24994 &  3.78355 &  1.25454 &  0.00000 \\
\end{tabular}
  \label{tablecov:Pb_Systematic_ThetaPi}
  \end{table}
\end{turnpage}

\begin{table}[h!]
  \centering
  \caption{Measured cross section as a function of $Q^{2}$ on CH, in units of
  $10^{-39}$ $\text{cm}^{2}$/[GeV/c]$^{2}$/$^{12}$CH, and the absolute and
  fractional cross section uncertainties.}
    \begin{tabular}{c||c||c|c||c|c}
      Bin edges ([GeV/c]$^{2}$)  &  $d\sigma/dQ^{2}$  &  Abs. Stat. Unc. & Abs. Tot. Unc.  &  Frac. Stat. Unc.  &  Frac. Tot. Unc. \\
      \hline\hline
         0 - 0.025 & 26.020 & 0.718 & 2.346 & 0.028 & 0.090 \\
      0.025 - 0.05 & 23.506 & 0.581 & 1.982 & 0.025 & 0.084 \\
      0.05 - 0.075 & 18.405 & 0.479 & 1.629 & 0.026 & 0.088 \\
       0.075 - 0.1 & 14.895 & 0.400 & 1.367 & 0.027 & 0.092 \\
        0.1 - 0.15 & 11.247 & 0.314 & 1.074 & 0.028 & 0.095 \\
        0.15 - 0.2 & 8.307 & 0.270 & 0.862 & 0.033 & 0.104 \\
         0.2 - 0.3 & 5.711 & 0.225 & 0.674 & 0.039 & 0.118 \\
         0.3 - 0.4 & 3.771 & 0.179 & 0.481 & 0.047 & 0.127 \\
         0.4 - 0.6 & 2.345 & 0.126 & 0.310 & 0.054 & 0.132 \\
         0.6 - 1.0 & 0.848 & 0.073 & 0.162 & 0.086 & 0.191 \\
    \end{tabular}
  \label{table:Q2XSec_CH}
\end{table}

\begin{table}[h!]
  \centering
  \caption{Statistical covariance matrix of the measured $d\sigma/dQ^{2}$ on CH, in units of $10^{-78}$ (cm$^{2}$/(GeV/c)$^{2}$/CH)$^{2}$.}
    \begin{tabular}{c||cccccccccc}
      Bin edges ([GeV/c]$^{2}$) & 0 - 0.025 & 0.025 - 0.05 & 0.05 - 0.075 & 0.075 - 0.1 & 0.1 - 0.15 & 0.15 - 0.2 & 0.2 - 0.3 & 0.3 - 0.4 & 0.4 - 0.6 & 0.6 - 1.0 \\
\hline \hline
         0 - 0.025 &  0.5156 &  0.0459 & -0.0416 & -0.0233 & -0.0069 & -0.0016 & -0.0004 & -0.0001 & -0.0000 & -0.0000 \\
      0.025 - 0.05 &  0.0459 &  0.3380 &  0.1158 & -0.0040 & -0.0178 & -0.0073 & -0.0018 & -0.0003 & -0.0001 & -0.0000 \\
      0.05 - 0.075 & -0.0416 &  0.1158 &  0.2292 &  0.1122 &  0.0034 & -0.0130 & -0.0050 & -0.0011 & -0.0002 & -0.0000 \\
       0.075 - 0.1 & -0.0233 & -0.0040 &  0.1122 &  0.1596 &  0.0662 & -0.0040 & -0.0089 & -0.0026 & -0.0004 & -0.0000 \\
        0.1 - 0.15 & -0.0069 & -0.0178 &  0.0034 &  0.0662 &  0.0987 &  0.0392 & -0.0037 & -0.0049 & -0.0012 & -0.0001 \\
        0.15 - 0.2 & -0.0016 & -0.0073 & -0.0130 & -0.0040 &  0.0392 &  0.0731 &  0.0280 & -0.0030 & -0.0031 & -0.0004 \\
         0.2 - 0.3 & -0.0004 & -0.0018 & -0.0050 & -0.0089 & -0.0037 &  0.0280 &  0.0507 &  0.0158 & -0.0022 & -0.0010 \\
         0.3 - 0.4 & -0.0001 & -0.0003 & -0.0011 & -0.0026 & -0.0049 & -0.0030 &  0.0158 &  0.0320 &  0.0092 & -0.0014 \\
         0.4 - 0.6 & -0.0000 & -0.0001 & -0.0002 & -0.0004 & -0.0012 & -0.0031 & -0.0022 &  0.0092 &  0.0159 &  0.0022 \\
         0.6 - 1.0 & -0.0000 & -0.0000 & -0.0000 & -0.0000 & -0.0001 & -0.0004 & -0.0010 & -0.0014 &  0.0022 &  0.0053 \\
    \end{tabular}
  \label{tablecov:CH_Statistical_Q2}
\end{table}

\begin{table}[h!]
  \centering
  \caption{Systematic covariance matrix of the measured $d\sigma/dQ^{2}$ on CH, in units of $10^{-78}$ (cm$^{2}$/(GeV/c)$^{2}$/CH)$^{2}$.}
    \begin{tabular}{c||cccccccccc}
      Bin edges ([GeV/c]$^{2}$) & 0 - 0.025 & 0.025 - 0.05 & 0.05 - 0.075 & 0.075 - 0.1 & 0.1 - 0.15 & 0.15 - 0.2 & 0.2 - 0.3 & 0.3 - 0.4 & 0.4 - 0.6 & 0.6 - 1.0 \\
\hline \hline
         0 - 0.025 &  4.9885 &  4.1479 &  3.2820 &  2.6012 &  1.8610 &  1.2879 &  0.7812 &  0.4455 &  0.2212 &  0.0983 \\
      0.025 - 0.05 &  4.1479 &  3.5895 &  2.9131 &  2.3747 &  1.7624 &  1.2641 &  0.8109 &  0.4847 &  0.2615 &  0.1164 \\
      0.05 - 0.075 &  3.2820 &  2.9131 &  2.4234 &  2.0130 &  1.5237 &  1.1189 &  0.7470 &  0.4564 &  0.2556 &  0.1154 \\
       0.075 - 0.1 &  2.6012 &  2.3747 &  2.0130 &  1.7087 &  1.3224 &  0.9877 &  0.6733 &  0.4156 &  0.2423 &  0.1119 \\
        0.1 - 0.15 &  1.8610 &  1.7624 &  1.5237 &  1.3224 &  1.0546 &  0.8101 &  0.5702 &  0.3568 &  0.2136 &  0.1002 \\
        0.15 - 0.2 &  1.2879 &  1.2641 &  1.1189 &  0.9877 &  0.8101 &  0.6694 &  0.5009 &  0.3237 &  0.1955 &  0.0908 \\
         0.2 - 0.3 &  0.7812 &  0.8109 &  0.7470 &  0.6733 &  0.5702 &  0.5009 &  0.4039 &  0.2731 &  0.1634 &  0.0755 \\
         0.3 - 0.4 &  0.4455 &  0.4847 &  0.4564 &  0.4156 &  0.3568 &  0.3237 &  0.2731 &  0.1989 &  0.1195 &  0.0538 \\
         0.4 - 0.6 &  0.2212 &  0.2615 &  0.2556 &  0.2423 &  0.2136 &  0.1955 &  0.1634 &  0.1195 &  0.0804 &  0.0384 \\
         0.6 - 1.0 &  0.0983 &  0.1164 &  0.1154 &  0.1119 &  0.1002 &  0.0908 &  0.0755 &  0.0538 &  0.0384 &  0.0209 \\
    \end{tabular}
  \label{tablecov:CH_Systematic_Q2}
\end{table}

\begin{table}[h!]
  \centering
  \caption{Measured cross section as a function of $Q^{2}$ on C, in units of
  $10^{-39}$ $\text{cm}^{2}$/[GeV/c]$^{2}$/$^{12}$C, and the absolute and
  fractional cross section uncertainties.}
    \begin{tabular}{c||c||c|c||c|c}
      Bin edges ([GeV/c]$^{2}$)  &  $d\sigma/dQ^{2}$  &  Abs. Stat. Unc. & Abs. Tot. Unc.  &  Frac. Stat. Unc.  &  Frac. Tot. Unc. \\
      \hline\hline
         0 - 0.025 & 22.794 & 4.605 & 6.511 & 0.202 & 0.286 \\
      0.025 - 0.05 & 24.134 & 4.070 & 6.173 & 0.169 & 0.256 \\
      0.05 - 0.075 & 18.964 & 3.519 & 5.132 & 0.186 & 0.271 \\
       0.075 - 0.1 & 11.425 & 2.800 & 4.210 & 0.245 & 0.368 \\
        0.1 - 0.15 & 6.534 & 2.161 & 3.362 & 0.331 & 0.515 \\
        0.15 - 0.2 & 5.486 & 1.942 & 3.031 & 0.354 & 0.552 \\
         0.2 - 0.3 & 6.298 & 1.828 & 2.709 & 0.290 & 0.430 \\
         0.3 - 0.4 & 6.217 & 1.812 & 2.354 & 0.291 & 0.379 \\
         0.4 - 0.6 & 1.727 & 1.091 & 1.389 & 0.632 & 0.804 \\
         0.6 - 1.0 & 0.877 & 0.709 & 0.771 & 0.809 & 0.880 \\
    \end{tabular}
  \label{table:Q2XSec_C}
\end{table}

\begin{table}[h!]
  \centering
  \caption{Statistical covariance matrix of the measured $d\sigma/dQ^{2}$ on C, in units of $10^{-78}$ (cm$^{2}$/(GeV/c)$^{2}$/C)$^{2}$.}
    \begin{tabular}{c||cccccccccc}
      Bin edges ([GeV/c]$^{2}$) & 0 - 0.025 & 0.025 - 0.05 & 0.05 - 0.075 & 0.075 - 0.1 & 0.1 - 0.15 & 0.15 - 0.2 & 0.2 - 0.3 & 0.3 - 0.4 & 0.4 - 0.6 & 0.6 - 1.0 \\
\hline \hline
         0 - 0.025 & 21.2085 &  3.2886 & -2.3019 & -1.4152 & -0.3826 & -0.0771 & -0.0201 & -0.0020 & -0.0013 & -0.0006 \\
      0.025 - 0.05 &  3.2886 & 16.5632 &  6.8748 &  0.1440 & -1.0024 & -0.5313 & -0.1513 & -0.0121 & -0.0039 & -0.0005 \\
      0.05 - 0.075 & -2.3019 &  6.8748 & 12.3855 &  6.6510 &  0.7848 & -0.7211 & -0.4276 & -0.0784 & -0.0072 & -0.0010 \\
       0.075 - 0.1 & -1.4152 &  0.1440 &  6.6510 &  7.8383 &  3.7387 &  0.3250 & -0.4746 & -0.2207 & -0.0297 & -0.0026 \\
        0.1 - 0.15 & -0.3826 & -1.0024 &  0.7848 &  3.7387 &  4.6694 &  2.3752 &  0.0629 & -0.3995 & -0.0611 & -0.0048 \\
        0.15 - 0.2 & -0.0771 & -0.5313 & -0.7211 &  0.3250 &  2.3752 &  3.7715 &  1.9818 & -0.3312 & -0.1747 & -0.0240 \\
         0.2 - 0.3 & -0.0201 & -0.1513 & -0.4276 & -0.4746 &  0.0629 &  1.9818 &  3.3407 &  1.0917 & -0.1886 & -0.0831 \\
         0.3 - 0.4 & -0.0020 & -0.0121 & -0.0784 & -0.2207 & -0.3995 & -0.3312 &  1.0917 &  3.2830 &  0.8365 & -0.0829 \\
         0.4 - 0.6 & -0.0013 & -0.0039 & -0.0072 & -0.0297 & -0.0611 & -0.1747 & -0.1886 &  0.8365 &  1.1897 &  0.2141 \\
         0.6 - 1.0 & -0.0006 & -0.0005 & -0.0010 & -0.0026 & -0.0048 & -0.0240 & -0.0831 & -0.0829 &  0.2141 &  0.5030 \\
    \end{tabular}
  \label{tablecov:C_Statistical_Q2}
\end{table}

\begin{table}[h!]
  \centering
  \caption{Systematic covariance matrix of the measured $d\sigma/dQ^{2}$ on C, in units of $10^{-78}$ (cm$^{2}$/(GeV/c)$^{2}$/C)$^{2}$.}
    \begin{tabular}{c||cccccccccc}
      Bin edges ([GeV/c]$^{2}$) & 0 - 0.025 & 0.025 - 0.05 & 0.05 - 0.075 & 0.075 - 0.1 & 0.1 - 0.15 & 0.15 - 0.2 & 0.2 - 0.3 & 0.3 - 0.4 & 0.4 - 0.6 & 0.6 - 1.0 \\
\hline \hline
         0 - 0.025 & 21.1782 & 20.8529 & 16.5509 & 13.0947 &  9.8807 &  7.2547 &  4.6254 &  1.1915 &  0.1300 & -0.0556 \\
      0.025 - 0.05 & 20.8529 & 21.5445 & 16.8088 & 12.9548 &  9.8018 &  7.3992 &  5.0006 &  1.4909 &  0.2115 &  0.0110 \\
      0.05 - 0.075 & 16.5509 & 16.8088 & 13.9511 & 11.3377 &  8.5919 &  6.4868 &  4.3881 &  1.7351 &  0.6126 &  0.1352 \\
       0.075 - 0.1 & 13.0947 & 12.9548 & 11.3377 &  9.8829 &  7.7727 &  6.0761 &  4.2276 &  2.0885 &  0.9718 &  0.2513 \\
        0.1 - 0.15 &  9.8807 &  9.8018 &  8.5919 &  7.7727 &  6.6324 &  5.6620 &  4.2262 &  2.2825 &  1.0771 &  0.2989 \\
        0.15 - 0.2 &  7.2547 &  7.3992 &  6.4868 &  6.0761 &  5.6620 &  5.4143 &  4.4157 &  2.5392 &  1.1839 &  0.3469 \\
         0.2 - 0.3 &  4.6254 &  5.0006 &  4.3881 &  4.2276 &  4.2262 &  4.4157 &  3.9978 &  2.5293 &  1.1517 &  0.3539 \\
         0.3 - 0.4 &  1.1915 &  1.4909 &  1.7351 &  2.0885 &  2.2825 &  2.5392 &  2.5293 &  2.2606 &  1.1786 &  0.3929 \\
         0.4 - 0.6 &  0.1300 &  0.2115 &  0.6126 &  0.9718 &  1.0771 &  1.1839 &  1.1517 &  1.1786 &  0.7394 &  0.2431 \\
         0.6 - 1.0 & -0.0556 &  0.0110 &  0.1352 &  0.2513 &  0.2989 &  0.3469 &  0.3539 &  0.3929 &  0.2431 &  0.0918 \\
    \end{tabular}
  \label{tablecov:C_Systematic_Q2}
\end{table}

\begin{table}[h!]
  \centering
  \caption{Measured cross section as a function of $Q^{2}$ on Fe, in units of
  $10^{-39}$ $\text{cm}^{2}$/[GeV/c]$^{2}$/$^{56}$Fe, and the absolute and
  fractional cross section uncertainties.}
    \begin{tabular}{c||c||c|c||c|c}
      Bin edges ([GeV/c]$^{2}$)  &  $d\sigma/dQ^{2}$  &  Abs. Stat. Unc. & Abs. Tot. Unc.  &  Frac. Stat. Unc.  &  Frac. Tot. Unc. \\
      \hline\hline
         0 - 0.025 & 62.420 & 7.342 & 14.558 & 0.118 & 0.233 \\
      0.025 - 0.05 & 60.968 & 5.938 & 12.758 & 0.097 & 0.209 \\
      0.05 - 0.075 & 46.116 & 5.092 & 10.833 & 0.110 & 0.235 \\
       0.075 - 0.1 & 32.553 & 4.390 & 9.614 & 0.135 & 0.295 \\
        0.1 - 0.15 & 20.223 & 3.681 & 8.842 & 0.182 & 0.437 \\
        0.15 - 0.2 & 12.698 & 3.209 & 7.940 & 0.253 & 0.625 \\
         0.2 - 0.3 & 9.747 & 2.727 & 5.854 & 0.280 & 0.601 \\
         0.3 - 0.4 & 8.747 & 2.402 & 3.957 & 0.275 & 0.452 \\
         0.4 - 0.6 & 5.949 & 1.779 & 2.728 & 0.299 & 0.459 \\
         0.6 - 1.0 & 1.995 & 1.040 & 1.374 & 0.521 & 0.689 \\
    \end{tabular}
  \label{table:Q2XSec_Fe}
\end{table}

\begin{table}[h!]
  \centering
  \caption{Statistical covariance matrix of the measured $d\sigma/dQ^{2}$ on Fe, in units of $10^{-78}$ (cm$^{2}$/(GeV/c)$^{2}$/Fe)$^{2}$.}
    %
  \label{tablecov:Fe_Statistical_Q2}
\end{table}

\begin{table}[h!]
  \centering
  \caption{Systematic covariance matrix of the measured $d\sigma/dQ^{2}$ on Fe, in units of $10^{-78}$ (cm$^{2}$/(GeV/c)$^{2}$/Fe)$^{2}$.}
    %
  \label{tablecov:Fe_Systematic_Q2}
\end{table}

\begin{table}[h!]
  \centering
  \caption{Measured cross section as a function of $Q^{2}$ on Pb, in units of
  $10^{-39}$ $\text{cm}^{2}$/[GeV/c]$^{2}$/$^{207}$Pb, and the absolute and
  fractional cross section uncertainties.}
    %
  \label{table:Q2XSec_Pb}
\end{table}

\begin{turnpage}
 \begin{table}[h!]
  \centering
  \caption{Statistical covariance matrix of the measured $d\sigma/dQ^{2}$ on Pb, in units of $10^{-78}$ (cm$^{2}$/(GeV/c)$^{2}$/Pb)$^{2}$.}
    %
  \label{tablecov:Pb_Statistical_Q2}
 \end{table}
\end{turnpage}

\begin{turnpage}
 \begin{table}[h!]
  \centering
  \caption{Systematic covariance matrix of the measured $d\sigma/dQ^{2}$ on Pb, in units of $10^{-78}$ (cm$^{2}$/(GeV/c)$^{2}$/Pb)$^{2}$.}
    %
  \label{tablecov:Pb_Systematic_Q2}
 \end{table}
\end{turnpage}

\begin{table}[h!]
  \centering
  \caption{Measured cross section as a function of $E_{\mu}$ on CH, in units of
  $10^{-39}$ $\text{cm}^{2}$/GeV/$^{12}$CH, and the absolute and
  fractional cross section uncertainties.}
    %
  \label{table:EmuXSec_CH}
\end{table}

\begin{table}[h!]
  \centering
  \caption{Measured cross section as a function of $E_{\mu}$ on C, in units of
  $10^{-39}$ $\text{cm}^{2}$/GeV/$^{12}$C, and the absolute and
  fractional cross section uncertainties.}
    %
  \label{table:EmuXSec_C}
\end{table}

\begin{table}[h!]
  \centering
  \caption{Measured cross section as a function of $E_{\mu}$ on Fe, in units of
  $10^{-39}$ $\text{cm}^{2}$/GeV/$^{56}$Fe, and the absolute and
  fractional cross section uncertainties.}
    %
  \label{table:EmuXSec_Fe}
\end{table}

\begin{table}[h!]
  \centering
  \caption{Measured cross section as a function of $E_{\mu}$ on Pb, in units of
  $10^{-39}$ $\text{cm}^{2}$/GeV/$^{207}$Pb, and the absolute and
  fractional cross section uncertainties.}
    %
  \label{table:EmuXSec_Pb}
\end{table}
\clearpage

\begin{turnpage}
 \begin{table}[h!]
  \centering
  \caption{Statistical covariance matrix of the measured $d\sigma/dE_{\mu}$ on CH, in units of $10^{-82}$ (cm$^{2}$/GeV/CH)$^{2}$.}
    %
  \label{tablecov:CH_Statistical_Emu}
  \end{table}
\end{turnpage}

\begin{turnpage}
 \begin{table}[h!]
  \centering
  \caption{Systematic covariance matrix of the measured $d\sigma/dE_{\mu}$ on CH, in units of $10^{-82}$ (cm$^{2}$/GeV/CH)$^{2}$.}
    %
  \label{tablecov:CH_Systematic_Emu}
  \end{table}
\end{turnpage}

\begin{turnpage}
 \begin{table}[h!]
  \centering
  \caption{Statistical covariance matrix of the measured $d\sigma/dE_{\mu}$ on C, in units of $10^{-82}$ (cm$^{2}$/GeV/C)$^{2}$.}
    %
  \label{tablecov:C_Statistical_Emu}
  \end{table}
\end{turnpage}

\begin{turnpage}
 \begin{table}[h!]
  \centering
  \caption{Systematic covariance matrix of the measured $d\sigma/dE_{\mu}$ on C, in units of $10^{-82}$ (cm$^{2}$/GeV/C)$^{2}$.}
    %
  \label{tablecov:C_Systematic_Emu}
  \end{table}
\end{turnpage}

\begin{turnpage}
 \begin{table}[h!]
  \centering
  \caption{Statistical covariance matrix of the measured $d\sigma/dE_{\mu}$ on Fe, in units of $10^{-82}$ (cm$^{2}$/GeV/Fe)$^{2}$.}
    %
  \label{tablecov:Fe_Statistical_Emu}
  \end{table}
\end{turnpage}

\begin{turnpage}
 \begin{table}[h!]
  \centering
  \caption{Systematic covariance matrix of the measured $d\sigma/dE_{\mu}$ on Fe, in units of $10^{-82}$ (cm$^{2}$/GeV/Fe)$^{2}$.}
    %
  \label{tablecov:Fe_Systematic_Emu}
  \end{table}
\end{turnpage}

\begin{turnpage}
 \begin{table}[h!]
  \centering
  \caption{Statistical covariance matrix of the measured $d\sigma/dE_{\mu}$ on Pb, in units of $10^{-82}$ (cm$^{2}$/GeV/Pb)$^{2}$.}
    %
  \label{tablecov:Pb_Statistical_Emu}
  \end{table}
\end{turnpage}

\begin{turnpage}
 \begin{table}[h!]
  \centering
  \caption{Systematic covariance matrix of the measured $d\sigma/dE_{\mu}$ on Pb, in units of $10^{-82}$ (cm$^{2}$/GeV/Pb)$^{2}$.}
    %
  \label{tablecov:Pb_Systematic_Emu}
  \end{table}
\end{turnpage}

\begin{table}[h!]
  \centering
  \caption{Measured cross section as a function of $\theta_{\mu}$ on CH, in units of
  $10^{-41}$ $\text{cm}^{2}$/Degrees/$^{12}$CH, and the absolute and
  fractional cross section uncertainties.}
    %
  \label{table:ThetaMuXSec_CH}
\end{table}

\begin{table}[h!]
  \centering
  \caption{Statistical covariance matrix of the measured $d\sigma/d\theta_{\mu}$ on CH, in units of $10^{-82}$ (cm$^{2}$/Degrees/CH)$^{2}$.}
    %
  \label{tablecov:CH_Statistical_ThetaMu}
\end{table}

\begin{table}[h!]
  \centering
  \caption{Systematic covariance matrix of the measured $d\sigma/d\theta_{\mu}$ on CH, in units of $10^{-82}$ (cm$^{2}$/Degrees/CH)$^{2}$.}
    %
  \label{tablecov:CH_Systematic_ThetaMu}
\end{table}

\begin{table}[h!]
  \centering
  \caption{Measured cross section as a function of $\theta_{\mu}$ on C, in units of
  $10^{-41}$ $\text{cm}^{2}$/Degrees/$^{12}$C, and the absolute and
  fractional cross section uncertainties.}
    %
  \label{table:ThetaMuXSec_C}
\end{table}

\begin{table}[h!]
  \centering
  \caption{Statistical covariance matrix of the measured $d\sigma/d\theta_{\mu}$ on C, in units of $10^{-82}$ (cm$^{2}$/Degrees/C)$^{2}$.}
    %
  \label{tablecov:C_Statistical_ThetaMu}
\end{table}

\begin{table}[h!]
  \centering
  \caption{Systematic covariance matrix of the measured $d\sigma/d\theta_{\mu}$ on C, in units of $10^{-82}$ (cm$^{2}$/Degrees/C)$^{2}$.}
    %
  \label{tablecov:C_Systematic_ThetaMu}
\end{table}

\begin{table}[h!]
  \centering
  \caption{Measured cross section as a function of $\theta_{\mu}$ on Fe, in units of
  $10^{-41}$ $\text{cm}^{2}$/Degrees/$^{56}$Fe, and the absolute and
  fractional cross section uncertainties.}
    %
  \label{table:ThetaMuXSec_Fe}
\end{table}

\begin{table}[h!]
  \centering
  \caption{Statistical covariance matrix of the measured $d\sigma/d\theta_{\mu}$ on Fe, in units of $10^{-82}$ (cm$^{2}$/Degrees/Fe)$^{2}$.}
    %
  \label{tablecov:Fe_Statistical_ThetaMu}
\end{table}

\begin{table}[h!]
  \centering
  \caption{Systematic covariance matrix of the measured $d\sigma/d\theta_{\mu}$ on Fe, in units of $10^{-82}$ (cm$^{2}$/Degrees/Fe)$^{2}$.}
    %
  \label{tablecov:Fe_Systematic_ThetaMu}
\end{table}

\begin{table}[h!]
  \centering
  \caption{Measured cross section as a function of $\theta_{\mu}$ on Pb, in units of
  $10^{-41}$ $\text{cm}^{2}$/Degrees/$^{207}$Pb, and the absolute and
  fractional cross section uncertainties.}
    %
  \label{table:ThetaMuXSec_Pb}
\end{table}

\begin{table}[h!]
  \centering
  \caption{Statistical covariance matrix of the measured $d\sigma/d\theta_{\mu}$ on Pb, in units of $10^{-82}$ (cm$^{2}$/Degrees/Pb)$^{2}$.}
    %
  \label{tablecov:Pb_Statistical_ThetaMu}
\end{table}

\begin{table}[h!]
  \centering
  \caption{Systematic covariance matrix of the measured $d\sigma/d\theta_{\mu}$ on Pb, in units of $10^{-82}$ (cm$^{2}$/Degrees/Pb)$^{2}$.}
    %
  \label{tablecov:Pb_Systematic_ThetaMu}
\end{table}

\clearpage
\pagebreak[4]

\section{Tables of cross section ratios}
\label{section:xsecRatiosTables}

\begin{table}[h!]
  \centering
  \caption{Measured cross section ratio of C with respect to CH as a function
  of $E_{\nu}$,  and the absolute and fractional uncertainties.}
    %
  \label{table:EnuXSecRatio_C}
\end{table}

\begin{table}[h!]
  \centering
  \caption{Statistical covariance matrix of the measured $\sigma(E_{\nu})_{C}/\sigma(E_{\nu})_{CH}$}
    %
  \label{tablecov:C_Statistical_Enu}
\end{table}

\begin{table}[h!]
  \centering
  \caption{Systematic covariance matrix of the measured $\sigma(E_{\nu})_{C}/\sigma(E_{\nu})_{CH}$}
    %
  \label{tablecov:C_Systematic_Enu}
\end{table}

\begin{table}[h!]
  \centering
  \caption{Measured cross section ratio of Fe with respect to CH as a function
  of $E_{\nu}$,  and the absolute and fractional uncertainties.}
    %
  \label{table:EnuXSecRatio_Fe}
\end{table}

\begin{table}[h!]
  \centering
  \caption{Statistical covariance matrix of the measured $\sigma(E_{\nu})_{Fe}/\sigma(E_{\nu})_{CH}$}
    %
  \label{tablecov:Fe_Statistical_Enu}
\end{table}

\begin{table}[h!]
  \centering
  \caption{Systematic covariance matrix of the measured $\sigma(E_{\nu})_{Fe}/\sigma(E_{\nu})_{CH}$}
    %
  \label{tablecov:Fe_Systematic_Enu}
\end{table}

\begin{table}[h!]
  \centering
  \caption{Measured cross section ratio of Pb with respect to CH as a function
  of $E_{\nu}$,  and the absolute and fractional uncertainties.}
    %
  \label{table:EnuXSecRatio_Pb}
\end{table}

\begin{table}[h!]
  \centering
  \caption{Statistical covariance matrix of the measured $\sigma(E_{\nu})_{Pb}/\sigma(E_{\nu})_{CH}$}
    %
  \label{tablecov:Pb_Statistical_Enu}
\end{table}

\begin{table}[h!]
  \centering
  \caption{Systematic covariance matrix of the measured $\sigma(E_{\nu})_{Pb}/\sigma(E_{\nu})_{CH}$}
    %
  \label{tablecov:Pb_Systematic_Enu}
\end{table}

\begin{table}[h!]
  \centering
  \caption{Measured cross section ratio of C with respect to CH as a function
  of $E_{\pi}$,  and the absolute and fractional uncertainties.}
    %
  \label{table:EpiXSecRatio_C}
\end{table}

\begin{table}[h!]
  \centering
  \caption{Statistical covariance matrix of the measured $\frac{d\sigma_{C}}{dE_{\pi}}/\frac{d\sigma_{CH}}{dE_{\pi}}$}
    %
  \label{tablecov:C_Statistical_Epi}
\end{table}

\begin{table}[h!]
  \centering
  \caption{Systematic covariance matrix of the measured $\frac{d\sigma_{C}}{dE_{\pi}}/\frac{d\sigma_{CH}}{dE_{\pi}}$}
    %
  \label{tablecov:C_Systematic_Epi}
\end{table}

\begin{table}[h!]
  \centering
  \caption{Measured cross section ratio of Fe with respect to CH as a function
  of $E_{\pi}$,  and the absolute and fractional uncertainties.}
    %
  \label{table:EpiXSecRatio_Fe}
\end{table}

\begin{table}[h!]
  \centering
  \caption{Statistical covariance matrix of the measured $\frac{d\sigma_{Fe}}{dE_{\pi}}/\frac{d\sigma_{CH}}{dE_{\pi}}$}
    %
  \label{tablecov:Fe_Statistical_Epi}
\end{table}

\begin{table}[h!]
  \centering
  \caption{Systematic covariance matrix of the measured $\frac{d\sigma_{Fe}}{dE_{\pi}}/\frac{d\sigma_{CH}}{dE_{\pi}}$}
    %
  \label{table:EpiXSecRatio_Pb}
\end{table}

\begin{table}[h!]
  \centering
  \caption{Statistical covariance matrix of the measured $\frac{d\sigma_{Pb}}{dE_{\pi}}/\frac{d\sigma_{CH}}{dE_{\pi}}$}
    %
  \label{table:ThetaPiXSecRatio_Pb}
\end{table}
\clearpage

\begin{turnpage}
 \begin{table}[h!]
  \centering
  \caption{Statistical covariance matrix of the measured $\frac{d\sigma_{C}}{d\theta_{\pi}}/\frac{d\sigma_{CH}}{d\theta_{\pi}}$}
    %
  \label{tablecov:Pb_Systematic_ThetaPi}
  \end{table}
\end{turnpage}

\begin{table}[h!]
  \centering
  \caption{Measured cross section ratio of C with respect to CH as a function
  of $Q^{2}$,  and the absolute and fractional uncertainties.}
    %
  \label{table:Q2XSecRatio_Pb}
\end{table}

\begin{turnpage}
 \begin{table}[h!]
  \centering
  \caption{Statistical covariance matrix of the measured $\frac{d\sigma_{Pb}}{dQ^{2}}/\frac{d\sigma_{CH}}{dQ^{2}}$}
    %
  \label{tablecov:Pb_Systematic_Q2}
 \end{table}
\end{turnpage}

\begin{table}[h!]
  \centering
  \caption{Measured cross section ratio of C with respect to CH as a function
  of $E_{\mu}$,  and the absolute and fractional uncertainties.}
    %
  \label{table:EmuXSecRatio_Pb}
\end{table}

\begin{turnpage}
 \begin{table}[h!]
  \centering
  \caption{Statistical covariance matrix of the measured $\frac{d\sigma_{C}}{dE_{\mu}}/\frac{d\sigma_{CH}}{dE_{\mu}}$}
    %
  \label{tablecov:Pb_Systematic_Emu}
  \end{table}
\end{turnpage}

\begin{table}[h!]
  \centering
  \caption{Measured cross section ratio of C with respect to CH as a function
  of $\theta_{\mu}$,  and the absolute and fractional uncertainties.}
    %
  \label{tablecov:Pb_Systematic_ThetaMu}
\end{table}


\begin{thebibliography}{00}
\bibitem{footnote2}
Diffractive scattering in H was not measured, and no correction
is applied for its contribution to the coherent cross section in 
CH because those events are removed by the vertex energy cut.
\end{thebibliography}
\end{document}